\documentclass[twocolumn,aps,prl,10pt,showpacs,dvips]{revtex4}
\usepackage{epsfig}
\begin{document}
\title{{\Large Analytical theory of mesoscopic Bose-Einstein condensation in ideal gas}}

\author{Vitaly V. Kocharovsky$^{1,2}$ and Vladimir V. Kocharovsky$^2$}

\affiliation{$^{1}$Department of Physics, Texas A\&M University, College Station, TX 77843-4242\\
$^{2}$Institute of Applied Physics, Russian Academy of Science,
603950 Nizhny Novgorod, Russia}

\date{\today}

\begin{abstract}
Paper is published in Phys. Rev. A {\bf 81}, 033615 (2010),  DOI: 10.1103/PhysRevA.81.033615

	We find the universal structure and scaling of the BEC statistics and thermodynamics (Gibbs free energy, average energy, heat capacity) for a mesoscopic canonical-ensemble ideal gas in a trap with arbitrary number of atoms, any volume, and any temperature, including the whole critical region. We identify a universal constraint-cut-off mechanism that makes BEC fluctuations strongly non-Gaussian and is responsible for all unusual critical phenomena of the BEC phase transition in the ideal gas. The main result is an analytical solution to the problem of critical phenomena. It is derived by, first, calculating analytically the universal probability distribution of the noncondensate occupation, or a Landau function, and then using it for the analytical calculation of the universal functions for the particular physical quantities via the exact formulas which express the constraint-cut-off mechanism. We find asymptotics of that analytical solution as well as its simple analytical approximations which describe the universal structure of the critical region in terms of the parabolic cylinder or confluent hypergeometric functions. The obtained results for the order parameter, all higher-order moments of BEC fluctuations, and thermodynamic quantities perfectly match the known asymptotics outside the critical region for both, low and high, temperature limits. 

	We suggest two-level and three-level trap models of BEC and find their exact solutions in terms of the cut-off negative binomial distribution (that tends to the cut-off gamma distribution in the continuous limit) and the confluent hypergeometric distribution, respectively. Also, we present an exactly solvable cut-off Gaussian model of BEC in a degenerate interacting gas. All these exact solutions confirm the universality and constraint-cut-off origin of the strongly non-Gaussian BEC statistics. We introduce a regular refinement scheme for the condensate statistics approximations on the basis of the infrared universality of higher-order cumulants and the method of superposition and show how to model BEC statistics in the actual traps. In particular, we find that the three-level trap model with matching the first four or five cumulants is enough to yield the remarkably accurate results for all interesting quantities in the whole critical region. We derive an exact multinomial expansion for the noncondensate occupation probability distribution and find its high temperature asymptotics (Poisson distribution) and corrections to it. Finally, we demonstrate that the critical exponents and a few known terms of the Taylor expansion of the universal functions, which were calculated previously from fitting the finite-size simulations within the phenomenological renormalization-group theory, can be easily obtained from the presented full analytical solutions for the mesoscopic BEC as certain approximations in the close vicinity of the critical point.
\end{abstract}

\maketitle

\section{I. THE PROBLEM OF MESOSCOPIC BEC AND UNIVERSAL STRUCTURE OF CRITICAL REGION}

    Statistical physics of Bose-Einstein condensation (BEC) in the mesoscopic, finite-size systems attracts a great interest in the recent years due to its intimate relation to still missing microscopic theory of BEC and other second order phase transitions as well as due to various modern experiments in the traps which contain usually a finite number of atoms $N \sim 10^2 \div 10^7$ (for a review, see \cite{PitString,Koch06} and references therein). One of the most important problems is to find the universal, common to the mesoscopic systems of any size, features in the behavior of an order parameter, its fluctuations, and thermodynamic quantities as the functions of the number of atoms in the trap and the temperature $T$ in the critical region, $T \sim T_c$ and $N \sim N_c$, as well as outside critical region. Finding the microscopic theory of fluctuations in the mesoscopic systems would yield a solution to the long-standing problem of the microscopic theory of critical phenomena in the phase transitions. In fact, the problem of finding the universal functions of the statistical and thermodynamic quantities in the critical region of BEC was not solved yet even for the ideal Bose gas which should be more than any interacting system amenable to analysis by rigorous analytical means. This is especially amazing since the problem itself is more than 80 years old and since after the works by Einstein, Bose, Gibbs, and many others (in particular, see early works on mesoscopic BEC \cite{London1938,Osborne1949,deGroot1950,Ziman1953,Pathria1974}) the BEC phase transition in the ideal gas is considered as a basic chapter of the statistical physics. It was studied by many authors and included literally in all textbooks on statistical physics, including the ones by Feynman \cite{Feynman}, Landau and Lifshitz \cite{LLV}, Abrikosov et al. \cite{AGD}, Pathria \cite{Pathria}, D. ter Haar \cite{Haar}, etc. A nice recent paper \cite{Kleinert2007} by Glaum, Kleinert, and Pelster clearly presents a modern status of this problem, including the problem of the structure of heat capacity near the $\lambda$-point as well as the first-quantized path-integral imaginary-time formalism, and, in particular, concludes that for the solution of this problem "analytical expression within the canonical ensemble could not be found, so we must be content with the numerical results". All previous attempts to solve the problem either gave a wrong answer, like the ones by Feynman \cite{Feynman} and by D. ter Haar \cite{Haar}, or failed to resolve the universal fine structure of the critical region, like a standard grand-canonical-ensemble approximation in the thermodynamic limit \cite{LLV,Pathria}, or did not arrive at the explicit analytical formulas for the universal functions in the whole critical region, like a phenomenological renormalization-group approach \cite{LLV,Fisher1974,Fisher1986,PatPokr,Kleinert1989,Gasparini}. In the present paper we obtain the exact simple analytical formulas for the universal functions of the statistical and thermodynamic quantities in the ideal gas in the whole critical region, including the universal structure of the heat capacity near the $\lambda$-point.  
    
    The modern theory of the second order phase transitions is based on the phenomenological renormalization-group approach and is focused on the calculation of the universal features of phase transitions for the macroscopic systems in the thermodynamic limit, such as the critical exponents, which are the same for all phase transitions within a given universality class (see reviews \cite{LLV,Fisher1974,Fisher1986,PatPokr,Kleinert1989,Gasparini} and references therein). However, whenever it comes to the specific calculations of critical exponents and universal functions for the particular models or systems, it uses Monte Carlo or other simulations for relatively small finite-size systems and fits that simulation data to some finite-size scaling ansatz (for the examples related to BEC see \cite{Gasparini,Pollock1992,Schultka1995,Ceperley1997,Holzmann1999,Svistunov2006,Campostrini2006,Wang2009}). Typically, the ansatz involves only one or a few first derivatives of the universal functions at the critical point. 
    
    Traditionally, in the whole statistical physics most studies were done for the macroscopic systems in the thermodynamic limit when both a volume $V$ and a number of atoms in the system tend to infinity \cite{PitString,Koch06,Fisher1974,Fisher1986,PatPokr,LLV,Kleinert1989,LL,AGD,Pathria,Ziff}. An opposite limit of a very few atoms in the trap ($N = 1, 2, 3,\ldots$) corresponds to a microscopic system studied by the methods of the standard quantum mechanics. 

     An intermediate case of a mesoscopic number of particles is the most difficult for it requires a solution that explicitly depends on the number of particles. Besides, for the mesoscopic systems an inapplicability of the standard in the statistical physics approaches, for example, a grand-canonical-ensemble method and a Beliaev-Popov diagram technique \cite{LL,AGD,Ziff,Shi}, becomes especially obvious, in particular, for the analysis of the anomalous fluctuations in the critical region. A simple example is given by a well-known grand-canonical catastrophe of the BEC fluctuations \cite{PitString,Koch06,Ziff}. In general, despite of its mathematical convenience, the grand-canonical-ensemble approximation, that was used starting from the very early works (see, e.g., \cite{London1938,Osborne1949,deGroot1950,Ziman1953,Pathria1974}), is not appropriate to describe the mesoscopic BEC phase transition neither in theory, nor in experiments \cite{Ziff,Balazs1998,PitString,Koch06,Sinner,Kleinert2007}. To get the solution of the BEC phase transition problem right, the most crucial issue is an exact account for a particle-number constraint ${\hat n}_{0} + \sum_{k\neq 0}{{\hat n}_{\vec{k}}} = N = const$ as an operator equation which is responsible for the very BEC phenomenon and is equivalent to an infinite set of the c-number constraints. It cannot be replaced by just one condition for the mean values, $N = {\bar n}_{0} + \bar n$, used in the grand-canonical-ensemble approach to specify an extra parameter, namely, a chemical potential $\mu$. Here ${\hat n}_{\vec{k}}$ is an occupation operator for a  $\vec{k}$-state of an atom in the trap and $\hat n = \sum_{k\neq 0}{{\hat n}_{\vec{k}}}$ is a total occupation of the excited states. 
     
     Thus, the problem is to find an explicit solution to a statistical problem of BEC for a finite number of atoms in the trap \textit{in a canonical ensemble}. Some results in this direction are known in the literature. However, a clear and full physical picture of the statistics and dynamics of BEC in the mesoscopic systems is absent until now not only in a general case of an interacting gas, but even in the case of an ideal gas (for a review, see e.g. \cite{Koch06,Ziff,Kleinert2007,Wang2009,Politzer,ww1997,Holthaus1997,Balazs1998,Holthaus1999,Wilkens2000,Baym2001,Sinner} and references therein). In particular, one of the most interesting in the statistical physics of BEC results, namely, a formula for the anomalously large variance of the ground-state occupation, $\left\langle \left({\hat n}_{0} - {\bar n}_{0}\right)^{2}\right\rangle \propto N^{4/3}(T/T_{c})^2$ , found both for the ideal gas \cite{Ziff,Dingle1949,Dingle1952,Dingle1973,Fraser1951,Reif,Hauge1969} and for the weakly interacting gas \cite{Pit98,KKS-PRL,KKS-PRA,Zwerger}, is valid only far enough from the critical point, where fluctuations of the order parameter are already relatively small, $\left\langle \left({\hat n}_0 - {\bar n}_{0}\right)^{2}\right\rangle \ll {\bar n}_{0}^2$. The same is relevant also to a known result on the analytical formula for all higher-order cumulants and moments of the BEC fluctuations, which demonstrates that the BEC fluctuations are essentially non-Gaussian even in the thermodynamic limit \cite{KKS-PRL,KKS-PRA}. Also, the probability distribution of the order parameter or the logarithm of that distribution, i.e. a Landau function \cite{Goldenfeld}, for the BEC in the ideal gas in the canonical ensemble was discussed in literature \cite{Koch06,ww1997,Holthaus1997,Balazs1998,KKS-PRL,KKS-PRA,Wilkens2000,Baym2001,Sinner}, however, its full analytical picture in the whole critical region was not found. Among fragments of that picture, we mention here a leading cubic term in the exponent of its asymptotics in the condensed phase correctly obtained in \cite{Sinner}. The universal structure of the mesoscopic BEC statistics in the ideal gas was found only recently \cite{KKD-RQE}, although the renormalization-group ansatz for the finite-size scaling variable both for the interacting and ideal gases was used earlier \cite{Pollock1992,Schultka1995,Ceperley1997,Holzmann1999,Svistunov2006,Campostrini2006,Wang2009,Baym2001}. In a whole, despite of the particular results, the problems of the origin, dynamics of formation, behavior, and universal scaling functions of the order parameter, moments of its fluctuations, and thermodynamic quantities for the mesoscopic system passing through the critical region remain open. 
     
     In the present paper we set forth a full analytical solution to this problem for the ideal gas. Namely, in the first part of the paper, we introduce a general method for the analysis of the second order phase transitions based on the universal constraint nonlinearity responsible for the phase transition through a reduction of the many-body Hilbert space (Sec. II). In Sec. III, we derive an exact multinomial expansion for the noncondensate occupation probability distribution which is especially useful for the analysis of the subtle nonuniversal finite-size effects. Then, in Sec. IV, we calculate an unconstrained probability distribution of the number of atoms in the condensate $n_0$  that is complimentary to the total number of atoms in the excited states (noncondensate) $n = N - n_0$ and find analytical formulas for its universal structure and asymptotics as well as elaborate on the grand-canonical-ensemble approximation which implies a very simplified exponential distribution. In Sec. V, we explain a remarkable constraint-cut-off mechanism that makes BEC fluctuations strongly non-Gaussian in the critical region and gives an origin to the nonanaliticity and all unusual critical phenomena of the BEC phase transition in the ideal gas. In particular, we rigorously prove that the cut-off distribution is the exact solution to a well-known recursion relation. Thus, in Sections IV and V we find analytically the Landau function \cite{Goldenfeld,Sinner}, that is the logarithm of the probability distribution of the order parameter, which plays a part of an effective fluctuation Hamiltonian and, due to an absence of interatomic interaction in the ideal gas, is the actual Hamiltonian for the mesoscopic BEC in the ideal gas in the canonical ensemble. On this basis we find the universal scaling and structure of the order parameter (Sec. VI) and all higher-order moments and cumulants (Sec. VII) of the BEC statistics for any number of atoms trapped in a box with any volume $V = L^3$ and temperature. We prove that our results perfectly match the known values of the statistical moments in the low-temperature region ($T \ll T_c$), where there is a well developed condensate, \cite{KKS-PRL,KKS-PRA} and their known asymptotics in the high-temperature region ($T \gg T_c$), where there is no condensate \cite{PitString,Pathria,Ziff}. 
     
          In the second part of the paper, we present the exactly solvable ideal gas models which allow us to study statistics of mesoscopic BEC in all details and to compare it with the predicted in \cite{KKD-RQE} and in Sections IV-VII universal behavior. First, we describe an exactly solvable Gaussian model that allows us to demonstrate both its insufficiency for an accurate description of BEC statistics as well as the universality and constraint-cut-off origin of the strongly non-Gaussian BEC statistics. Namely, we demonstrate that the constraint-cut-off mechanism does yield the strongly non-Gaussian BEC fluctuations, similar to the ones found for the ideal gas in the box, even if one employs a pure Gaussian model for an unconstrained probability distribution of the noncondensate occupation that corresponds to an exactly solvable model of BEC in a degenerate interacting gas (Sec. VIII). Then we introduce the two-level (Sec. IX) and three-level (Sec. X) trap models of BEC which can be used as the basic blocks in the theory of BEC and have an analogy with very successful two-level and three-level atom models in quantum optics. Namely, we consider the two- and three-energy-level traps with arbitrary degeneracy of the upper level(s) and find their analytical solutions for the condensate statistics in a mesoscopic ideal gas with arbitrary number of atoms and any temperature, including a critical region. The solution of the two-level trap model is a cut-off negative binomial distribution that tends to a cut-off gamma distribution in the thermodynamic limit. In particular, we demonstrate that a quasithermal ansatz, suggested in \cite{CNBII}, is a solution for some effective two-level trap and, thus, we explain why and to what extent it gives a good approximation for real traps. The solution of the three-level trap model is given via a confluent hypergeometric distribution. We compare the results of all these models against BEC statistics in an actual box trap. In Sec. XI, we introduce a regular refinement scheme for the condensate statistics approximations based on the infrared universality \cite{Koch06,KKS-PRL,KKS-PRA} of higher-order cumulants and the method of superposition. Remarkably, we find that a superposition of the two-level trap model with shifted average (Pirson distribution of the III type) and the Gaussian model (Sec. XI) yields the same universal statistics in the critical region as the three-level trap model with matching the first four cumulants (Sec. X). These two models as well as the three-level trap model with the shifted average are enough to yield the remarkably accurate results for all interesting quantities in the whole critical region. Finally, we obtain the thermodynamic-limit asymptotics for these exact solutions in terms of the parabolic cylinder and confluent hypergeometric functions and, thus, find a remarkably simple analytical solution to the problem of the universal structure of the critical region.
          
          In the third part of the paper, on the basis of the developed analytical theory of BEC statistics, we find universal scaling, structure, and asymptotics of the main thermodynamic quantities of the mesoscopic ideal gas in the canonical ensemble, such as a Gibbs free energy (Sec. XII), an average energy (Sec. XIII) and a heat capacity (Sec. XIV). Finally, in Sec. XV, we demonstrate that the critical exponents and a few first terms of the Taylor expansion for the universal functions, which were calculated previously from fitting the finite-size simulations within the phenomenological renormalization-group theory, follow from the presented full analytical solutions of the mesoscopic BEC as certain approximations in the vicinity of the critical point. In Sec. XVI, we conclude with a general discussion of the obtained results. 
          
     In order to present the universal scaling of the BEC statistics and its constraint-cut-off origin as clear and visual as possible, we consider here only the case of the ideal gas. An important motivation for the publication of this work is the fact that a more involved problem of critical fluctuations in a weakly interacting gas can be solved on the basis of the same method and in terms of the same functions, as we use here for the ideal gas, since the constraint-cut-off origin of critical behavior in the second order phase transitions is universal. In other words, a unique property and relative simplicity of the mesoscopic ideal gas system, that demonstrates phase transition even without interparticle interaction, allows us to find the analytical solution and the universal constraint-cut-off origin for the nonanaliticity and critical phenomena in all other interacting systems demonstrating second order phase transition. Of course, in addition one has to take into account a deformation of the statistical distribution due to a feedback of the order parameter on the quasiparticle energy spectrum and interparticle correlations, that can be done in a non-perturbative-in-fluctuations way using a theorem on the nonpolynomial averages in statistical physics and appropriate diagram technique \cite{KochLasPhys2007,KochJMO2007}. The solution for the weakly interacting gas will be presented in a separate paper. 

\section{II. CONSTRAINT NONLINEARITY AND 
MANY-BODY FOCK SPACE CUT OFF 
IN THE CANONICAL ENSEMBLE}

Let us consider an equilibrium ideal gas of $N$ Bose atoms trapped in a cubic box with the periodic boundary conditions and discrete one-particle energies $\epsilon_{\vec{k}} = \hbar^2 k^{2}/(2m)$, where $m$ is a mass of an atom and $\vec{k} = 2\pi \vec{q}/L$ is a wavevector with the components $k_i = 2\pi q_{i}/L, q_i = 0, \pm 1, \pm 2, \ldots$. This mesoscopic system is described by a Hamiltonian $H = \sum^{\infty}_{k=0}{\epsilon_{\vec{k}} {\hat n}_{\vec{k}}}$. As we discussed in \cite{KochLasPhys2007,KochJMO2007}, the only reason for the BEC of atoms on the ground state $\vec{k} = 0$ is the conservation of the total number of the Bose particles in the trap, $N = {\hat n}_0 + \hat n$. Hence, the occupation operators ${\hat n}_{\vec{k}}$ are not independent and the many-body Hilbert space is strongly constrained. A more convenient equivalent formulation of the problem can be given if one introduces a constraint nonlinearity in the dynamics and statistics, even for the ideal, noninteracting gas, on the basis of the particle number constraint. In most previous studies, an actual (e.g., canonical or microcanonical) quantum ensemble was substituted by an artificial grand canonical ensemble, where only the mean number of particles is fixed by the appropriate choice of the chemical potential $\mu$ and, therefore, most quantum effects in dynamics and statistics of BEC were lost or misunderstood. 

     Following our general approach \cite{Koch06,KKS-PRL,KKS-PRA,KochLasPhys2007,KochJMO2007}, we solve for the constraint from the very beginning through the proper reduction of the many-body Hilbert space. In the present case of the ideal gas in the canonical ensemble, we have to consider as independent only noncondensate Fock states $\left| \left\{ n_{\vec{k}}, \vec{k} \neq 0\right\} \right\rangle$  which uniquely specify the ground-state Fock state $\left| n_0 = N - \sum_{k\neq 0}{n_{\vec{k}}}\right\rangle$. However, the remaining noncondensate Fock space should be further cut off by the boundary $\sum_{k\neq 0}{n_{\vec{k}}} \leq N$. The latter is equivalent to an introduction of a step-function $\theta (N-{\hat n})$, i.e., 1 if $n \leq N$ or 0 if $n > N$, in all operator equations and under all trace operations. That $\theta (N-{\hat n})$ factor is the constraint nonlinearity that, being accepted, allows us to consider the noncondensate many-body Fock space formally as an unconstrained one. We adopt this point of view from now on. 
     
     The BEC fluctuations are described by a "mirror" image $\rho_{cond}(n_{0}) = \rho_{n=N-n_0}$ of the probability distribution of the total number of excited (noncondensed) atoms $n$, 
\begin{equation}
\rho_n = \frac{1}{2\pi} \int_{-\pi}^{\pi} e^{-iun} \Theta (u)du ,
\label{rho}
\end{equation}
\begin{equation}
\Theta (u) = Tr\left\{e^{iun}{\hat \rho}\theta (N-{\hat n})\right\} = \sum^{N}_{n=0}{e^{iun}\rho_{n}},
\label{Thu}
\end{equation}
where $\Theta (u)$ is a characteristic function for the stochastic variable $n$, 
\begin{equation}
\hat \rho = e^{-H/T}\theta (N-{\hat n})/Z  
\label{hatrho}
\end{equation}
is an equilibrium density matrix,
\begin{equation}
Z = Tr\left\{e^{-H/T}\theta (N-{\hat n})\right\} 
\label{Z}
\end{equation}
is a partition function, and the temperature is measured in the energy units, so that the Boltzmann constant is set to be 1. Thus, the exact solution for an actual probability distribution of the total noncondensate occupation in the canonical ensemble 
\begin{equation}
\rho_n = \rho_{n}^{(\infty)} \theta (N-n)/\sum_{n=0}^{N}{\rho^{(\infty)}_{n}}
\label{rhocut}
\end{equation}
is merely a $\theta (N-{\hat n})$ cut off of the unconstrained probability distribution $\rho^{(\infty)}_{n}$ for an infinite interval of the noncondensate occupations $n \in \left[ 0, \infty \right)$, as is shown in Fig. 1. In other words, for the ideal Bose gas in the mesoscopic trap in the canonical ensemble the Landau function \cite{Goldenfeld,Sinner}, $-\ln \rho_n$, that is the effective fluctuation Hamiltonian, has an infinite potential wall at $n=N$ due to the constraint nonlinearity.  

     The unconstrained probability distribution 
\begin{equation}   
\rho^{(\infty)}_{n} = \frac{1}{2\pi} \int_{-\pi}^{\pi} e^{-iun} \Theta^{(\infty)} (u)du
\label{rhoinfinity}
\end{equation}
was analytically calculated in \cite{KKS-PRL,KKS-PRA} for arbitrary trap via its characteristic function 
\begin{equation}
\Theta^{(\infty)} (u) = \prod_{\vec{k}\neq 0}{\left(e^{\epsilon_{\vec{k}}/T} - 1\right)/\left(e^{\epsilon_{\vec{k}}/T} - e^{iu}\right)} ,
\label{char}
\end{equation}
i.e. via all its moments and cumulants. Thus, only a straightforward calculation of the moments and cumulants of the cut off probability distribution, given in Eq.~(\ref{rhocut}) and depicted as a curve OAN in Fig. 1, remains to fulfil in order to find the actual BEC statistics in the mesoscopic system for all numbers of atoms and temperatures, including a critical region. 

\begin{figure}
\center{\epsfig{file=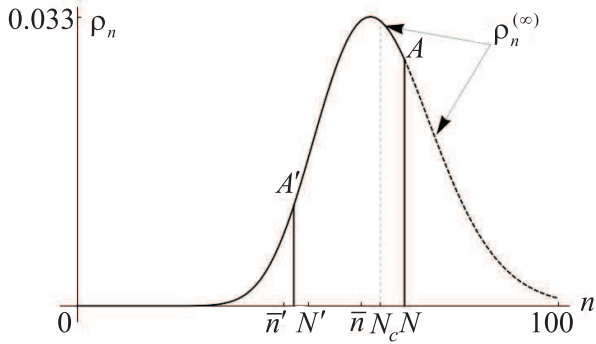,width=8cm}}
\caption{Unconstrained probability distribution $\rho_{n}^{(\infty)}$ of the noncondensate occupation $n$ (dashed line) and its constraint cuts off (solid lines) for a small number of atoms $N' < N_c$ (OA'N' - there is no condensate) and for a large number of atoms $N > N_c$ (OAN - there is condensate) for the trap with a given volume and temperature: the trap-size parameter in Eq.~(\ref{Nv}) is $N_v = 100$.}
\end{figure}

     The solution in Eq.~(\ref{rhocut}) is amazingly simple, powerful, and exact. It allows us to solve the problem of critical fluctuations in the mesoscopic BEC in the ideal gas. The rest of the paper is devoted to a detailed analysis of that solution. 
     
     The Taylor expansion of the characteristic function, $\Theta (u) = \sum_{m=0}^{\infty}{\alpha_m u^m /m!}$, gives the initial moments $\alpha_m$. They are related to the central moments $\mu_m =\left\langle (n - {\bar n})^m \right\rangle$ as well as to the cumulants $\kappa_m$ by the following explicit formulas \cite{a}: 
\begin{equation}
\alpha_r = \sum_{k=0}^{r}{C_{r}^{k} \mu_{r-k} {\bar n}^k} , \qquad \mu_r = \sum_{k=0}^{r}{(-1)^k C_{r}^{k} \alpha_{r-k} {\bar n}^k} ,
\label{moments}
\end{equation}
$$\kappa_r = \sum_{m=1}^{r}{\frac{(m-1)!}{(-1)^{m-1}}\sum^{(r,m)}{(r;a_{1},\ldots, a_{r})' \alpha_{1}^{a_1} \ldots \alpha_{r}^{a_r}}},$$
\begin{equation}
\alpha_m = \sum_{r=1}^{m}{\sum^{(m,r)}{(m;a_{1}, \ldots , a_{m})' \kappa_{1}^{a_1} \ldots \kappa_{m}^{a_m}}} ,
\label{cumulants}
\end{equation}
where $C_{r}^{k} = r!/(k!(r-k)!)$, $(m;a_{1}, \ldots , a_{m})' = m!/\left((1!)^{a_1}a_{1}!(2!)^{a_2}a_{2}! \ldots (m!)^{a_m}a_{m}!\right)$ is a multinomial coefficient, and the sum $\sum^{(m,r)}$ in Eq.~(\ref{cumulants}) runs over the nonnegative integers $a_{1}, \ldots ,a_{r}$ which satisfy the following two conditions: $a_1 + 2a_2 + \ldots +ra_r = r$ and $a_1 + \ldots + a_r = m$. The cumulants $\kappa_m$ and the generating cumulants $\tilde{\kappa}_m$ are determined by the Taylor expansion of the logarithm of the characteristic function, 
\begin{equation}
\ln \Theta (u) = \sum_{m=1}^{\infty}{\kappa_m \frac{(iu)^{m}}{m!}} = \sum_{m=1}^{\infty}{\tilde{\kappa}_m \frac{(e^{iu} - 1)^{m}}{m!}}. 
\label{lnTh}
\end{equation}
They are related by means of the Stirling numbers of the first and second kinds \cite{a}, 
\begin{equation}
\tilde{\kappa}_m = \sum_{r=1}^{m}{S_{m}^{(r)}\kappa_r}; \quad \kappa_r = \sum_{m=1}^{r}{\sigma_{r}^{(m)}\tilde{\kappa}_m};
\label{cgc}
\end{equation}
$$\sigma_{r}^{(m)} = \frac{1}{m!}\sum_{k=0}^{m}{(-1)^{m-k}C_{m}^{k}k^r} ,$$
\begin{equation}
S_{m}^{(r)} = \sum_{k=0}^{m-r}{(-1)^{k}C_{m-1+k}^{m-r+k}C_{2m-r}^{m-r-k}\sigma_{m-r+k}^{(k)}};
\label{sS}
\end{equation}
in particular, $\kappa_1 = \tilde{\kappa}_1 , \quad \kappa_2 = \tilde{\kappa}_2 + \tilde{\kappa}_1$, $\kappa_3 = \tilde{\kappa}_3 + 3
\tilde{\kappa}_2 + \tilde{\kappa}_1$, $\kappa_4 = \tilde{\kappa}_4 + 6 \tilde{\kappa}_3 + 7 \tilde{\kappa}_2 + \tilde{\kappa}_1$. The generating cumulants of $\rho_{n}^{(\infty)}$ for the ideal gas in arbitrary trap are known \cite{KKS-PRL,KKS-PRA} 
\begin{equation}
\tilde{\kappa}_{m}^{(\infty)} = (m-1)! \sum_{\vec{k}\neq 0}{\left(e^{\epsilon_{\vec{k}}/T} - 1\right)^{-m}} .
\label{igcumulants}
\end{equation}
 
     We calculate the quantities which are the most important and convenient for the analysis of the BEC statistics, namely, the mean value $\bar n$ (which is complimentary to the BEC order parameter ${\bar n}_0 = N - \bar n$) as well as the central moments $\mu_m$ and cumulants $\kappa_m$ of the total noncondensate occupation. The point is that the ground-state (condensate) occupation fluctuates complimentary to a sum of many, to a large extent independent occupations of the excited states in the noncondensate, conditioned by the particle-number constraint. The first four cumulants are related to the central moments as follows $\kappa_1 = {\bar n}, \kappa_2 = \mu_{2}, \kappa_3 = \mu_{3}, \kappa_4 = \mu_{4} - 3\mu_{2}^2$. The central moments of the condensate fluctuations differ from the corresponding central moments of the noncondensate fluctuations only by the sign for the odd orders, $\left\langle (n_0 - {\bar n}_{0})^{m}\right\rangle = (-1)^{m}\left\langle (n-{\bar n})^{m}\right\rangle$.

\section{III. MULTINOMIAL EXPANSION FOR THE NONCONDENSATE OCCUPATION PROBABILITY DISTRIBUTION}

    On the basis of the formulated above statistical constraint-cut-off approach, let us start the analysis with a derivation of important general expansion for the noncondensate occupation probability distribution that will be especially useful for the analysis of finite-size effects in the small number of atoms region where a discreteness of the noncondensate occupation $n$ is essential (in particular, see the end of Sec. IV and the end of Sec. XIV). The definition in Eq.~(\ref{Thu}) means that the probability $\rho_n$ to find $n$ excited atoms in the noncondensate is equal to the n-th coefficient in the Taylor series of the characteristic function $\tilde{\Theta}(z)=\Theta (u)$ viewed as a function of a complex variable $z=e^{iu}$, namely,
\begin{equation}
\rho_n = \frac{\tilde{\Theta}^{(n)}(z=0)}{n!}, \Theta (u)=\tilde{\Theta} (z)=\sum_{n=0}^{N}{\frac{\tilde{\Theta}^{(n)}(z=0)}{n!}z^n},
\label{Taylor}
\end{equation}    
where $\tilde{\Theta}^{(n)}=d^{n}\tilde{\Theta}/dz^n$. The same is true for the unconstrained probability 
$$\rho_{n}^{(\infty)} = \frac{1}{n!} \tilde{\Theta}^{(\infty)(n)}(z=0),$$
\begin{equation}
\Theta^{(\infty)} (u)=\tilde{\Theta}^{(\infty)} (z)=\sum_{n=0}^{\infty}{\frac{1}{n!} \tilde{\Theta}^{(\infty)(n)}(z=0)z^n}.
\label{Taylor1}
\end{equation}

     The latter Taylor series for the unconstrained characteristic function in Eq.~(\ref{char}) can be evaluated as follows:
$$\tilde{\Theta}^{(\infty)}(z)=\exp \left\{\sum_{\vec{k}\neq 0}{\left[\ln \left(1-e^{-\frac{\epsilon_{\vec{k}}}{T}}\right)-\ln \left(1-ze^{-\frac{\epsilon_{\vec{k}}}{T}}\right)\right]}\right\}$$
\begin{equation}
=\rho_{n=0}^{(\infty)}\exp \left(\sum_{n=1}^{\infty}\frac{B_n}{n}z^{n}\right), \quad B_n = \sum_{\vec{k}\neq 0}{e^{-n\epsilon_{\vec{k}}/T}}.
\label{Taylor2}
\end{equation}
Here the unconstrained probability to find zero atoms in the noncondensate is equal to 
\begin{equation}
\rho_{n=0}^{(\infty)}=\exp \left[\sum_{\vec{k}\neq 0}{\ln \left(1-e^{-\epsilon_{\vec{k}}/T}\right)}\right] \to \exp\left[-\frac{\zeta(5/2)N_{v}}{\zeta(3/2)}\right],
\label{rho0}
\end{equation}
where we give also its thermodynamic-limit value at $N_{v} \to \infty$. Hence, Eqs.~(\ref{Taylor1}) and (\ref{Taylor2}) yield 
\begin{equation}
\rho_{n}^{(\infty)} = \frac{\rho_{n=0}^{(\infty)}}{n!}\left[\frac{d^n}{dz^n}\exp \left(\sum_{n=1}^{\infty}{\frac{B_n}{n}z^n}\right)\right]_{z=0}.
\label{rhon}
\end{equation}

     Finally, using a generating function \cite{a} 
\begin{equation}
\frac{1}{m!}\left[\sum_{k=1}^{\infty}{\frac{x_k}{k}t^{k}}\right]^m = \sum_{n=m}^{\infty}{\frac{t^n}{n!}\sum^{(m,n)}{(n;a_{1},\ldots,a_{n})^{*}x_{1}^{a_1}\ldots x_{n}^{a_n}}}
\label{MGF}
\end{equation}
for the well-known multinomial coefficients 
\begin{equation}
(n;a_{1},a_{2},\ldots,a_{n})^{*} = n!/\left(1^{a_1}a_{1}!2^{a_2}a_{2}!\ldots n^{a_n}a_{n}!\right)
\label{multinom}
\end{equation}
in the Taylor expansion of the exponential function 
$$\exp \left(\sum_{n=1}^{\infty}{B_{n}z^{n}/n}\right) = \sum_{m=0}^{\infty}{\frac{1}{m!}\left(\sum_{n=1}^{\infty}{B_{n}z^{n}/n}\right)^{m}}$$
in Eq.~(\ref{rhon}), we obtain very powerful and exact multinomial expansion for the unconstrained probability distribution of the total noncondensate occupation 
\begin{equation}
\rho_{n}^{(\infty)} = \frac{\rho_{n=0}^{(\infty)}}{n!}\sum_{m=0}^{n}{\sum^{(m,n)}{(n;a_{1},a_{2},\ldots,a_{n})^{*}B_{1}^{a_1}B_{2}^{a_2}\ldots B_{n}^{a_n}}},
\label{multinomrho}
\end{equation}
where the sum $\sum^{(m,n)}$ runs over all nonnegative integers $a_{1},\ldots,a_{n}$ which satisfy the following two conditions: $a_{1}+a_{2}+\ldots+a_{n}=m$ and $a_{1}+2a_{2}+\ldots+na_{n}=n$. In particular, it immediately yields the unconstrained probability to find one atom in the noncondensate 
\begin{equation}
\rho_{n=1}^{(\infty)} = \rho_{n=0}^{(\infty)}B_1 \quad \to  \frac{N_v}{\zeta(3/2)}\exp\left[-\frac{\zeta(5/2)N_{v}}{\zeta(3/2)}\right],
\label{rho1}
\end{equation}
where we again give also its thermodynamic-limit value.

\section{IV. UNIVERSAL STRUCTURE OF THE UNCONSTRAINED PROBABILITY DISTRIBUTION OF THE NONCONDENSATE OCCUPATION}

The best way to analyze BEC statistics in the mesoscopic systems is to study the central moments and cumulants of the noncondensate occupation as the functions of the number of atoms in the trap since these functions are more physically instructive and more directly related to the intrinsic quantum statistics in a finite system than less transparent temperature dependences. The maximum number of the noncondensed atoms ${\bar n}^{(\infty)} = N_c$ is achieved in the limit of an infinite number of atoms loaded in the trap, $N \to \infty$, and is given by a discrete sum \cite{KKS-PRL,KKS-PRA} 
\begin{equation}
N_c = \sum_{\vec{k}\neq 0}{\left(e^{\epsilon_{\vec{k}}/T} - 1\right)^{-1}} .
\label{Nc}
\end{equation}
In the standard analysis in the thermodynamic limit this sum is approximated by a continuous integral that yields a little bit larger number 
\begin{equation}
N_v = \zeta(3/2)\left(\frac{mT}{2\pi \hbar^2}\right)^{3/2} V = N \left(\frac{T}{T_c}\right)^{3/2} ,
\label{Nv}
\end{equation}
where $\zeta$ is the zeta function of Riemann, $\zeta(3/2) \approx 2.612$. Let us note also that a ratio of an energy scale for the box trap to the temperature is determined by precisely the same trap-size parameter $N_v$ in Eq.~(\ref{Nv}), namely, for the energy of the first excited state one has 
\begin{equation}
\epsilon_{1} /T = \pi \left[\zeta(3/2)/N_{v}\right]^{2/3} . 
\label{e1}
\end{equation}
Hence, the sum in Eq.~(\ref{Nc}) over the energy spectrum of the trap as well as all other similar sums, like the one in Eq.~(\ref{sigma}) below, actually depend only on a single combination of the trap parameters given by Eq.~(\ref{Nv}).  Thus, the mesoscopic system of the ideal gas atoms in the finite box is completely specified by two parameters, $N_v$ and $N$. It is convenient to study a development of the BEC phase transition with an increase of the number of atoms $N$ assuming that the volume and the temperature of the mesoscopic system are fixed, that is the trap-size parameter $N_v$ given by Eq.~(\ref{Nv}) is fixed. The critical number of the loaded in the trap atoms is equal to the close to $N_v$ number $N_c$ given by Eq.~(\ref{Nc}). Hence, when we increase the number of atoms from $N < N_c$ to $N > N_c$ the system undergoes the same BEC phase transition phenomenon as the one observed when we decrease the temperature around the critical temperature $T_c$ from $T > T_c$ to $T < T_c$. 

\subsection{A. Critical Region: \\
Universality of Critical BEC Fluctuations}

     Following the approach formulated in Sec. II and Fig. 1, we immediately find that the analytically calculated in \cite{KKS-PRL,KKS-PRA} unconstrained probability distribution $\rho_{n}^{(\infty)}$ for different sizes and temperatures of the trap tends, with an increase of the trap-size parameter $N_v$, to a universal function 
\begin{equation}
\rho^{(univ)}(x')=\frac{1}{2\pi i}\int_{-i \infty}^{i \infty}{e^{px}\Theta^{(univ)}(ip)dp}
\label{rhouniv}
\end{equation}
$$\equiv \frac{1}{2\pi}\int_{-\infty}^{\infty}{e^{-iu'x'}\Theta^{(univ)}(u')du'},$$
\begin{equation}
\Theta^{(univ)}(u') = \exp \left[\sum_{m=2}^{\infty}{\frac{s_m}{m}(iu')^{m}}\right] ,
\label{Thetauniv}
\end{equation}
if it is considered for the scaled stochastic variable centered to have zero mean value, 
\begin{equation}   
x' = (n - N_{c})\epsilon_{1}/T .
\label{xuniv}
\end{equation}
The corresponding argument of the characteristic function in Eq.~(\ref{Thetauniv}) is related to the one in Eqs.~(\ref{rho}), (\ref{char}), and (\ref{lnTh}) via the same but inverse factor, namely, $u' = uT/\epsilon_{1}$. The result in Eqs.~(\ref{rhouniv}) and (\ref{Thetauniv}) follows from the definitions in Eqs.~(\ref{rho}) and (\ref{lnTh}) in the thermodynamic limit $N_c \to \infty$, $\sigma^{(\infty)} \to \infty$ because $\tilde{\kappa}_{1}^{(\infty)} = N_c$ and in the thermodynamic limit all higher-order cumulants in Eq.~(\ref{igcumulants}) scale as the powers of the dispersion, $\tilde{\kappa}_{m}^{(\infty)} \to (m-1)! (\sigma^{(\infty)}/s_{2}^{1/2})^m s_m$, since the main contribution in Eq.~(\ref{igcumulants}) for $m\geq 2$ comes from the energies much lower than temperature, $\epsilon_{\vec{k}} \ll T$, where $(e^{\epsilon_{\vec{k}}/T}-1)^m \approx (\epsilon_{1}/T)^{m}\vec{k}^{2m} , \quad \vec{k} = 2\pi \vec{q}/L$, and vector $\vec{q}=\{q_{j}, j=x,y,z\}$ has integer components, $q_{j}=0,\pm 1,\pm 2,\ldots$. Here the universal numbers
\begin{equation}
s_m = \sum_{\vec{q}\neq 0}{\frac{1}{q^{2m}}}
\label{sm}
\end{equation}
are given by the generalized Einstein function (see \cite{KochPhysicaA2001} and Fig. 10 in Sec. XI) and the dispersion of the BEC fluctuations $\sigma^{(\infty)}$ is an independent on the number of atoms $N$ quantity calculated for the unconstrained probability distribution ($N \to \infty$) in \cite{KKS-PRL,KKS-PRA} as a function of the trap-size parameter $N_v$, 
\begin{equation}
\sigma^{(\infty)} \equiv \sqrt{\kappa_{2}^{(\infty)}}=\left[\sum_{\vec{k}\neq 0}{\frac{1}{(e^{\epsilon_{\vec{k}}/T}-1)^{2}}+N_{c}}\right]^{1/2}.
\label{sigma}
\end{equation}
In the thermodynamic limit the last discrete sum can be approximated as a continuous integral and one has 
\begin{equation}
\sigma^{(\infty)} \approx \frac{s_{2}^{1/2}N_{v}^{2/3}}{\pi \left[\zeta(3/2)\right]^{2/3}} = \frac{s_{2}^{1/2}T}{\epsilon_1} , 
\label{sigmainfinity}
\end{equation}
\begin{equation}
s_2 = \sum_{\vec{q}\neq 0}{\frac{1}{q^{4}}} \approx 16.533,\quad s_{3}=8.4019, s_{4}=6.9458.
\label{s234}
\end{equation}
The dispersion of the BEC fluctuations (Eqs.~(\ref{sigma}) and (\ref{sigmainfinity})) is anomalously large and scales as $\sigma^{(\infty)} \sim N_{v}^{2/3}$, contrary to a much smaller value $\sim N_{v}^{1/2}$, which one could naively expect from a standard analysis based on the grand-canonical or thermodynamic theory of fluctuations. 

\begin{figure}
\center{\epsfig{file=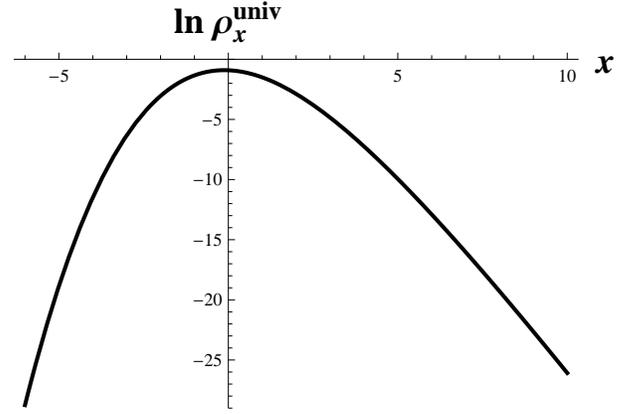,width=8cm}}
\caption{Logarithm of the universal unconstrained probability distribution $\rho_{x}^{(univ)}$ as a function of the scaled noncondensate occupation $x=(n-N_{c})/\sigma^{(\infty)}$.} 
\end{figure}

     Thus, we find the exact analytical formula for the universal unconstrained probability distribution $\rho_{x'}^{(univ)}$ in Eq.~(\ref{rhouniv}) via its Laplace transform, i.e. its characteristic function in Eq.~(\ref{Thetauniv}). It is presented in Fig. 2. In the analysis and figures that follow we use also somewhat more physically transparent rescaled stochastic variable 
\begin{equation}
x = (n-N_{c})/\sigma^{(\infty)} \approx x'/\sqrt{s_2} ,
\label{x}
\end{equation}
that represents noncondensate fluctuations measured in the units of dispersion and has rescaled universal probability distribution
\begin{equation}
\rho_{x}^{(univ)} = \sqrt{s_2}\rho_{x'}^{(univ)} .
\label{rhox}
\end{equation}
Please note that in order to pin the critical point to zero $x=0$ (or $x'=0$) we have to measure the scaled variables $x$ and $x'$ in Eqs.~(\ref{x}) and (\ref{xuniv}) relative to the exact mesoscopic critical value $N_c$ given by the discrete sum in Eq.~(\ref{Nc}) that cannot be replaced here by its continuous approximation $N_v$ in Eq.~(\ref{Nv}). Otherwise, all universal functions for the stochastic and thermodynamic quantities would acquire a trap-size dependent shift $\eta_{c}=(N_{v}-N_{c})/\sigma^{(\infty)}$ that does not tend to zero, but instead slowly increases like $\sim N_{v}^{\delta}$ with a power index $\delta \sim 0.1$. That shift is not addressed in the usual grand-canonical-ensemble approximation in the thermodynamic limit but is important to resolve correctly the universal structure of the critical region as is clearly seen from an example of the heat capacity discussed below in Sec. XIV and Figs. 14, 15.  

	 Below we derive also analytical approximations for the universal probability distribution of the total noncondensate occupation, the most accurate of which are given in terms of the Kummer's confluent hypergeometric function 
$$\rho_{x}^{(univ)}\approx \frac{e_{1}^{g_1}e_{2}^{g_2}X^{g_{1}+g_{2}-1}e^{-e_{2}X}}{\Gamma (g_{1}+g_{2})}M(g_{1},g_{1}+g_{2},(e_{2}-e_{1})X),$$
\begin{equation}
X = x + g_{1}/e_{1} + g_{2}/e_{2} ,
\label{rhounivKummer}
\end{equation}
$$e_{1}\approx 4.303,\quad e_{2}\approx 29.573,\quad g_{1}\approx 8.504,\quad g_{2}\approx 473,$$	 
and in terms of the parabolic cylinder function 
\begin{equation}
\rho_{x}^{(univ)} \approx \frac{c^{g}e^{c^{2}/2-Y^2}}{\sqrt{2\pi (1-s_{3}^{2}/(s_{2}s_{4}))}}D_{-g}[2(c-Y)],
\label{rhounivPC}
\end{equation}
$$Y=\frac{x+s_{3}\sqrt{s_2}/s_4}{2\sqrt{1-s_{3}^{2}/(s_{2}s_{4})}},\quad g=\frac{s_{3}^{4}}{s_{4}^{3}},\quad c=\frac{s_{3}}{s_{4}}\sqrt{s_{2}-\frac{s_{3}^{2}}{s_4}}.$$ 
They are derived by means of an exact analytical solution (\ref{rho3infty}) for a three-level-trap model with matching the first five (Eq.~(\ref{r3buniv})) or four (Eq.~(\ref{limitparameters})) cumulants, respectively. Amazingly, we obtain absolutely the same asymptotics (\ref{rhounivPC}) from the exact analytical solution (\ref{PsG}) for a completely different model (\ref{aPsGb2}) that is a superposition of the two-level trap model and the Gaussian model (see sections VII, IX, XI). The analytical results in Eq.~(\ref{rhounivKummer}) and in Eq.~(\ref{rhounivPC}) are remarkably accurate in the whole central part of the critical region (namely, in the intervals $-4 < x < 10$ and $-3 < x < 6$, respectively), not only near the critical point, and allow us to calculate analytically the universal functions for all moments of BEC statistics, including the order parameter, and other physical quantities (see Sections VI, VII, and XII-XIV) via the exact formulas which express the constraint-cut-off mechanism described in Sec. V. 

	Note that the universal probability distribution $\rho_{x}^{(univ)}$ for the box trap with periodic boundary conditions, according to Eqs.~(\ref{rhouniv}) and (\ref{Thetauniv}), does not include any parameters and, in a sense, is a pure mathematical special function. Similar universal probability distribution of the total noncondensate occupation can be derived for any other trap, e.g. a box with the Dirichlet boundary conditions, following exactly the same scheme starting from the known unconstrained probability distribution $\rho_{n}^{(\infty)}$ for arbitrary trap \cite{KKS-PRL,KKS-PRA}. 

\begin{figure}
\center{\epsfig{file=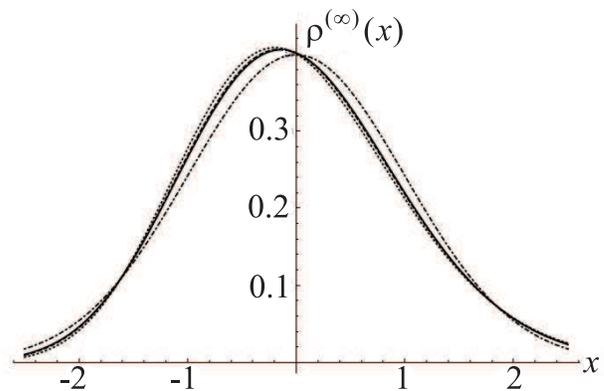,width=8.5cm}}
\caption{Scaled unconstrained probability distribution of the stochastic variable $x = (n - N_{c})/\sigma^{(\infty)}$ for different finite traps: $N_v = 10^2$ (dotted line), $N_v = 10^3$ (dashed line), $N_v = 10^4$ (solid line). The Gaussian distribution $\exp (-x^{2}/2)/\sqrt{2\pi}$ is depicted by a dotted-dashed line.} 
\end{figure}

     That remarkable universality is valid in the whole critical region, $\left|n-N_{c}\right|/\sigma^{(\infty)} \ll N_{v}^{1/3}$, that tends to an infinite interval of values $x\in (-\infty , \infty)$ in the thermodynamic limit $N_v \to \infty$. It includes very large values $|x| \gg 1$ and is much wider than a relatively narrow vicinity of the maximum of the distribution, $\left|n - N_{c}\right| \leq \sigma^{(\infty)}$, where a Gaussian approximation always works well. In particular, from Figs. 3 and 4 we see that this is true for the right tail at least up to $n - N_{c} \sim 8\sigma^{(\infty)}$ for $N_v = 10^3$ and up to $n - N_{c} \sim 10\sigma^{(\infty)}$ for $N_v = 10^4$ and also for the left tail, of course, except close to $n \approx 0$ part, where the probability should become zero. Thus, the actual mesoscopic probability distribution becomes very close to the universal, thermodynamic-limit probability distribution already starting from quite moderate values of the trap-size parameter $N_v \sim 10^2 \div 10^3$. This result is more clearly shown in Fig. 4, where a logarithmic scale allows us to see the behavior of the tails in more details. The universal probability distribution has very fat and long right tail (\ref{rap}) of the large occupation values, $n - N_{c} \gg \sigma^{(\infty)}$, whereas the left tail (\ref{arhox}) of small occupation values, $N_{c} - n \gg \sigma^{(\infty)}$, is strongly suppressed as compared to the Gaussian distribution. 

\begin{figure}
\center{\epsfig{file=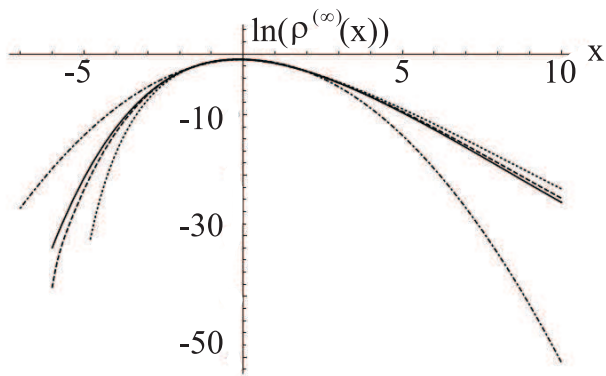,width=8.5cm}}
\caption{Logarithm of the scaled unconstrained probability distribution of the stochastic variable $x = (n - N_{c})/\sigma^{(\infty)}$ for different finite traps: $N_v = 10^2$ (dotted line), $N_v = 10^3$ (dashed line), $N_v = 10^4$ (solid line). Logarithm of the Gaussian distribution, $-x^{2}/2- \ln (2\pi) /2$, is depicted by a dotted-dashed line.} 
\end{figure}
     
     The most crucial point is that the universal probability distribution does not collapse to a kind of $\delta$-function or a pure Gaussian distribution but remains finite, smooth, and nontrivial in the thermodynamic limit, so that an intrinsic critical structure of the BEC phase transition clearly reveals itself in the already quite small mesoscopic systems with the critical number of atoms $N_c \sim 10^2$. We find that a Taylor series for the logarithm of the universal probability distribution, i.e. the negative Landau function, 
$$\ln \left[\rho_{x}^{(univ)}\right] = a_0 + a_{2}(x-\Delta x)^{2}/2!$$
\begin{equation}
+ a_{3}(x - \Delta x)^{3}/3! + a_{4}(x - \Delta x)^{4}/4! + \ldots 
\label{lnrho}
\end{equation}
contains very essential third ($a_3 \approx 0.26$) and forth ($a_4 \approx -3$) order terms which provide the same or larger contributions at the tails as compared to that of the quadratic part ($a_2 \approx -1.04$). There is also a relatively small but finite shift $\Delta x \approx -0.12$ of the maximum of the probability distribution to the left of the mean value $\bar x =0$, i.e. ${\bar n} = N_c$, due to the discussed above asymmetric tails. The normalization coefficient is $a_0 \approx -\ln \sqrt{2\pi} + 0.013$. Note that a pure Gaussian distribution has only two nonzero coefficients, $a_{0}^{(Gauss)} = -\ln \sqrt{2\pi} \approx -0.919$ and $a_{2}^{(Gauss)} = -1$, which are very close to their counterparts in Eq.~(\ref{lnrho}). 

     All exact numerical simulations for the BEC statistics in the box trap with a finite number of atoms $N$ presented in this paper were obtained by direct calculation of the characteristic function in Eq.~(\ref{Thetauniv}) for the mesoscopic ideal gas using the analytical formulas \cite{KKS-PRL,KKS-PRA} and then the probability distribution in Eq.~(\ref{rho}) by means of a Fast Fourier Transform technique (FFT). Also, the standard simulation technique based on the recursion relation \cite{Landsberg,Borrmann1993,Brosens1997,ww1997,Holthaus1997,Balazs1998,recursion1999,Borrmann1999,Kleinert2007,Wang2009} was used. Both techniques yield the same results and allow us to calculate the BEC statistics in the mesoscopic systems up to relatively large critical number of atoms, $N_c < 10^5$.
     
     The exact universal probability distribution in Eqs.~(\ref{rhouniv}), (\ref{Thetauniv}), (\ref{rhox}) can be approximated very efficiently by exact taking into account contributions $g_{j}/q^{2m}$ to $s_m$ in Eq.~(\ref{sm}) for the generating cumulants (\ref{igcumulants}) from a few first energy levels $\epsilon_{j}, j= 1, 2, 3, \ldots, J$, as well as by exact taking into account the remaining parts of a few first cumulants via sums $s_{m}- \sum_{j=1}^{J}{g_{j}/Q_{j}^{2m}}, m= 1, 2, \ldots, m^{*},$ and omitting only contributions to the higher order cumulants $s_{m}, m>m^{*},$ from all energy levels with high energies $\epsilon_{\vec{k}}>\epsilon_J$. Here a degeneracy $g_j$ is equal to the number of the atomic $\vec{k}$-states which have the same energy $\epsilon_{\vec{k}} = \epsilon_j \equiv \epsilon_{1}Q_{j}^{2}$, where $Q_{j}^{2} = q_{x}^{2}+q_{y}^{2}+q_{z}^{2}$ is a dimensionless energy of the j-th energy level, $Q_{j}^{2} = 1, 2, 3, 4, 5, 6, 8, 9, 12, \ldots; g_1 =6, g_2 =12, g_3 =8, g_4 =6, g_5 =24, g_6 =24, \ldots$. Thus, since $\sum_{m=1}^{\infty}{(iy)^{m}/m} = - \ln (1-iy)$, we can approximate the universal characteristic function in Eq.~(\ref{Thetauniv}) as follows
$$\Theta^{(univ)}(u') \approx \left[\prod_{j=1}^{J}{\left(1-\frac{iu'}{Q_{j}^{2}}\right)^{-g_j}}\right] \exp \Big[-iu'\sum_{j=1}^{J}{\frac{g_j}{Q_{j}^{2}}}$$
\begin{equation}
+\sum_{m=2}^{m^{*}}{\left(s_{m}-\sum_{j=1}^{J}{\frac{g_j}{Q_{j}^{2m}}}\right)\frac{(iu')^m}{m}} \Big] ,
\label{mJTh}
\end{equation}
where the numbers $J$ and $m^{*}$ of the taking into account energy levels and cumulants, respectively, specify an accuracy. Straightforward numerical calculation of Eq.~(\ref{mJTh}) and corresponding Eq.~(\ref{rhox}) show that with increasing numbers $J$ and $m^{*}$ these approximations nicely converge to the exact universal functions (\ref{Thetauniv}) and (\ref{rhox}) in the proportionally increasing intervals of values $\left|u'\right|\leq u'^{*}(J, m^{*})$ and $\left|x\right|\leq x^{*}(J, m^{*})$. The main reason for that fact is that the residual parts of the cumulants $\propto s_{m}-\sum_{j=1}^{J}{g_{j}/Q_{j}^{2m}}$ becomes, after subtracting contributions from the first J energy levels, very small with increasing $m$ and $J$. 

     In fact, already the approximation with $J=6$ and $m^{*} = 14$ is more than enough for all practical purposes, can be efficiently used to plot all statistical and thermodynamic quantities by means of the standard Mathematica or similar elementary code packages, and works perfectly well in the whole critical region, $-6 < x < 15$, including even a large part of the asymptotics region $\left|x\right| \gg 1$. 
     
     For the simplest nontrivial approximation $(J=1, m^{*}=2)$ in Eq.~(\ref{mJTh}) we find 
\begin{equation}
\Theta^{(univ)}(u') \approx \frac{e^{-6iu'}}{(1-iu')^6} \exp \left[-\frac{s_{2}-6}{2}(u')^2\right],
\label{21Th}
\end{equation} 
that takes into account all contributions from the first energy level and the Gaussian part of the remaining contributions from all higher energy levels. In this case the integral in Eq.~(\ref{rhouniv}) can be calculated analytically if we represent it as a convolution 
\begin{equation}
\rho_{x'}^{(univ)} = \int_{0}^{\infty}{f_{6}(y)R(x'+6-y)dy},
\label{21int}
\end{equation} 
where 
\begin{equation}
f_{6}(y) = \frac{1}{2\pi i} \int_{-i\infty}^{i\infty}{\frac{e^{py}}{(1+p)^6}dp}=\frac{y^5}{5!}e^{-y}
\label{21f6}
\end{equation}
is a cut-off gamma distribution for the first energy level (see Sec. IX) and 
$$R(y)=\frac{1}{2\pi i} \int_{-i\infty}^{i\infty}{\exp \left(py+\frac{s_{2}-6}{2}p^{2}\right)dp}$$
\begin{equation}
= \frac{\exp \left[-y^{2}/(2(s_{2}-6))\right]}{\sqrt{2\pi (s_{2}-6)}}
\label{21R}
\end{equation}
is a Gaussian distribution. The integral in Eq.~(\ref{21int}) yields the simplest nontrivial approximation for the universal probability distribution of the total noncondensate occupation 
$$\rho_{x}^{(univ)} = \frac{s_{2}^{1/2}D_{-6}(X')}{\sqrt{2\pi}(s_{2}-6)^{-5/2}}\exp\left[ -\frac{1}{2}\frac{(x'+6)^2}{s_{2}-6}+\frac{X'^2}{4}\right],$$
\begin{equation}
X' = \sqrt{s_{2}-6}\left(1-\frac{x'+6}{s_{2}-6}\right),
\label{21rhox}
\end{equation}
where $D_{-g_1}(X')$ is a parabolic cylinder function \cite{a} with an index specified by the degeneracy of the first energy level $g_1 =6$, $x'=\sqrt{s_2}x$, and $s_2$ is given in Eq.~(\ref{s234}). That simple approximation works reasonably well in the central part of the critical region, $-1 < x < 5$, and is essentially better than the plain polynomial approximation in Eq.~(\ref{lnrho}) that has narrow interval of validity even with four or six terms taken into account. 

\subsection{B. Asymptotics of the Universal Probability Distribution in the Critical Region}

   \textbf{Noncondensed Phase:} Asymptotics of the universal probability distribution (\ref{rhouniv}) or (\ref{rhox}) at the left tail of the critical region, $x \to -\infty$, i.e. in the noncondensed phase, is determined by a contribution accumulated near a complex stationary point $u'=u_{s}'$ in the inverse Laplace integral (\ref{rhouniv}) or (\ref{rhox}) and cannot be found directly from any approximation (\ref{mJTh}) keeping only a finite number of terms in it since the stationary point tends to infinity, $|u_{s}'| \to \infty$, when $x \to -\infty$. Hence, we first have to find explicitly the asymptotics of the logarithm of the characteristic function 
\begin{equation}
f(u') \equiv \ln \Theta^{(univ)}(u') = \sum_{m=2}^{\infty}{\frac{s_m}{m}(iu')^m}
\label{f}
\end{equation}
at $|u'| \to \infty$. It can be done by calculating its first derivative directly from Eq.~(\ref{Thetauniv}) in the continuous approximation as follows 
$$f'(u') = i\sum_{m=2}^{\infty}{s_{m}(iu')^{m-1}} = -\sum_{\vec{q}\neq 0}{\frac{u'}{q^{2}(q^{2}-iu')}}$$
\begin{equation}
\sim -\int_{0}^{\infty}{\frac{4\pi u'dq}{q^{2}-iu'}} = -2\pi^2 \sqrt{iu'}.
\label{Df}
\end{equation}
Another way to do this is to calculate the derivative of the logarithm of the original characteristic function (\ref{char}) in the continuous approximation: 
\begin{equation}
\frac{d\ln \Theta^{(\infty)}}{du'} \sim \frac{2\pi i}{\sqrt{\epsilon_{1}/T}}\int_{0}^{\infty}{\frac{\sqrt{y}dy}{e^{y-iu'\frac{\epsilon_{1}}{T}}-1}} = \frac{i\pi^{\frac{3}{2}}Li_{\frac{3}{2}}(e^{\frac{iu'\epsilon_{1}}{T}})}{\sqrt{\epsilon_{1}/T}}.
\label{lnThPolyLog}
\end{equation}
Using an expansion of a well-known polylogarithm, or Bose (\ref{Bosefun}), function \cite{Bateman} for a pure imaginary argument $y=iu'\epsilon_{1}/T\to 0$ for any finite, even large $u'$,
\begin{equation}
Li_{\frac{3}{2}}(e^{y})\sim 2i\sqrt{\pi y} + \sum_{m=0}^{\infty}{\frac{\zeta (\frac{3}{2}-m)y^{m}}{m!}},
\label{PolyLogseries}
\end{equation}
we find at $|u'| \gg 1$ the asymptotics
$$\frac{d\ln \Theta^{(\infty)}}{du'} \sim -2\pi^{2}\sqrt{iu'} + \frac{iN_{v}\epsilon_1}{T} $$
\begin{equation}
+ \frac{i\pi^{3/2}}{\sqrt{\epsilon_{1}/T}} \sum_{m=1}^{\infty}{\frac{\zeta (3/2-m)}{m!}\left(\frac{iu'\epsilon_1}{T}\right)^{m}},
\label{DlnThPolyLog}
\end{equation}
the leading term of which is exactly the same as in Eq.~(\ref{Df}). Note that the zeroth-order ($m=0$) term in Eq.~(\ref{DlnThPolyLog}), which is responsible for the average noncondensate occupation $N_c \to N_v$, has been already exactly accounted for in the universal distribution (\ref{rhouniv}), or (\ref{rhox}), via the mean value $N_c$. The logarithm of the universal characteristic function in Eq.~(\ref{f}) contains no zeroth-order or linear in $u'$ terms which, besides, cannot be given correctly by the continuous approximation since already the linear term is proportional to the discrete corrections to the variable $x$ in Eq.~(\ref{x}) via the exact critical number $N_c$ which is not universal as we discussed in the beginning of Sec. IV. Thus, namely the asymptotics of the second derivative of the logarithm of the universal characteristic function (\ref{f}) is given correctly by Eqs.~(\ref{Df}) and (\ref{DlnThPolyLog}) as 
\begin{equation}
f''(u') \approx -\pi^2 \sqrt{i/u'} ,
\label{D2f}
\end{equation}
that numerically perfectly coincides with the exact values of $f''(u')$ for all $|u'| > 1$ starting already with $|u'| \approx 1.2$. The aforementioned zeroth-order and linear terms in the asymptotics of $f(u')$ can be easily found by comparison with the exact function in Eq.~(\ref{f}) at some finite point $u'$, for example, using approximation (\ref{mJTh}), or by calculation of the appropriate discrete sums. These terms are equal to $f_{0}\approx 3.3$ and $iu'f_{01}\approx iu' 2.2 \sqrt{s_2}$, respectively. 

     Now we can find the asymptotics of the universal probability distribution in Eq.~(\ref{rhouniv}) or (\ref{rhox}) as follows 
$$\rho_{x}^{(univ)} \approx \frac{\sqrt{s_2}}{2\pi}\int_{-\infty}^{\infty}{e^{-iu'x'+f_{0}+if_{01}u'+f''(u')}du'}$$
\begin{equation}
= \frac{\sqrt{s_2}e^{f_0}}{2\pi^{\frac{7}{3}}3^{-\frac{2}{3}}} Re\int_{0}^{\infty}{e^{F(t)}dt},\quad F(t)=-i(t\tilde{x}+t^{\frac{3}{2}})-t^{\frac{3}{2}},
\label{rhof}
\end{equation}
where $\tilde{x}=3^{2/3}(\sqrt{s_{2}}x-f_{01})/(2\pi^{4/3})<0$. The complex stationary point of the latter integral is determined by the equation $F'(t=t_{s})=0$ and is equal to $t_{s}=2i\tilde{x}^{2}/9$. Complex Gaussian approximation of the function $F(t) \approx -2|\tilde{x}/3|^{3} - 9(t-t_{s})^{2}/(8|\tilde{x}|)$ in the stationary point vicinity, that provides the major contribution to the inverse Laplace integral in Eq.~(\ref{rhof}), allows us to calculate that integral explicitly and yields the asymptotics of the left tail $(-x\gg 1)$ of the universal probability distribution as follows 
\begin{equation}
\rho_{x}^{(univ)} \approx \frac{s_{2}^{3/4}}{2\pi^{5/2}} \sqrt{x_{0}-x} \exp \left[f_{0}+\frac{s_{2}^{3/2}}{12\pi^4}(x-x_{0})^{3}\right] ,
\label{arhox}
\end{equation}
where $x_{0}=f_{01}/\sqrt{s_2} \approx 2.2$, $f_{0}\approx 3.3$, and an accuracy is excellent starting already from $x < -3$. That result is very nontrivial and unusual for statistical physics. Indeed, the unconstrained universal probability distribution in the critical region at $-x \gg 1$, i.e. in the noncondensed phase, decays with a cubic exponent, that is much faster than both a decay with a linear exponent at the right tail (see Eq.~(\ref{rap}) below) and a standard Gaussian, quadratic exponential decay. 

     \textbf{Condensed Phase:} Asymptotics of the universal probability distribution (\ref{rhouniv}), or (\ref{rhox}), at the right tail of the critical region, $x \to +\infty$, i.e. in the condensed phase, is completely different from that at the left tail since for positive values $x' \gg 1$ a frequency of oscillations, i.e. an imaginary part of the derivative of the exponent in the integrand of Eq.~(\ref{rhouniv}), always increases with increasing $|u'|$, that is there is no stationary point near an integration path anymore, contrary to the case $x' < 0$ at the left tail. Asymptotics at the right tail is determined mainly by the first energy level contribution with a finite shift and renormalization due to background of the higher energy levels. It can be calculated explicitly as a residue of the integrand in the integral (\ref{rhouniv})  (along a counterclockwise contour closed through $-\infty$) at the pole $p=p_{1}=-1$ corresponding to the pole $u'=u_{1}=-i\epsilon_{1}/T$ of the characteristic function (\ref{char}) that is related to the first energy level and has an order equal to the first energy level degeneracy $g_{1}=6$. Its contribution is proportional to $\exp (-x')$ and in the asymptotics $x' \to \infty$ becomes exponentially large compared with the contributions from the poles $p_{k}=-\epsilon_{k}/\epsilon_{1}$ of all higher energy levels which go as $\exp [-(\epsilon_{k}/\epsilon_{1})x']$, where the decay rate $\epsilon_{k}/\epsilon_{1}= 2, 3, \ldots$ is larger than 1 for all higher energy levels. 
     
     To implement this approach, we rewrite Eq.~(\ref{rhouniv}) in the equivalent form, similar to Eq.~(\ref{mJTh}), using a new integration variable $z=1+p$, 
$$\rho_{x'}^{(univ)}=\int_{-i\infty}^{i\infty}{\frac{e^{p(x'+g_{1})}}{(1+p)^{g_1}}\exp \left(\sum_{m=2}^{\infty}{\frac{s_{m}-g_{1}}{m}(-p)^m}\right)\frac{dp}{2\pi i}}$$
\begin{equation}
= \frac{e^{-x'-g_1}}{2\pi i}\int_{-i\infty}^{i\infty}{e^{f_{1}(z)}\frac{dz}{z^{g_1}}},
\label{rho11}
\end{equation}
where the exponent is determined by the sum over cubic lattice of all dimensionless wavevectors $\vec{q} = \{q_{x}, q_{y}, q_{z}\}$ with integer components $q_{x,y,z} = 0, \pm 1, \pm 2, \ldots$ of all atomic states, excluding states on the ground and first excited energy levels ($|\vec{q}|\neq 0, 1$), as follows 
\begin{equation}
f_{1}(z) = z(x'+g_{1}) - \sum_{\left\{\vec{q}: |\vec{q}|>1\right\}}{\left[\ln \left(1-\frac{1}{q^2}+\frac{z}{q^2}\right)+\frac{1-z}{q^2}\right]}.
\label{f1}
\end{equation}
The exact Taylor series of the latter function is given by a simple formula
\begin{equation}
f_{1}(z) = s_{0}' + \sum_{j=1}^{\infty}{\frac{x_j}{j}z^j},
\label{f1T}
\end{equation}
where $x_{j}=(-1)^{j}s_{j}'$ for $j \geq 2$ and 
\begin{equation}
x_1 = x' + g_1 - s_{0}'' ,
\label{x1}
\end{equation}
\begin{equation}
s_{0}' = \sum_{\left\{\vec{q}: |\vec{q}|>1\right\}}{\left[\ln \left(\frac{q^2}{q^{2}-1}\right)-\frac{1}{q^2}\right]} \equiv \sum_{m=2}^{\infty}{\frac{s_{m}-6}{m}} \approx 6.45,
\label{s01}
\end{equation}
\begin{equation}
s_{0}'' = \sum_{\left\{\vec{q}: |\vec{q}|>1\right\}}{\frac{1}{q^{2}(q^{2}-1)}} \approx 14.7,
\label{s02}
\end{equation}
\begin{equation}
s_{j}' = \sum_{\left\{\vec{q}: |\vec{q}|>1\right\}}{\frac{1}{(q^{2}-1)^j}} , \quad j = 2, 3, 4, \ldots .
\label{sj1}
\end{equation}
The constants $s_{j}'$ tend to the degeneracy of the second energy level $g_2 = 12$ with increasing index $j$, namely, $s_{2}' = 22.24, s_{3}' = 14.06, s_{4}' = 12.73, s_{5}' = 12.31, \ldots$. The required residue at the pole $z=0$ in Eq.~(\ref{rho11}) is obviously determined by the coefficient $c_{-1}$ of the Laurent expansion of the function $\exp [f_{1}(z)]/z^6 = \sum_{l}{c_{l}z^l}$, that is by the coefficient 
\begin{equation}
\alpha_{5}(x') = \frac{1}{5!}\sum_{m=0}^{5}{\sum^{(5,m)}{(5;a_{1},\ldots,a_{5})^{*}x_{1}^{a_1}\ldots x_{5}^{a_5}}}
\label{a5}
\end{equation}
in the Taylor series of $\exp \left[f_{1}(z)\right]=\sum_{n=0}^{\infty}{\alpha_{n}z^n}$. The latter was found by means of an expansion
\begin{equation}
e^{\sum_{k=1}^{\infty}{\frac{x_{k}z^k}{k}}}=\sum_{n=0}^{\infty}{\sum_{m=0}^{n}{\frac{z^n}{n!}\sum^{(n,m)}{(n; a_{1},\ldots ,a_{n})^{*}x_{1}^{a_1}\ldots x_{n}^{a_n}}}},
\label{T1}
\end{equation}
where we used a generating function (\ref{MGF}) for the multinomial coefficients \cite{a} in Eq.~(\ref{multinom}).

     Thus, we find the asymptotics of the universal unconstrained probability distribution of the total noncondensate occupation (see Eqs.~(\ref{rhouniv}) and (\ref{rhox})) in the critical region at $x \to \infty$, i.e. in the condensed phase, in the following analytical form
\begin{equation}
\rho_{x}^{(univ)} \approx \sqrt{s_2}\alpha_{5}(x')e^{-x'-6+s_{0}'} , \quad x' = \sqrt{s_2}x,
\label{rap}
\end{equation}
\begin{equation}
\alpha_{5}(x') = \frac{x_{1}^5}{5!}+\frac{s_{2}'x_{1}^3}{12}-\frac{s_{3}'x_{1}^2}{6}+\left(\frac{s_{2}'^2}{2}+s_{4}'\right)\frac{x_1}{4}-\frac{s_{2}'s_{3}'}{6}-\frac{s_{5}'}{5},
\label{a5s}
\end{equation}
where $x_1$ and constants $s_{0}', s_{0}''$, and $s_{j}'$ are defined in Eqs.~(\ref{x1})-(\ref{sj1}). Please note that both the leading exponent and the pre-exponential polynomial in the asymptotics (\ref{rap}) are found exactly. Numerically that asymptotics works excellent for all $x > 3$. 

     The most striking result of the analysis of the asymptotics of the universal unconstrained probability distribution $\rho_{x}^{(univ)}$ is its highly pronounced asymmetry with an incredibly fast, cubic (\ref{arhox}) and very slow, linear (\ref{rap}) exponential decays at the left tail (noncondensed phase) and at the right tail (condensed phase), respectively. Both of them are quite different from a standard in statistical physics Gaussian, quadratic exponential decay. 

\subsection{C. Outside Critical Region in the Condensed Phase: Asymptotics in the Large Number of Atoms Region}

    Let us consider the condensed phase of the fully developed condensate outside critical region in the thermodynamic limit at very low temperatures and very large numbers of atoms in the trap, including region $N-N_c \geq N_c$. Here the values of the noncondensate occupation $n$ are so large that the universal variable (\ref{x}) is larger than any finite value, that is $x \equiv \left[\pi(\zeta (3/2))^{2/3}/\sqrt{s_2}\right] (n-N_{c})/N_{v}^{2/3} \sim N_{v}^{\delta} \to \infty$ where $\delta > 0$, including the value $\delta = 1/3$ when $n-N_c \sim N_v \to \infty$ and $n-N_c > N_v$. In that whole region, outside the critical region, in addition to the universal dependence on $x$, the probability distribution $\rho_{n}^{(\infty)}$ acquires a non-universal extra dependence on the trap-size parameter $N_v$ and on $n$ which is irreducible to any $x$-dependence. We can find the asymptotics of the noncondensate occupation probability distribution $\rho_{n}^{(\infty)}$ starting from the exact result in Eqs.~(\ref{rhoinfinity}) and (\ref{char}), namely,
$$\rho_{n}^{(\infty)}=\frac{1}{2\pi}\int_{-\pi}^{\pi}{e^{-iun}\Theta^{(\infty)}(u)du}$$
\begin{equation}
=\frac{1}{2\pi i}\oint_{|z|=1}{\frac{\exp \left[-\sum_{\left\{\vec{q}: |\vec{q}|>1\right\}}{\ln \left(1-\frac{z-1}{e^{q^{2}\epsilon_{1}/T}-1}\right)}\right]dz}{z^{n+1}\left[1-(z-1)/(e^{\epsilon_{1}/T}-1)\right]^{6}}},
\label{rexact}
\end{equation}
and proceeding similar to the derivation of the asymptotics (\ref{rap}) from Eq.~(\ref{rho1}). Here we start with the variable $z=e^{iu}-1$. Changing it to a new variable $s =(e^{\epsilon_{1}/T}-z)/(e^{\epsilon_{1}/T}-1)$, we can rewrite the integral in Eq.~(\ref{rexact}) in an equivalent form 
\begin{equation}
\rho_{n}^{(\infty)}=-\frac{e^{\frac{\epsilon_1}{T}}-1}{2\pi i e^{b}}\oint{\frac{e^{\tilde{f}_{1}(s)}}{s^{g_1}}ds},\quad \tilde{f}_{1}(s)=\tilde{s}_{0}' +\sum_{j=1}^{\infty}{\frac{\tilde{x}_{j}s^j}{j}},
\label{rdiscrete}
\end{equation}
where $g_1 =6$ is the degeneracy of the first energy level,
\begin{equation}
\tilde{x}_{j}=(-1)^{j}\tilde{s}_{j}' + (n+1)\left(1-e^{-\epsilon_{1}/T}\right)^j ,
\label{xj}
\end{equation}
\begin{equation}
\tilde{s}_{j}' = \sum_{\left\{\vec{q}: |\vec{q}|>1\right\}}{\left[\left(e^{\epsilon_{1}/T}-1\right)/\left(e^{q^{2}\epsilon_{1}/T}-e^{\epsilon_{1}/T}\right)\right]^{j}} ,
\label{sjNv}
\end{equation}
\begin{equation}
\tilde{s}_{0}'=-\sum_{\left\{\vec{q}: |\vec{q}|>1\right\}}{\left[\frac{e^{\epsilon_{1}/T}-1}{e^{q^{2}\epsilon_{1}/T}-1}+\ln \left(1-\frac{e^{\epsilon_{1}/T}-1}{e^{q^{2}\epsilon_{1}/T}-1}\right)\right]},
\label{s0Nv}
\end{equation}
$$b = \frac{\epsilon_1}{T}(n+1) -\sum_{\left\{\vec{q}: |\vec{q}|>1\right\}}{\frac{e^{\epsilon_{1}/T}-1}{e^{q^{2}\epsilon_{1}/T}-1}}$$
\begin{equation}
\equiv \frac{\epsilon_1}{T}(n-N_{c})+g_{1}+\frac{\epsilon_1}{T} -\left(e^{\epsilon_{1}/T}-1-\frac{\epsilon_1}{T}\right)N_{c}.
\label{bNv}
\end{equation}

     Keeping and calculating contribution only from the residue at the pole $s=0$ related to the first energy level, similar to Eqs.~(\ref{a5}), (\ref{T1}), (\ref{MGF}), and (\ref{multinom}), we find the required asymptotics with the exact analytical formulas both for the leading exponent and for its pre-exponential polynomial: 
$$\rho_{n}^{(\infty)}\approx \left(e^{\epsilon_{1}/T}-1\right)\tilde{\alpha}_{5}(n)\exp \{-\frac{\epsilon_1}{T}(n-N_{c})-6+\tilde{s}_{0}'$$
\begin{equation}
-\frac{\epsilon_1}{T}+\left(e^{\epsilon_{1}/T}-1-\frac{\epsilon_1}{T}\right)N_{c} \},
\label{rapn}
\end{equation}
\begin{equation}
\tilde{\alpha}_{5}(n) = \frac{\tilde{x}_{1}^5}{5!}+\frac{\tilde{x}_{2}\tilde{x}_{1}^3}{12}+\frac{\tilde{x}_{3}\tilde{x}_{1}^2}{6}+\left(\frac{\tilde{x}_{2}^2}{2}+\tilde{x}_{4}\right)\frac{\tilde{x}_1}{4}+\frac{\tilde{x}_{3}\tilde{x}_{2}}{6}+\frac{\tilde{x}_{5}}{5}.
\label{a5x}
\end{equation}

    It is remarkable that in the thermodynamic limit inside the critical region, when we can neglect by all small terms of the order of $\epsilon_{1}/T \sim N_{v}^{-2/3} \to 0$ and higher orders, this result is obviously reduced to the universal asymptotics in Eqs.~(\ref{rap}), (\ref{a5s}) since $\tilde{x}_{1}\to x'+6-s_{0}'',\quad \tilde{x}_{j}\to x_j$ for $j \geq 2$, and $\tilde{s}_{j}'\to s_{j}',\quad \tilde{s}_{0}'\to s_{0}',\quad b\to x'+6$. Outside the critical region, in particular, for relatively small mesoscopic systems, the result in Eq.~(\ref{rapn}) describes deviations from the universal behavior due to finite-size, mesoscopic effects.
     
\subsection{D. Outside Critical Region in the Noncondensed Phase}

  \textbf{Asymptotics in the Small Number of Atoms Region: Poisson Distribution and Corrections.} Outside critical region, when very small number of atoms is loaded into the trap, $N_c - N \sim N_c$, and, hence, the temperature is very high compared to the critical BEC temperature, $T-T_c \sim T_c$ or $T-T_c > T_c$, one has very dilute ideal gas without condensate, but with strongly pronounced finite-size and discreteness effects. These effects are important for the quantum statistics of the noncondensate as well as condensate fluctuations in that small number of atoms region, where the probability distribution $\rho_{n}^{(\infty)}$ does not follow anymore the universal asymptotics (\ref{arhox}) of the left tail of the critical region, but instead has a completely different, not self-similar structure which is attached to the end point $n=0$ of the probability distribution $\rho_{n}^{(\infty)}$. As we will see, for small enough noncondensate occupation $n \ll 3\sqrt{N_v}$ it tends to the Poisson distribution. 
    
    In order to reveal that structure and its asymptotics we use the multinomial expansion, Eq.~(\ref{multinomrho}). First, let us consider more simple case of the thermodynamic limit when the sums $B_n$ for $n \geq 1$ in Eq.~(\ref{Taylor2}) are equal to 
\begin{equation}
B_n = N_{v}/\left(n^{3/2} \zeta (3/2)\right) , \quad N_v \gg 1 .
\label{Bntl}
\end{equation}
Then, if we introduce new, generalized multinomial coefficients and their sums over all nonnegative integers $a_{1}, \ldots , a_{n}$ which satisfy two conditions, $a_{1}+a_{2}+\ldots +a_{n}=m$ and $a_{1}+2a_{2}+\ldots +na_{n}=n$, as follows 
\begin{equation}
(n; a_{1}, a_{2},\ldots , a_{n})_{x}^{*} = \frac{n!}{1^{a_{1}x}a_{1}!2^{a_{2}x}a_{2}!\ldots n^{a_{n}x}a_{n}!},
\label{newmnc}
\end{equation}
\begin{equation}
K_{n}^{(m)}(x) = \sum^{(m,n)}{(n; a_{1}, a_{2},\ldots , a_{n})_{x}^{*}} ,
\label{Kmn}
\end{equation}
we find the thermodynamic limit of the unconstrained probability distribution (\ref{rhoinfinity}) as follows
\begin{equation}
\rho_{n}^{(\infty)} = \frac{\rho_{n=0}^{(\infty)}}{n!}\sum_{m=0}^{n}{K_{n}^{(m)}(x=5/2)B_{1}^{m}} ,\quad N_{v}\gg 1.
\label{rhoB}
\end{equation}
Some, necessary for us properties of the coefficients $K_{n}^{(m)}(x)$ for $n>0$ are summarized below:
$$K_{n+1}^{(0)}(x)\equiv 0; K_{1}^{(1)}(x)=K_{n}^{(n)}(x)=1;\quad K_{2}^{(1)}(x)=2^{1-x};$$
$$K_{n}^{(n-1)}(x)=\frac{n!}{(n-2)!2^x};$$
\begin{equation}
K_{n}^{(n-2)}(x)=\frac{n!}{(n-4)!2!2^x}+\frac{n!}{(n-3)!3^x};
\label{Kmn1}
\end{equation}
$$K_{n}^{(n-3)}(x)=\frac{n!}{(n-6)!3!2^{3x}}+\frac{n!}{(n-5)!2^{x}3^x}+\frac{n!}{(n-4)!2^{2x}}.$$

    A general case of arbitrary finite trap-size parameter $N_v$ can be considered similarly. In particular, we can find exact analytical formulas for the unconstrained as well as actual (via the constraint-cut-off Eq.~(\ref{rhocut}), if $N$ is also small) probabilities to have any small number of atoms $n$ in the noncondensate (for $n$ and $N$ up to a few tens). The first six of them are given in Eqs.~(\ref{rho0}) and (\ref{rho1}) (for $\rho_{n=0}^{(\infty)}$ and $\rho_{n=1}^{(\infty)}$) and below,
$$\rho_{n=2}^{(\infty)}=\rho_{n=0}^{(\infty)}\left(\frac{B_2}{2}+\frac{B_{1}^2}{2}\right),$$
$$\rho_{n=3}^{(\infty)}=\rho_{n=0}^{(\infty)}\left(\frac{B_3}{3}+\frac{B_{2}B_{1}}{2}+\frac{B_{1}^3}{3!}\right),$$
$$\rho_{n=4}^{(\infty)}=\rho_{n=0}^{(\infty)}\left(\frac{B_4}{4}+\frac{B_{3}B_{1}}{3}+\frac{B_{2}^{2}}{8}+\frac{B_{2}B_{1}^{2}}{4}+\frac{B_{1}^4}{4!}\right),$$
\begin{eqnarray}
\rho_{n=5}^{(\infty)}=\rho_{n=0}^{(\infty)} \Big(\frac{B_5}{5}+\frac{B_{4}B_{1}}{4}+\frac{B_{3}B_{2}}{6}+\frac{B_{2}^{2}B_{1}}{8}\nonumber\\
+\frac{B_{3}B_{1}^{2}}{6}+\frac{B_{2}B_{1}^3}{3!2}+\frac{B_{1}^5}{5!}\Big).
\label{r2345}
\end{eqnarray}
Note that the probabilities $\rho_{n}^{(\infty)}$ in Eqs.~(\ref{r2345}) are written in the form that is valid for arbitrary trap-size parameter $N_v$, not only in the thermodynamic limit. Analysis of actual, constraint-cut-off probabilities $\rho_n$ for small $n$ and $N$ as functions of the trap-size parameter $N_v$ is straightforward. In the thermodynamic limit they become rational functions of only first of the sums $B_j$ in Eq.~(\ref{Taylor2}), $B_{1}\to N_{v}/\zeta (3/2)$. Namely, they have a form $\rho_{n}=R_{n}(B_{1})/Q_{N}(B_{1})$, where $R_n$ and $Q_N$ are polynomials of orders $n$ and $N$, respectively, with definite numerical coefficients which are universal. The explicit dependence of probabilities and, hence, of all statistical and thermodynamic quantities on the trap-size parameter constitutes a strong finite-size effect and cannot be cast in a form of a universal function of some self-similar version of the variable $n$, contrary to the universality in the critical region (see Eqs.~(\ref{rhouniv}), (\ref{Thetauniv}), (\ref{xuniv}), (\ref{rhox}), and (\ref{x})). Here we skip that analysis and proceed to the analysis of the asymptotics for small noncondensate occupations $n \ll 3\sqrt{N_v}$. 

     In the latter case the leading term in the asymptotics of $\rho_{n}^{(\infty)}$ in Eq.~(\ref{multinomrho}) is the one with $m=n$, the next order term comes from $m=n-1$, and so on. Thus, we find the asymptotics
\begin{eqnarray}
\rho_{n}^{(\infty)}=\rho_{n=0}^{(\infty)}\frac{B_{1}^n}{n!}\Big\{ 1+\frac{n!B_2}{(n-2)!2B_{1}^2}+\frac{n!}{(n-4)!B_{1}^2}\Big[\frac{B_{2}^2}{8B_{1}^2}\nonumber\\
+\frac{B_3}{3(n-3)B_1}\Big]+\frac{n!}{(n-6)!B_{1}^3} \Big[\frac{B_{2}^3}{3!8B_{1}^3}+\frac{B_{2}B_3}{6(n-5)B_{1}^2}\nonumber\\
+\frac{B_4}{4(n-4)(n-5)B_1}\Big]+\ldots\Big\},\quad   n\ll\sqrt{N_v}.\quad
\label{rhoBa}
\end{eqnarray}

     The leading term in the asymptotics (\ref{rhoBa}) of the noncondensate occupation statistics,
\begin{equation}
\rho_{n}^{(P)(\infty)} = \rho_{n=0}^{(\infty)} B_{1}^{n}/n! ,
\label{Poisson0}
\end{equation}
is the same function of $n$ as a well-known Poisson distribution $e^{-B_1}B_{1}^{n}/n!$, but a normalization factor is essentially different since the unconstrained probability distribution $\rho_{n}^{(\infty)}$ is not Poissonian at $n >3\sqrt{N_v}$. The next to leading terms in asymptotics (\ref{rhoBa}) describe corrections to the Poisson distribution. Thus, we come to a general conclusion that in a very dilute ideal gas ($N\ll 3\sqrt{N_c}, T\gg T_c$) in the canonical ensemble the noncondensate occupation statistics is Poissonian and is not exponential, as the grand-canonical-ensemble approximation suggests (see the next subsection for details).

    Now we apply the constrain-cut-off solution in Eq.~(\ref{rhocut}) to the result in Eq.~(\ref{Poisson0}). That yields the cut-off Poisson distribution
\begin{equation}
\rho_{n}^{(P)} = \frac{\Gamma(N+1)e^{-B_{1}}B_{1}^{n}}{\Gamma(N+1,B_{1})n!} , \quad B_{1}=\sum_{\vec{k}\neq 0}{e^{-\epsilon_{\vec{k}}/T}},
\label{cutoffPoisson}
\end{equation}   
where $\Gamma(N+1,B_{1})$ is an incomplete gamma function \cite{a}, as well as the mean value and moments of the actual noncondensate occupation statistics in the trap with a small number of loaded atoms, $N\ll 3\sqrt{N_v}$. The cumulative distribution function for the cut-off Poisson distribution (\ref{cutoffPoisson}) is determined by a complimentary cumulative distribution function $Q(\chi^2|\nu)= \Gamma(N+1,B_{1})/\Gamma(N+1)$ of the chi-square $\chi^2$-distribution \cite{a} with $\nu=2(N+1)$ degrees of freedom and $\chi^2 =2B_1$, namely, 
\begin{equation}
P_{n}^{(P)} = \sum_{m=0}^{n}{\rho_{m}^{(P)}} = \frac{\Gamma(N+1)\Gamma(n+1,B_{1})}{\Gamma(N+1,B_{1})\Gamma(n+1)} .
\label{PPoisson}
\end{equation}
Its properties are well-known. All initial moments are given by the following formula
\begin{equation}
\left\langle n^m\right\rangle^{(P)} = \frac{\partial^{m}\left[e^{B_1}\Gamma(N+1,B_{1})\right]/\partial(\ln B_{1})^m}{e^{B_1}\Gamma(N+1,B_{1})}.
\label{mPoisson}
\end{equation}
In particular, the mean noncondensate occupation is equal to 
\begin{equation}
\left\langle n\right\rangle^{(P)} = B_1 - \frac{B_{1}^{N+1}e^{-B_1}}{\Gamma(N+1,B_{1})}
\label{meanPoisson}
\end{equation}
and the second cumulant (variance) $\kappa_{2}^{(P)}=\left\langle n^2\right\rangle^{(P)}-(\left\langle n\right\rangle^{(P)})^2$ is equal to 
\begin{equation}
\kappa_{2}^{(P)} = B_1 + \frac{(B_{1}-N-1)B_{1}^{N+1}}{e^{B_1}\Gamma(N+1,B_{1})} - \left[\frac{B_{1}^{N+1}}{e^{B_1}\Gamma(N+1,B_{1})}\right]^{2}.
\label{c2Poisson}
\end{equation}

    In the thermodynamic limit, when $N_v \gg 1$ and $B_{1}\approx N_{v}/\zeta (3/2)$, we find the mean noncondensate occupation to be close to the number of atoms in the trap $N$ and the variance to be much less than unity:
\begin{equation}
\left\langle n\right\rangle^{(P)}\approx N\left(1-\frac{1}{B_1}\right),\quad \kappa_{2}^{(P)}\approx \frac{N}{B_1}\left(N-1-\frac{N}{B_1}\right).
\label{TlimPoisson}
\end{equation}   

     Thus, the asymptotics of the actual probability distribution $\rho_n$ of the total noncondensate occupation in the small number of atoms region, $N\ll 3 \sqrt{N_v}$, is the cut-off Poisson distribution (\ref{cutoffPoisson}) with very steep slope rising to the sharp peak adjacent to the cut-off point $n=N$ as is shown schematically by the curve OA'N' in Fig. 1.

   \textbf{Grand-Canonical-Ensemble Approximation: Exponential Distribution.} The limit of the small number of atoms $N$, namely, $N_c - N \gg \sigma^{(\infty)}$, corresponds to a high-temperature regime of a classical gas without condensate and is well studied in the grand-canonical-ensemble approximation \cite{PitString,Koch06,LLV,LL,AGD,Pathria,Ziff}. In that approximation the occupations of all states, both in the condensate, $n_0$, and in the noncondensate, $n_{\vec{k}}$ , are treated as the independent stochastic variables with the probability distributions 
\begin{equation}
\rho_{n_{\vec{k}}} = \frac{\exp \left[-n_{\vec{k}}(\epsilon_{\vec{k}} - \mu )/T\right]}{\left\{1 - \exp \left[-(\epsilon_{\vec{k}} - \mu )/T\right]\right\}}
\label{gce-distr}
\end{equation}
and the particle-number constraint is satisfied only on average, $N = {\bar n}_0 + \sum_{k\neq 0}{{\bar n}_{\vec{k}}}$. This is achieved by bringing in an extra term $-\mu {\hat n}_{\vec{k}}$ into the Hamiltonian $H = \sum_{\vec{k}=0}^{\infty}{(\epsilon_{\vec{k}} - \mu ) {\hat n}_{\vec{k}}}$ and by choosing the chemical potential $\mu$ to satisfy the mean particle-number constraint $\sum_{k\neq 0}^{\infty}{\left(e^{(\epsilon_{\vec{k}}-\mu )/T} - 1\right)^{-1}} = N$. The chemical potential is negative, $\mu < 0$, and is directly related to the ground-state ($\epsilon_{\vec{k}=0} = 0$) occupation $\bar{n}_0 = \left(e^{-\mu /T} - 1\right)^{-1}$. 

	The condensate occupation distribution in Eq.~(\ref{gce-distr}), $\rho_{n_0} = e^{\mu n_{0}/T}/(1 - e^{\mu /T})$, implies a pure exponential approximation, 
\begin{equation}
\rho_{n} = e^{-\mu n/T} \frac{e^{\mu N/T}}{1 - e^{\mu /T}},\quad n \leq N, N_c - N \gg \sigma^{(\infty)},
\label{gce-ndistr}
\end{equation}
for the related to this case cut-off probability distribution $\rho_n$, represented by the curve OA'N' in Fig. 1. Although we know from the previous subsections that the left tail of the unconstrained probability distribution $\rho_{n}^{(\infty)}$ in Figs. 1-4 is not purely exponential (in fact, it is exponential with the cubic exponent (\ref{arhox}) at the left wing of the critical region and almost Poissonian (\ref{rhoBa}) near the far left end beyond the critical region), the grand-canonical-ensemble approximation is reasonable since the main contribution to the condensate statistics comes in this case from a relatively narrow (with a width of the order of a few dispersions) region, adjacent to the left of the point A' in Fig. 1. It is instructive to check explicitly how good the pure exponential grand-canonical-ensemble approximation (\ref{gce-ndistr}) fits the actual Poisson asymptotics (\ref{cutoffPoisson}) of the noncondensate occupation distribution near the cut-off point $n=N$. Basically, only the exponents matter for that comparison. For the Poisson distribution the exponential function near the cut-off point $n=N$ is $\exp \left\{n\left[\ln \left(N_{v}/(N\zeta (3/2))\right)\right]+1\right\}$, that corresponds to the following effective scaled chemical potential $\alpha^{(P)}=-\mu /T = 1+\ln \left[N_{v}/(N\zeta (3/2))\right]$ in Eq.~(\ref{gce-distr}). In the grand-canonical-ensemble method the value of $\alpha$ is determined from the self-consistency equation (see, e.g., \cite{Ziff,Wang2004})
\begin{equation}
N = N_{v}g_{3/2}(\alpha)/\zeta(3/2) + \bar{n}_0 ,
\label{mugce}
\end{equation}
where the mean condensate occupation that far from the critical region is infinitesimal, $\bar{n}_{0}\approx 0$, and the Bose function, or polylogarithm, \cite{Bateman,Robinson}
\begin{equation}
g_{m}(\alpha)\equiv Li_{m}(e^{-\alpha})=\frac{1}{\Gamma(m)}\int_{0}^{\infty}{\frac{t^{m-1}dt}{e^{t+\alpha}-1}}=\sum_{j=1}^{\infty}{\frac{e^{-\alpha j}}{j^m}}
\label{Bosefun}
\end{equation}
can be approximated by its asymptotics $e^{-\alpha}$ for $\alpha \gg 1$. Hence, the grand-canonical-ensemble method yields the exponent $\alpha^{(GC)}=-\mu /T = \ln \left[N_{v}/(N\zeta (3/2))\right]$ which, indeed, makes only relatively small difference with the Poisson value, $(\alpha^{(P)}-\alpha^{(GC)}) \ll \alpha^{(P)}$.

     Obviously, the smaller is the interval of the allowed noncondensate occupations $\left[0, N\right]$, i.e., the smaller is the number of atoms in the trap, the better is the grand-canonical-ensemble approximation in Eq.~(\ref{gce-ndistr}). Besides, all calculations, utilizing the pure exponential distribution in Eq.~(\ref{gce-ndistr}), are elementary \cite{Koch06,Pathria}. 

	The result is the explicit asymptotics for the average condensate occupation and central moments and cumulants of the total noncondensate occupation as well as for the thermodynamic quantities, which are discussed below in Sections VI-VII and XII-XIV. An agreement with the exact numerical simulations in the region of application of this approximation, $N_c - N \gg \sigma^{(\infty)}$, is very good. However, of course, in the whole critical region, $\left|N_c - N\right| \leq 2\sigma^{(\infty)}$, and for the region of the well developed BEC, $N - N_c \gg 2\sigma^{(\infty)}$, the grand-canonical-ensemble approximation fails.

\section{V. UNIVERSAL CONSTRAINT-CUT-OFF MECHANISM OF STRONGLY NON-GAUSSIAN BEC FLUCTUATIONS}

\subsection{A. Cut-Off Distribution: Origin of Nonanaliticity and Strong Non-Gaussian Effects in Critical Fluctuations}
   
     Let us apply now the remarkable universality and general constraint-cut-off approach to the analysis of various effects of BEC in the mesoscopic systems with a finite number of atoms in the trap in the critical region as well as below and above the critical region. To this end, we have to introduce a finite number of atoms $N \sim N_c$, $N < N_c$, and $N > N_c$, respectively, and to perform a cut off of the probability distribution $\rho_{n}^{(\infty)}$ dictated by the particle-number constraint as was formulated in Eq.~(\ref{rhocut}) in Sec. II. An immediate result is that the actual, cut off probability distribution (OAN or OA'N' in Fig. 1) is strongly asymmetric and peculiar for all $N < N_c$ and $N \sim N_c$, including the critical region. In terms of the Landau function \cite{Goldenfeld,Sinner}, $-\ln \rho_n$, this result means that, even without any interatomic interaction in the gas, the constraint nonlinearity, that originates from many-body Fock space cut off in the canonical ensemble, produces the infinite potential wall at $n=N$ in the effective fluctuation Hamiltonian and makes it highly asymmetric. We find that the outlined constraint-cut-off mechanism is responsible for all unusual critical phenomena of the BEC phase transition in the ideal gas and, in particular, makes the BEC statistics strongly non-Gaussian. In the deeply condensed region, $N - N_c \gg \sigma^{(\infty)}$, the non-Gaussian behavior is less pronounced but remains finite even in the thermodynamic limit due to the discussed above non-Gaussian asymmetric tails found in \cite{KKS-PRL,KKS-PRA}. 

\subsection{B. Cut-Off Distribution As the Exact Solution to the Recursion Relation: Rigorous Proof}

     There is another way to prove that the exact solution for the noncondensate occupation probability distribution is the constraint-cut-off distribution (\ref{rhocut}). Namely, we can directly prove that (\ref{rhocut}) is the solution to the well-known exact recursion relation \cite{Landsberg,Borrmann1993,Brosens1997,ww1997,Holthaus1997,Balazs1998,recursion1999,Borrmann1999,Kleinert2007,Wang2009}
\begin{equation}
Z_{n}(T) = \frac{1}{n}\sum_{p=1}^{n}{Z_{1}(T/p)Z_{n-p}(T)} 
\label{recursion}
\end{equation}
for the cumulative distribution function multiplied by an independent on $n$ factor $\sum_{m=0}^{N}{\rho_{m}^{(\infty)}/\rho_{n=0}^{(\infty)}}$,
\begin{equation}
Z_{n}(T) = \sum_{m=0}^{n}{\rho_{m}^{(\infty)}/\rho_{n=0}^{(\infty)}} . 
\label{ZnT}
\end{equation}
In the recursion relation (\ref{recursion}), it is assumed that the lower order functions $Z_{n}(T), n= -1, 0, 1$, are as follows
\begin{equation}
Z_{1}(T/p) = 1+B_p ,\quad  Z_{0}(T) = 1,\quad Z_{-1}(T) = 0,
\label{Z-101}
\end{equation}
where $B_p$ is defined in Eq.~(\ref{Taylor2}).

    We start the proof with calculation of the function (\ref{ZnT}) for the distribution (\ref{rhocut}) using Eq.~(\ref{rhon}) as follows
$$Z_{n}(T) = \Big[\frac{1}{n!}\frac{d^n}{dz^n}\exp \left(\sum_{i=1}^{n}{\frac{B_{i}z^{i}}{i}}\right)$$
\begin{equation}
+\sum_{m=0}^{n-1}{\frac{1}{m!}\Big[\frac{d^m}{dz^m}\exp \left(\sum_{i=1}^{n-1}{B_{i}z^{i}/i}\right)\Big]e^{B_{n}z^{n}/n}}\Big]_{z=0}. 
\label{ZnT1}
\end{equation}
The second term in Eq.~(\ref{ZnT1}) is equal to $Z_{n-1}(T)$, while the first term is equal to
\begin{equation}
\frac{1}{n!}\left[\frac{d^{(n-1)}}{dz^{(n-1)}}\left(\sum_{p=1}^{n}{B_{p}z^{p-1}}\right)e^{\sum_{i=1}^{n}{B_{i}z^{i}/i}}\right]_{z=0} 
\label{ZnT2}
\end{equation}
$$=\sum_{p=1}^{n}{\frac{B_p}{n(n-p)!}\left[\frac{d^{n-p}}{dz^{n-p}}e^{\sum_{i=1}^{\infty}{B_{i}z^{i}/i}}\right]_{z=0}} =\sum_{p=1}^{n}{\frac{B_{p}\rho_{n}^{(\infty)}}{n\rho_{n=0}^{(\infty)}}},$$    
where we used a well-known formula for the n-th derivative of a product of two functions, $d^{n}(fg)/dz^{n}=\sum_{m=0}^{n}{C_{n}^{m}f^{(m)}g^{(n-m)}}$, and Eq.~(\ref{rhon}).

    Then, using definition (\ref{ZnT}) for the right side of Eq.~(\ref{ZnT2}) and combining both terms in Eq.~(\ref{ZnT1}), we find that the function (\ref{ZnT}) for the distribution (\ref{rhocut}) satisfies the following recursive relation 
$$Z_{n}(T) = Z_{n-1}(T) + \frac{1}{n}\sum_{p=1}^{n}{B_{p}\left[Z_{n-p}(T)-Z_{n-p-1}(T)\right]}$$
\begin{equation}
= Z_{n-1} +\sum_{p=1}^{n}{\frac{1+B_{p}}{n}\left[Z_{n-p}-Z_{n-p-1}\right]}-\frac{Z_{n-1}}{n}.
\label{ZnT3}
\end{equation}
It can be rewritten in the form $f_{n}=f_{n-1}$, where $f_{n} =nZ_{n}-\sum_{p=1}^{n}{(1+B_{p})Z_{n-p}}$, that means that the quantity $f_n$ is an independent on $n$ constant. Moreover, the latter constant is equal to zero, $f_{n}=0$, since for $n=1$ one has $f_{1}=Z_{1}-(1+B_{1})Z_{0}=0$ due to definition (\ref{Z-101}).

     Thus, we find that the normalized cumulative distribution function (\ref{ZnT}) for the constraint-cut-off distribution (\ref{rhocut}) satisfies the recursion relation $nZ_{n}(T) =\sum_{p=1}^{n}{(1+B_{p})Z_{n-p}(T)}$ which is precisely the same as the renowned recursion relation (\ref{recursion}). That completes the proof.

\section{VI. UNIVERSAL SCALING AND STRUCTURE OF THE BEC ORDER PARAMETER}

In accord with the constraint-cut-off mechanism, depicted in Fig. 1, the mean noncondensate occupation $\bar n$ almost linearly follows the cut-off value $N$ of the number of loaded in the trap atoms until its value $\bar n$ saturates at the critical level $N_c$, when $N$ passes through the critical value $N_c$ by an amount about $2\sigma^{(\infty)}$ (Fig. 5). The complimentary, mean condensate occupation ${\bar n}_0$, i.e. the BEC order parameter, has a similar, but upside-down pattern that, with an increase of trap-size parameter $N_v$, becomes a degenerate straight-line angle OCB, which represents the BEC behavior before and after the phase transition as it is approximated by the standard Landau mean-field theory in the thermodynamic limit. From this point of view, a universal fine structure of the critical region in the BEC phase transition is missing.
 
\begin{figure}
\center{\epsfig{file=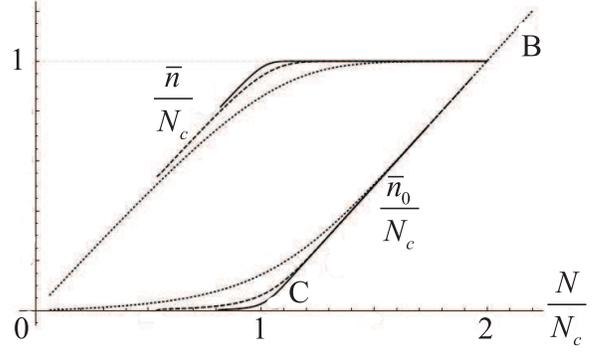,width=8cm}}
\caption{The mean occupations of the noncondensate, ${\bar n}/N_c$, and the condensate, ${\bar n}_{0}/N_c$, as the functions of the number of atoms, $N/N_c$, loaded in the trap; all quantities are normalized by the critical number of atoms $N_c$ from Eq.~(\ref{Nc}): $N_v = 10^2$ - dotted line,  $N_v = 10^3$ - dashed line, $N_v = 10^4$ - solid line.} 
\end{figure}
 
     To unveil and resolve the universal scaling and structure of the BEC order parameter near a critical point we divide both the function and the argument by the dispersion of the BEC fluctuations $\sigma^{(\infty)}$, Eq.~(\ref{sigma}), and calculate a scaled condensate occupation ${\bar n}_{0}' = {\bar n}_0 /\sigma^{(\infty)}$ as a function of a scaled deviation from the critical point 
\begin{equation}
\eta = (N- N_c )/\sigma^{(\infty)}. 
\label{eta}
\end{equation}
We find that with an increase of the trap-size parameter $N_v$ the function ${\bar n}_{0}'(\eta )$ quickly converges to a universal regular function 
\begin{equation}
F_{0}(\eta ) = \eta - \int_{-\infty}^{\eta}{x\rho_{x}^{(univ)}dx}/\int_{-\infty}^{\eta}{\rho_{x}^{(univ)}dx}, 
\label{n0eta}
\end{equation} 
\begin{figure}
\center{\epsfig{file=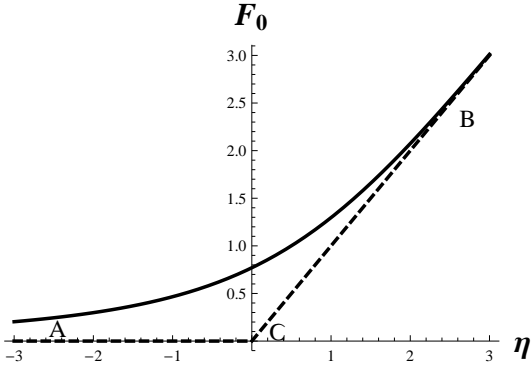,width=7cm}}
\caption{Universal function $F_{0}(\eta)$, Eq.~(\ref{n0eta}), of the scaled BEC order parameter $\bar{n}_{0}'=\bar{n}_{0}/\sigma^{(\infty)}$as a function of $\eta =(N-N_{c})/\sigma^{(\infty)}$ in the critical region. The angle ACB represents the prediction of the standard Landau mean-field theory.} 
\end{figure}
which describes the universal structure of the BEC order parameter in the critical region in the thermodynamic limit $N_v \to \infty$, as is shown in Figs. 6 and 7. Presented above explicit formula for the universal function $F_{0}(\eta )$ immediately follows from the  formula for the exact, cut-off probability distribution (\ref{rhocut}) (see Sections II and V) and from the universal probability distribution (\ref{rhouniv}), or (\ref{rhox}), analyzed in all details in Sec. IV. The cumulative distribution function 

\begin{figure}
\center{\epsfig{file=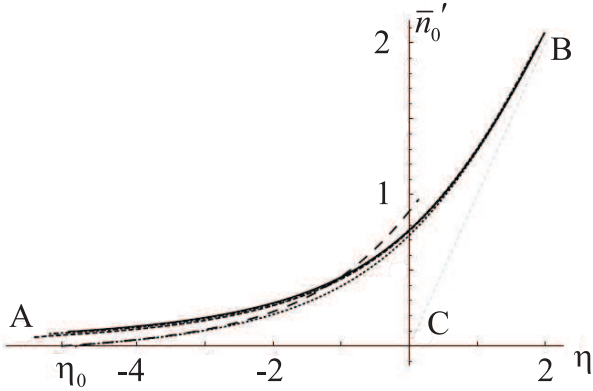,width=8.5cm}}
\caption{The structure of the scaled BEC order parameter ${\bar n}_{0}' = {\bar n}_0 /\sigma^{(\infty)}$ as a function of $\eta = (N- N_c )/\sigma^{(\infty)}$ in the critical region: the solid line is the function ${\bar n}_{0}'(\eta )$ for the mesoscopic system with the trap-size parameter $N_v = 10^4$, the dashed line - for $N_v = 10^3$, the dotted line - for $N_v = 10^2$. The dashed-dotted line is the simple fit, Eq.~(\ref{univ-eta}), of the universal function $F_{0}(\eta )$, Eq.~(\ref{n0eta}). The long-dashed line represents the result within the grand-canonical-ensemble approximation in Eq.~(\ref{gce-ndistr}) for $N_v = 10^2$. The angle ACB represents the prediction of the standard Landau mean-field theory.} 
\end{figure}

\begin{equation}
P_{\eta}^{(univ)}  = \int_{-\infty}^{\eta}{\rho_{x}^{(univ)}dx} ,  
\label{Puniv}
\end{equation}
which is required for proper normalization of the cut-off probability distribution and, hence, is present in the denominator in Eq.~(\ref{n0eta}), determines the universal Gibbs free energy and is analyzed in Sec. XII (see Fig. 12). 

     The exact analytical result in Eq.~(\ref{n0eta}) for the universal structure of the BEC order parameter in the critical region can be easily written in terms of the polynomial, exponential, Kummer's confluent hypergeometric, and parabolic cylinder functions if we use the explicit formulas for $\rho_{x}^{(univ)}$ in Eqs.~(\ref{rhounivKummer}) and (\ref{rhounivPC}) for the central part of the critical region and in Eqs.~(\ref{arhox}) and (\ref{rap}) for the left (condensed) and right (noncondensed) wings of the critical region to calculate an explicit integral in Eq.~(\ref{n0eta}). We skip these straightforward expressions in order do not overload the paper with formulas. The universal function of the BEC order parameter $F_{0}(\eta)$ is depicted in Fig. 6 and is truly universal since it contains no free or any physical parameters of the system at all and involve only pure mathematical numbers $\pi$ and $s_m$ defined in Eq.~(\ref{sm}).

    The result (\ref{n0eta}) is very different from the prediction of the Landau mean-field theory shown by the broken line ACB in Fig. 7. We can immediately conclude that even for the small mesoscopic systems with $N_v \sim 10^2$ the difference between the universal order-parameter and the mesoscopic order-parameter functions is relatively small, $\left|F_{0}(\eta ) \sigma^{(\infty)} - {\bar n}_{0}(\eta )\right| \ll {\bar n}_{0} (\eta )$. This statement is true everywhere except the very beginning of the curve ${\bar n}_{0}(\eta )$, where the system is not mesoscopic anymore, there are only a few atoms in the trap $N \leq 10$, and, obviously, the number of atoms in the condensate should become exactly zero, ${\bar n}_0 = 0$, at the end point $n=0$, i.e. at 
\begin{equation}
\eta_0 (N_{v}) = -N_{c}/\sigma^{(\infty)} \sim -N_{v}^{1/3} ,
\label{endpoint}
\end{equation}  
where there are no atoms in the trap, $N=0$, as is seen in Fig. 6 at $\eta_0 = -5.1$ for $N_v = 10^2$. At the critical point, where the number of atoms in the trap is critical, $N = N_c$, we find that the order parameter just reaches a level of fluctuations, ${\bar n}_0 \approx 0.77 \sigma^{(\infty)}$. 

	We find an elementary fit for the universal function in Eq.~(\ref{n0eta}) that is good in the critical region at $\left|\eta \right| < 5$ with an accuracy of order of few percents as is shown in Fig. 7. That fit involves only the elementary functions if we consider an inverse function $\eta = g_{0}(y)$, where $y = {\bar n}_{0}/(2^{1/2}\sigma^{(\infty)})$ is a scaled condensate occupation. Namely, the fit is  
\begin{equation}
\eta = g_{0}(y) \approx 2^{1/2} \left(1 - \frac{e^{-5y/3}}{y^{3/2}}\right) y .
\label{univ-eta}
\end{equation}
A small difference between the universal and actual order-parameter curves for the finite mesoscopic system with the trap-size parameter $N_v = 10^4$ is even hardly seen in Fig. 7. 

    The standard Landau mean-field theory does not resolve the smooth, regular universal structure in Figs. 5-7.
    
       We stress again that the well-known grand-canonical-ensemble approximation fails \cite{Koch06,Pathria} in the whole critical region, $\left|N_c - N\right| \leq 2\sigma^{(\infty)}$, and in the region of the well developed BEC, $N - N_c \gg 2\sigma^{(\infty)}$. It is valid only in the limit of the small number of atoms, $N_c - N \gg \sigma^{(\infty)}$, which corresponds to a high-temperature regime of a classical gas without condensate, as it was discussed in Sec. IV. The main excuse for the grand-canonical-ensemble approximation is simplicity of all its calculations utilizing the pure exponential distribution in Eq.~(\ref{gce-ndistr}). The result is the explicit asymptotics for the average condensate occupation, ${\bar n}_0$, which is depicted in Fig. 7 by the long-dashed lines for the mesoscopic system with the trap-size parameter $N_v = 10^2$. An agreement with the exact numerical simulations is very good only far from the critical point, namely, in the region of application of this approximation, $N_c - N \gg \sigma^{(\infty)}$. 

\section{VII. UNIVERSAL SCALING AND STRUCTURE OF ALL HIGHER-ORDER CUMULANTS AND MOMENTS OF THE BEC FLUCTUATIONS}

     As a direct consequence of the universality of the noncondensate occupation probability distribution formulated in Sec. IV, we find that all higher-order moments and cumulants of the BEC fluctuations also have the universal scaling and smooth nontrivial structure. The analysis is similar to the one developed in Sec. VI for the order parameter and is based on the calculation of the scaled central moments $\mu_{m}' = \mu_{m}/(\sigma^{(\infty)})^m$ and scaled cumulants $\kappa_{m}' = \kappa_{m}/(\sigma^{(\infty)})^m$ as the functions of the scaled deviation from the critical point, $\eta = (N- N_c )/\sigma^{(\infty)}$. We find that with an increase of the trap-size parameter $N_v$ the functions $\mu_{m}'(\eta )$ and $\kappa_{m}'(\eta )$ quickly converge to the universal functions 
\begin{equation}
M_{m}(\eta ) = \int_{-\infty}^{\eta}{(x-\bar{x})^{m} \rho_{x}^{(univ)}dx}/\int_{-\infty}^{\eta}{\rho_{x}^{(univ)}dx}, 
\label{Meta} 
\end{equation}   
and 
\begin{equation}
C_{m}(\eta ) = \sum_{r=1}^{m}{\frac{(r-1)!}{(-1)^{r-1}}\sum^{(m,r)}{(m;a_{1},\ldots,a_{m})'\alpha_{1}'^{a_1}\ldots \alpha_{m}'^{a_m}}}, 
\label{Ceta} 
\end{equation} 
respectively, where we used Eqs.~(\ref{moments}) and (\ref{cumulants}) and the universal initial moments
\begin{equation}
\alpha_{m}' =\int_{-\infty}^{\eta}{x^{m} \rho_{x}^{(univ)}dx}/\int_{-\infty}^{\eta}{\rho_{x}^{(univ)}dx},\quad \bar{x}=\alpha_{1}'.
\label{meaneta} 
\end{equation}

\begin{figure}
\center{\epsfig{file=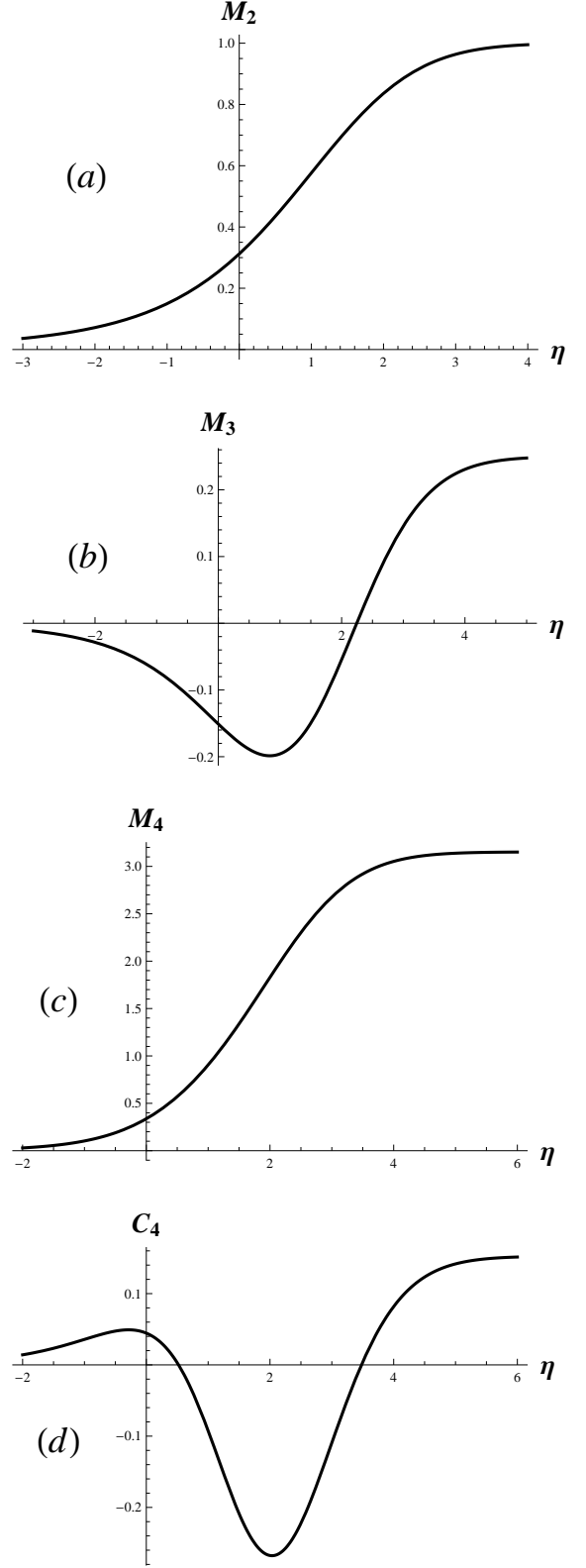,width=7.5cm}}
\caption{Universal functions $M_{m}(\eta)$, Eq.~(\ref{Meta}), and $C_{m}(\eta)$, Eq.~(\ref{Ceta}), of the scaled central moments and cumulants (a) $\mu_{2}' \equiv \kappa_{2}' = \mu_{2}/(\sigma^{(\infty)})^2$, (b) $\mu_{3}' \equiv \kappa_{3}' = \mu_{3}/(\sigma^{(\infty)})^3$, (c) $\mu_{4}' = \mu_{4}/(\sigma^{(\infty)})^4$, and (d) $\kappa_{4}' \equiv \mu_{4}' - 3(\mu_{2}')^2$ of the total noncondensate occupation in the critical region as the functions of $\eta = (N- N_c )/\sigma^{(\infty)}$.} 
\end{figure}
They describe the universal structure of the BEC critical fluctuations in the thermodynamic limit, $N_v \to \infty$, as is shown in Figs. 8 and 9 for the second, third, and forth moments and cumulants of the noncondensate occupation. Presented above explicit formulas for these universal functions immediately follow from exact formulas in Eq.~(\ref{rhocut}) and Eq.~(\ref{rhouniv}), or (\ref{rhox}), similar to the derivation of Eq.~(\ref{n0eta}). The functions $M_{m}(\eta )$ and $C_{m}(\eta )$ do not involve any physical parameters of the system and, therefore, are truly universal. They are depicted in the separate Fig. 8 (for $m = 2, 3, 4$) since these thermodynamic-limit functions practically coincide with the corresponding functions for the mesoscopic system with the trap-size parameter $N_v = 10^4$. Corresponding central moments and cumulants for the mesoscopic systems with different finite values of the trap-size parameter are depicted in Fig. 9. The universal behavior is clearly observed starting from very small mesoscopic systems with a typical number of atoms $N_v \sim 10 \div 100$. 

\begin{figure}
\center{\epsfig{file=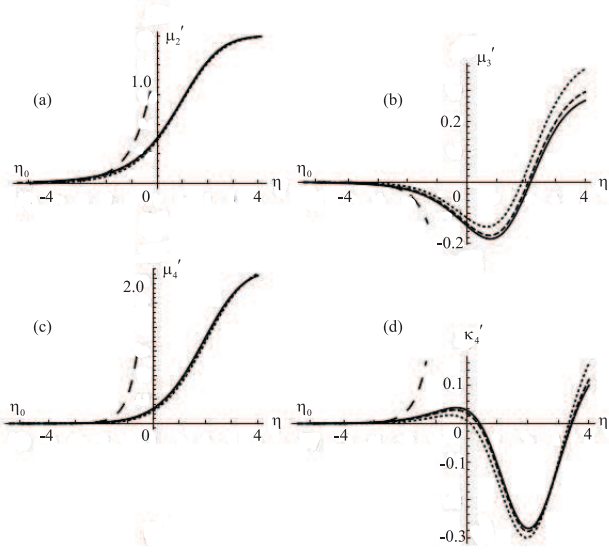,width=8.5cm}}
\caption{Scaled central moments and cumulants (a) $\mu_{2}' \equiv \kappa_{2}' = \mu_{2}/(\sigma^{(\infty)})^2$, (b) $\mu_{3}' \equiv \kappa_{3}' = \mu_{3}/(\sigma^{(\infty)})^3$, (c) $\mu_{4}' = \mu_{4}/(\sigma^{(\infty)})^4$, and (d) $\kappa_{4}' \equiv \mu_{4}' - 3(\mu_{2}')^2$ of the total noncondensate occupation in the critical region calculated as the functions of $\eta = (N- N_c )/\sigma^{(\infty)}$ for the mesoscopic system with the trap-size parameter $N_v = 10^2$ (dotted line), $N_v = 10^3$ (dashed line), $N_v = 10^4$ (solid line). The long-dashed line represents the result within the grand-canonical-ensemble approximation in Eq.~(\ref{gce-ndistr}) for $N_v = 10^2$.} 
\end{figure}

     An essential deviation from the universal curves takes place only at the very beginning of each curve near the end point (\ref{endpoint}), $\eta_0$, where the number of atoms in the trap is zero and, hence, all fluctuations $\mu_{m}'(\eta )$ and $\kappa_{m}'(\eta )$ should be exactly zero, as is seen in Fig. 7 at $\eta_0 \approx -5.1$ for $N_v = 10^2$. In this limit the system loses its mesoscopic status and can be studied quantum mechanically as a microscopic system of a few atoms $N = 1, 2, 3, \ldots$. 
     
     Qualitatively, behavior of the moments and cumulants, depicted in Figs. 8 and 9, can be immediately predicted on the basis of the constraint-cut-off mechanism using Fig. 1. The variance $\mu_2 \equiv \kappa_2$ has to grow monotonically with increasing $N$ and to have a maximum derivative at $N \approx N_c$ because the width of the cut-off probability distribution OAN increases when the cut-off boundary AN moves to the right and the maximum width's derivative is achieved at the center of the critical region. That behavior, indeed, is found in the universal function $M_{2}(\eta)$ and in our numerical simulations depicted in Figs. 8a and 9a, respectively. 
     
     The third central moment, or the third cumulant, $\mu_3 \equiv \kappa_3 =\left\langle (n - {\bar n})^{3}\right\rangle$, is the main characteristic of an asymmetry of the probability distribution relative to the mean value $\bar n$. For small enough numbers of atoms in the trap $N$, when the probability distribution has a strongly asymmetric, "curved-triangle" shape OA'N' in Fig. 1, the value of the asymmetry $\mu_3$ is negative due to a large contribution from the left tail and increases in magnitude with increasing $N$ until some maximum-in-magnitude negative value is reached. When the number of atoms $N$ enters the central part of the critical region, the absolute value of the asymmetry $\left|\mu_3\right|$ decreases and after passing through the critical point $N = N_c$ approaches zero, since the shape of the cut-off probability distribution OAN in Fig. 1 becomes more and more symmetric. Finally, the asymmetry coefficient $\mu_{3}' = \mu_{3}/(\sigma^{(\infty)})^3$ changes the sign and tends to a finite positive value $\mu_{3}'^{(\infty)} = 0.41, 0.32, 0.29$ for the values $N_v = 10^{2}, 10^{3}, 10^4$, respectively, that is a characteristic feature of the unconstrained probability distribution $\rho_{n}^{(\infty)}$ due to a large positive contribution of the fat and wide right tail, discussed in Sec. IV and Figs. 3 and 4. The predicted behavior of the asymmetry $\mu_3 = \kappa_3$ is precisely revealed in the universal function $M_{3}(\eta )=C_{3}(\eta )$ and in the simulations presented in Figs. 8b and 9b, respectively. 
     
     In a similar way, one can explain the depicted in Figs. 8c,d and 9c,d behavior of the forth moment $\mu_4$ and the forth cumulant, the excess $\kappa_4$. The latter, in general, characterizes a positive excess (if $\kappa_4 > 0$) or a deficit (if $\kappa_4 < 0$) of the flatness of the "plateau" of the probability distribution relative to the flatness of the plateau of the Gaussian distribution. Again, one has to take into account that the unconstrained probability distribution $\rho_{n}^{(\infty)}$, according to Figs. 3 and 4, is more flat than the Gaussian distribution, that is it has a positive excess coefficient $\kappa_{4}'^{(\infty)} \approx 0.29, 0.22$, and $0.19$ for the values $N_v = 10^2, 10^3$, and $10^4$, respectively. 
     
     Again, as is shown in Fig. 8, the grand-canonical-ensemble approximation is valid only in the limit of the small number of atoms, $N_c - N \gg \sigma^{(\infty)}$, that is in the high-temperature regime of a classical gas without condensate as it was discussed in Sections IV and VI. 
     
     The other limit, opposite to the high-temperature case, is the limit when the number of atoms is large, $N - N_c \gg 2\sigma^{(\infty)}$. It corresponds to a low-temperature regime of the fully developed condensate. In this limit, the cut-off part of the probability distribution in Fig. 1 contains only an unimportant end piece of the right tail. Thus, the mean value as well as all moments and cumulants of the noncondensate occupation tend to the constants, which are precisely their unconstrained values analytically calculated in \cite{KKS-PRL,KKS-PRA}. In particular, the limiting values of the scaled cumulants are equal to $\kappa_{m}'^{(\infty)} =\tilde{\kappa}_{m}^{(\infty)}/(\sigma^{(\infty)})^{m} =(m-1)!s_{m}/s_{2}^{m/2}$ so that the asymmetry and excess coefficients tend to $\mu_{3}'^{(\infty)} =2s_{3}/s_{2}^{3/2} \approx 0.25$ and $\kappa_{4}'^{(\infty)} = 6s_{4}/s_{2}^{2} \approx 0.15$, respectively. We find that this is indeed true, as is clearly seen in Figs. 5-9. 
     
     It is a straightforward exercise to write down explicit formulas for the universal functions $F_{0}(\eta ), M_{m}(\eta )$, and $C_{m}(\eta )$ of the order parameter, central moments, and cumulants as the simple integrals in Eqs.~(\ref{n0eta}), (\ref{Meta}), and (\ref{Ceta}) via the universal unconstrained probability distribution $\rho_{x}^{(univ)}$ given by the explicit analytical formulas in Eqs.~(\ref{rhounivKummer}) and (\ref{rhounivPC}) (the central part of the critical region) and in Eqs.~(\ref{arhox}) and (\ref{rap}) (asymptotics of the left and right wings of the critical region). These formulas will be presented elsewhere. 

\section{VIII. EXACTLY SOLVABLE CUT-OFF GAUSSIAN MODEL OF BEC STATISTICS}

In the critical region, the universal scaling and structure of the BEC statistics found in Sections IV-VII can be qualitatively explained within a pure Gaussian model for the unconstrained probability distribution of the total noncondensate occupation $n \in \left[0, \infty \right)$, 
\begin{equation}
\rho_{n}^{(\infty)} = \exp\left[-\frac{(n - N_{c})^2}{2\sigma^2}\right]/\sum_{m=0}^{\infty}{\exp\left[-\frac{(m - N_{c})^2}{2\sigma^2}\right]}.
\label{gauss}
\end{equation}
It is depicted in Figs. 3 and 4. That model corresponds to a degenerate interacting gas of $N$ trapped atoms with a very degenerate interaction between the excited atoms in the noncondensate and the ground-state ($\epsilon_{\vec{k}=0} = 0$), condensed atoms, described by the Hamiltonian $H = \left(\sum_{\vec{k}\neq 0}{{\hat n}_{\vec{k}}} - N_{c}\right)^{2} T/\left(2\sigma^{2}\right)$ and the equilibrium density matrix in Eq.~(\ref{hatrho}). 

     The two parameters of the model, $\sigma$ and $N_c$, correspond, respectively, to the dispersion $\sigma^{(\infty)}$ and the critical number of atoms $N_c$ used for the ideal gas in the box in the previous sections. In order to compare the results for the Gaussian model with the results for the ideal gas in the box, we assume, following Eqs.~(\ref{sigma}) and (\ref{sigmainfinity}), that $\sigma = \sigma^{(\infty)} \approx \left(s_{2}^{1/2}/\pi\right)\left[N_{v}/\zeta(3/2)\right]^{2/3}$, where $N_v$ depends on $N_c$ in accord with Eq.~(\ref{Nc}). 
     
     The mean value and all moments and cumulants of the noncondensate occupation within the Gaussian model can be calculated exactly. We find their universal structures in the thermodynamic limit, $N_c \to \infty$, in terms of the error function $erf(x)$ and the related special functions, since the probability distribution of the scaled variable $x = (n - N_{c})/\sigma$ becomes a standard continuous unrestricted Gaussian distribution 
\begin{equation}  
\rho_{x}^{(Gauss)(\infty)} = \frac{\exp (-x^{2}/2)}{\sqrt{2\pi}}, \quad x \in (-\infty, \infty), 
\label{puregauss}
\end{equation}
and a continuous approximation of the discrete sums by the integrals is applied. For simplicity, we extend an allowable interval of the variable $x$ until $-\infty$ since the negative values $n < 0$, i.e. $x < -N_{c}/\sigma$, make exponentially small contribution in the only interesting for us case of relatively large critical number of atoms $N_c \gg \sigma$. All cumulants of the unconstrained Gaussian distribution $\rho_{x}^{(\infty)}$ are zero, except the variance $\kappa_{2}'^{(\infty)} = \mu_{2}'^{(\infty)} = 1$, that is $\kappa_{m}'^{(\infty)} = 0$ for $m \neq 2$. 
     
     However, the actual physical system of $N$ atoms in the trap is described by the constraint-cut-off probability distribution 
\begin{equation}
\rho_{x}^{(Gauss)} = \frac{\exp (-x^{2}/2) \theta(\eta - x)}{\int_{-\infty}^{\eta}{\exp (-x^{2}/2)dx}}, \eta = \frac{N - N_{c}}{\sigma} ,
\label{cutoffgauss}
\end{equation}
as is discussed in Sec. II and Fig. 1. This actual distribution, in a general case, is essentially non-Gaussian and, hence, all cumulants are nonzero, $\kappa_{m}'^{(\infty)} \neq 0$ for $m \geq 2$. Nevertheless, to find the mean value and all moments is easy, in particular, 
\begin{equation}
\bar x = -(2/\pi )^{1/2}\exp(-\eta^{2}/2)/\left[1 + erf(\eta /2^{1/2})\right] . 
\label{meanx}
\end{equation}
Thus, the universal structure of the order parameter ${\bar n}_0 = N - \bar n$ in the Gaussian model is given by the following analytical formula 
\begin{equation}
\frac{{\bar n}_0}{\sigma} = \eta + \left(\frac{2}{\pi}\right)^{1/2} \frac{\exp(-\eta^{2}/2)}{1 + erf(\eta /2^{1/2})} .
\label{n0gauss}
\end{equation}

     Following a tradition of the previous sections, we skip all elementary derivations and proceed to the results. 
     
     In the whole critical region the result for the universal structure of the scaled order parameter ${\bar n}_{0}'(\eta ) = {\bar n}_{0}/\sigma^{(\infty)}$ in the Gaussian model, given by the exact analytical solution in Eq.~(\ref{n0gauss}), is very close to the universal structure of the order parameter in the ideal gas in the box. 
     
     Comparison of the universal structures of the higher-order moments and cumulants (the variance $\kappa_2$, the asymmetry $\kappa_3$, and the excess $\kappa_4$) of the Gaussian model with the corresponding functions of the ideal gas in the box proves that in the whole critical region they have qualitatively similar structures, which are governed by the universal constraint-cut-off mechanism as it is explained in Sections IV and VII. Of course, the details of these structures, especially far from the critical region, are different since the tails of the unconstrained probability distribution $\rho_{n}^{(\infty)}$ in the ideal gas are essentially non-Gaussian and asymmetric. The latter fact is the reason why all cumulants $\kappa_m$, except the variance $\kappa_2$, vanish in the deeply condensed region, $N \to \infty$, in the Gaussian model and remain finite, even in the thermodynamic limit, in the ideal gas in the box. 
     
     A remarkable general conclusion is that in the whole critical region all cumulants are essentially nonzero (i.e., the BEC statistics is essentially non-Gaussian) for the mesoscopic systems of any size as well as for the macroscopic systems in the thermodynamic limit, both for the pure Gaussian model and for the ideal gas in the trap. 
     
\section{IX. EXACTLY SOLVABLE TWO-LEVEL TRAP MODEL OF BEC}

Let us consider the BEC of $N$ atoms in a trap with just two energy levels, the ground level $\epsilon_0 = 0$ and one excited level $\epsilon > 0$, but allow the excited level to contain arbitrary number $g \geq 1$ of degenerate states. Our idea behind this model is to isolate and study a contribution of a subset of closely spaced one-particle energy levels in the trap to the BEC phenomenon. It is similar to modeling of an inhomogeneously broaden optical transition in quantum optics by a homogeneously broaden two-level atoms. 

\subsection{A. Exact Discrete Statistics: Cut-Off Negative Binomial Distribution}

We can easily find the unconstrained probability distribution and characteristic function of the total noncondensate occupation as a superposition of $g$ identical random variables,
\begin{equation}
\rho_{n}^{(2)(\infty)} = \frac{(g-1+n)!}{n! (g-1)!} (1-q)^{g}q^n ,
\label{rho2infty}
\end{equation}
$$\Theta^{(2)(\infty)}(u) = \left(\frac{1-q}{1-zq}\right)^g , \quad z = e^{iu} , \quad q = e^{-\epsilon /T}.$$
It is a well-known negative binomial distribution \cite{a} which has the following generating cumulants $\tilde{\kappa}_{m}^{(2)(\infty)} = g\left[q/(1-q)\right]^{m}(m-1)!$. Its cumulative probability distribution 
\begin{equation}
P_{N}^{(2)(\infty)}=\sum_{n=0}^{N}{\rho_{n}^{(2)(\infty)}} \equiv \left\langle \theta(N-n)\right\rangle^{(2)(\infty)} = I_{1-q} (g,N+1)
\label{theta2infty}
\end{equation}
is given by the incomplete beta function and yields, via Eq.~(\ref{rhocut}), the explicit formulas for the cut-off negative binomial distribution as well as its characteristic function and cumulants:
\begin{equation}
\rho_{n}^{(2)} = \frac{\Gamma(n+g)(1-q)^{g}q^n}{\Gamma(g)\Gamma(n+1)I_{1-q}(g,N+1)} ,
\label{rho2}
\end{equation}
\begin{equation}
\Theta^{(2)} = \left(\frac{1-q}{1-zq}\right)^g \frac{I_{1-zq}(g,N+1)}{I_{1-q}(g,N+1)} ; 
\label{theta2}
\end{equation}
\begin{equation}
\frac{\tilde{\kappa}_{m+1}^{(2)}}{q^{m+1}} = \frac{d^m}{dq^m} \frac{\tilde{\kappa}_{1}^{(2)}}{q} , \quad
\tilde{\kappa}_{1}^{(2)} = \frac{gq}{1-q}-Q ,
\label{kappam}
\end{equation}
\begin{equation}
\kappa_{2}^{(2)} = \frac{gq}{(1-q)^2} + \left[\frac{g-1}{1-q} - N - g - Q\right] Q ;
\label{kappa2}
\end{equation}
\begin{equation}
Q = \frac{(1-q)^{g-1}q^{N+1}}{B(g,N+1)I_{1-q}(g,N+1)} ,
\label{Q}
\end{equation}
where $\tilde{\kappa}_{1}^{(2)} \equiv \bar n$ is a mean number of noncondensed atoms, $B(a,b) = \Gamma(a)\Gamma(b)/\Gamma(a+b)$ is the beta function, $\Gamma(a)$ is the gamma function. 

\subsection{B. Continuous Approximation: Cut-Off Gamma Distribution}

The most interesting is a case when an energy difference between levels in the trap is less than the temperature, $\epsilon \ll T$. The latter implies that $1-q \ll 1$ and $N_c \equiv \tilde{\kappa}_{1}^{(2)(\infty)} = gq/(1-q) \gg g$, that is the critical number of atoms $N_c$ for the distribution (\ref{rho2infty}) is much larger than the number of levels $g$. In that case, in the whole interesting for BEC region $n \gg g, \quad N \gg g$, we can neglect by the discreteness of the random variable $n$ and replace the discrete distribution (\ref{rho2infty}) with a continuous gamma distribution 
\begin{equation}
\rho_{n}^{(\Gamma)(\infty)} = \frac{\epsilon\left[\epsilon(n+g-1)/T\right]^{g-1}}{T\Gamma(g)\exp\left[\epsilon(n+g-1)/T\right]} ,
\label{rhogammainfty}
\end{equation}
\begin{equation}
\Theta^{(\Gamma)(\infty)} \equiv \int_{1-g}^{\infty}{e^{itn}\rho_{n}^{(\Gamma)(\infty)}dn} = e^{-i(g-1)t}\left(1-\frac{itT}{\epsilon}\right)^{-g} ,
\label{thetagammainfty}
\end{equation}
for which the mean value $\left\langle n\right\rangle^{(\Gamma)(\infty)} = gT/\epsilon -g+1$ and all cumulants of orders  $m \geq 2$, $\kappa_{m}^{(\Gamma)(\infty)} = g\left[T/\epsilon\right]^{m}(m-1)!$, are equal to the corresponding generating cumulants of the distribution (\ref{rho2infty}). We derive Eq.~(\ref{rhogammainfty}) from Eq.~(\ref{rho2infty}) using the Stirling formula $n! \approx \sqrt{2\pi}n^{n+1/2}e^{-n}$ and an approximation $\sqrt{(n+g-1)/n} \approx 1$. The cumulative distribution function of the distribution (\ref{rhogammainfty}) 
\begin{equation}
P_{\eta}^{(\Gamma)(\infty)}\equiv\left\langle \theta(N-n)\right\rangle^{(\Gamma)(\infty)} = \gamma(g,(N+g-1)\epsilon/T)/\Gamma(g)
\label{Pgammainfty}
\end{equation}
(see Eq.~(\ref{eta})) is given by the incomplete gamma function $\gamma(a,x) = \int_{0}^{x}{t^{a-1}e^{-t}dt}$ and yields the explicit formulas for the probability density function of the cut-off gamma distribution and all its initial moments ($m = 1, 2, \ldots$) 
\begin{equation}
\rho_{n}^{(\Gamma)} = \frac{\epsilon E_{n}^{g-1} e^{-E_{n}}}{T\gamma(g,E_{N})} , \quad E_{n} = \frac{\epsilon(n+g-1)}{T} ,
\label{rhogamma}
\end{equation}
\begin{equation}
\left\langle (n+g-1)^{m}\right\rangle^{(\Gamma)} = \left(\frac{T}{\epsilon}\right)^{m} \frac{\gamma(g+m,E_{N})}{\gamma(g,E_{N})} .
\label{meangamma}
\end{equation}
 
    The cut-off gamma distribution (\ref{rhogamma}) approximates the discrete distribution (\ref{rho2}) so good that any differences between the two distributions as well as between their cumulants cannot be even seen in the whole critical region. (Of course, the properly renormalized function $\rho_{n}^{(\Gamma)(\infty)} \int_{0}^{\infty}{\rho_{n}^{(\infty)}dn} \approx q\rho_{n}^{(\Gamma)(\infty)}$ should be compared against $\rho_{n}^{(\infty)}$, not the function $\rho_{n}^{(\Gamma)(\infty)}$ itself.)

\subsection{C. Two-Level Trap Model with Shifted Average: \\
Pirson Distribution of the III Type}

     The two-level trap model can be nicely generalized by an overall shift of the variable $n$. Thus, instead of the gamma distribution (\ref{rhogammainfty}) we arrive at a model described by the Pirson distribution of the III type 
\begin{equation}
\rho_{n}^{(Ps)(\infty)} = \frac{\epsilon\left[\epsilon(n-\Delta n)/T\right]^{g-1}}{T\Gamma(g)\exp\left[\epsilon(n-\Delta n)/T\right]},\quad n\in [\Delta n,\infty ),
\label{Ps}
\end{equation}
where we have one more free parameter $\Delta n$ to model an actual trap. Cumulants, moments, characteristic function, and all corresponding cut-off quantities for the model (\ref{Ps}) are the same as for the gamma distribution with the only modification, namely, a plain shift of the variable $n$ and its mean value $\bar{n}$ by the amount $\Delta n + g - 1$. 

\subsection{D. Modeling BEC in an Actual Trap}

BEC statistics in an actual trap is essentially the constraint-cut-off statistics of a sum of the populations of all excited states with inhomogeneously broaden spectrum of energies $\epsilon_{\vec{k}}$ ranging from the first level $\epsilon_1$ through all levels up to the energies $\sim T$. We can describe it analytically by using the exact solution for the two-level trap as a building block. In fact, we need just to find the unconstrained distribution of the total noncondensate occupation, that is the sum of the independent random occupations of the excited states, and then to cut off it as is explained in Sec. II. There are different ways to implement this program.

     First of all, we can model the whole energy spectrum of a trap $\epsilon_{\vec{k}}$ by means of just one effective energy level $\epsilon$ with the degeneracy $g$ and choose the two parameters $\epsilon$ and $g$ in Eq.~(\ref{rho2infty}) to ensure matching the first two cumulants of the model with their corresponding values in the actual trap, $\kappa_{1}^{(\infty)} \equiv \tilde{\kappa}_{1}^{(\infty)} = N_c$, $\kappa_{2}^{(\infty)} \equiv \tilde{\kappa}_{2}^{(\infty)} + \tilde{\kappa}_{1}^{(\infty)} = \sigma^{(\infty)2}$. According to the negative binomial distribution (\ref{rho2infty}), we find $g = N_{c}^{2}/(\sigma^{(\infty)2} - N_{c})$ and $q = 1 - N_{c}/\sigma^{(\infty)2}$, that is $\epsilon/T = g/N_c$ if $\epsilon \ll T$. The result is given by the cut-off negative binomial distribution (\ref{rho2}) and exactly coincides with the well-known quasithermal ansatz which we suggested in \cite{CNBII}. Thus, the quasithermal ansatz is not only a good guess anymore, but a rigorously justified effective two-level trap model of BEC statistics. 
     
     This fact explains why the quasithermal ansatz was so successful and close to the BEC statistics in the actual mesoscopic traps for the low order moments and, at the same time, reveals its main drawback. Namely, in the thermodynamic limit, $N_v \to \infty$, for the box trap one has $g \approx cN_{v}^{2/3}$ and $\epsilon/T \approx c/N_{v}^{1/3} \gg \epsilon_{1}/T = \pi \left[\zeta(3/2)/N_{v}\right]^{2/3}$, where $c = \pi^{2}(\zeta(3/2))^{4/3}/s_{2} \approx 2.15$. So, all higher order cumulants in Eq.~(\ref{igcumulants}) $\tilde{\kappa}_m \approx (m-1)! 6 (e^{\epsilon_{1}/T} - 1)^{-m}$, $m \geq 3$, which are dominated in the box trap by the most long-wavelength 6-fold degenerate first excited state with the energy $\epsilon_1$ and wavenumber $k_1 = 2\pi /L$, are modeled incorrectly. That means that the quasithermal ansatz does not describe the long-ranged correlations and anomalies in the BEC statistics \cite{Koch06} in the deep condensed regime ($N \gg N_c$) and, in particular, predicts vanishing non-Gaussian coefficients 
$$\tilde{\kappa}_{m}^{(\infty)}/\tilde{\kappa}_{2}^{(\infty)m/2} = (m-1)!/g^{m/2-1} \propto N_{v}^{-(m+2)/3} \to 0.$$ 
The distribution (\ref{rho2infty}) for the effective two-level trap tends to the pure Gaussian distribution (\ref{gauss}) with increasing trap parameter $N_v$ and does not coincide with the one for the actual box trap. In the actual box trap all non-Gaussian coefficients remain nonzero and do not depend on $N_v$ in the thermodynamic limit: $\tilde{\kappa}_{m}^{(\infty)}/\tilde{\kappa}_{2}^{(\infty)m/2} = (m-1)!/6^{m/2-1}$. Thus, the formulated above choice for the parameters of the effective two-level trap model and, hence, the quasithermal ansatz \cite{CNBII} fail to give correct higher order cumulants in the condensed phase outside the critical region. However, in the critical region, which is the most interesting for us, this effective two-level trap model yields all cumulants and all universal, strongly non-Gaussian functions of the constraint-cut-off BEC statistics \cite{KKD-RQE} qualitatively right, namely, similar to the cut-off Gaussian model (\ref{cutoffgauss}). 
     
     In order to describe correctly all higher order cumulants, one has to take into account exactly a dominant contribution from the lowest energy level. For the box trap, it could be done by setting $\epsilon = \epsilon_1$ and $g = g_1 = 6$ in the two-level trap model since the lowest level is 6-fold degenerate. However, that model would not give the correct values for the most important lower order cumulants $\kappa_{1}, \kappa_{2}$, and $\kappa_{3}$ which have large contributions from the higher energy levels. Introduction of the additional shift $\Delta n$ in the model (\ref{Ps}) would not help much, except for correcting the mean value. 
     
     Nevertheless, the modeling of the actual trap with the two-level trap solution can be essentially improved if we use another choice of the free parameters of the model (\ref{Ps}). Namely, let us match exactly the first three cumulants for the unconstrained noncondensate occupation,
\begin{equation}
N_{c} =\Delta n +gT/\epsilon,\quad \kappa_{2}^{(\infty)}=g(T/\epsilon)^{2},\quad \kappa_{3}^{(\infty)}=2g(T/\epsilon)^{3},
\label{Psb1}
\end{equation}
and leave whatever mismatch remains for all other higher order cumulants. Solution of the Eq.~(\ref{Psb1}) for the model parameters is straightforward:
\begin{equation}
\Delta n=N_{c}-\frac{2\sigma^{(\infty)4}}{\kappa_{3}^{(\infty)}},\quad \frac{\epsilon}{T}=\frac{2\sigma^{(\infty)2}}{\kappa_{3}^{(\infty)}},\quad g=\frac{4\sigma^{(\infty)6}}{\kappa_{3}^{(\infty)2}},   
\label{Psb2}
\end{equation}
where $\kappa_{3}^{(\infty)}=\tilde{\kappa}_{3}^{(\infty)}+3\sigma^{(\infty)2}-2N_c$. Thus, for the scaled noncondensate occupation $x=(n-N_{c})/\sigma^{(\infty)}$ the Pirson distribution of the III type yields 
\begin{equation}
\rho_{x}^{(Ps)(\infty)} = \frac{g^{g/2}}{\Gamma(g)}(x+\sqrt{g})^{g-1} e^{-(x+\sqrt{g})\sqrt{g}}
\label{Psb}
\end{equation}
for $x\in [-\sqrt{g},\infty )$ and $\rho_{x}^{(Ps)(\infty)} = 0$ for $x<-\sqrt{g}$. The universal unconstrained probability distribution $\rho_{x}^{(Ps)(univ)}$ is given by the same formula (\ref{Psb}) if one uses the thermodynamic-limit value of the effective degeneracy of the excited energy level in the two-level trap model (\ref{Psb2}), that is  $g=s_{2}^{3}/s_{3}^{2}\approx 64$. 
It is immediate to find also corresponding to it cumulative distribution function (see Eq.~(\ref{Pgammainfty}))
\begin{equation}
P_{\eta}^{(Ps)(univ)} = \int_{-\sqrt{g}}^{\eta}{\rho_{x}^{(Ps)(univ)}dx} = \frac{\gamma\left(g,(\eta +\sqrt{g})\sqrt{g}\right)}{\Gamma(g)}.
\label{PPsbuniv}
\end{equation}

     The result in Eq.~(\ref{Psb}) is quite remarkable in two respects. First of all, it is extremely simple for involves only elementary functions. Second, it works reasonably well in the most nontrivial, central part of the critical region (for $-2 < x < 4$) and exactly matches the asymmetry cumulant $\kappa_{3}^{(\infty)}$. 
     
\section{X. EXACTLY SOLVABLE THREE-LEVEL TRAP MODEL OF BEC: CONFLUENT HYPERGEOMETRIC DISTRIBUTION}
     
     A natural way to get more accurate approximation is to adopt a three-level trap model where the total noncondensate occupation is a sum of occupations of two excited levels: $\epsilon_1$ and $\epsilon_2$ with degeneracies $g_1$ and $g_2$, respectively, so that $n = n_1 + n_2$. The unconstrained distribution $\rho_{n}^{(3)(\infty)}$ for that three-level trap model is just a probability distribution of a superposition of two independent stochastic variables of the two-level trap models. Hence, in the continuous approximation we have to calculate the following convolution  
$$\rho_{n}^{(3)(\infty)} = \int_{1-g_{2}}^{n+g_{1}-1}{\rho_{n-k}^{(\Gamma)(\infty)(1)}\rho_{k}^{(\Gamma)(\infty)(2)}dk}$$     
of two continuous gamma distributions of the two-level trap models in Eq.~(\ref{rhogammainfty}) with the parameters $g_1 , \epsilon_1$ and $g_2 ,
\epsilon_2$, respectively. In a result, we find   
\begin{eqnarray}
\rho_{n}^{(3)(\infty)} = \frac{\left(\frac{\epsilon_1}{T}\right)^{g_1}\left(\frac{\epsilon_2}{T}\right)^{g_2} (n+g_1 +g_2 -2)^{g_{1}+g_{2}-1}}{\Gamma(g_{1}+g_{2})\exp\left[\epsilon_{2}(n+g_{1}+g_{2}-2)/T\right]}\nonumber\\
\times \quad M\left(g_{1},g_{1}+g_{2},(n+g_{1}+g_{2}-2)\frac{\epsilon_2 - \epsilon_1}{T}\right),\quad
\label{rho3infty}
\end{eqnarray}
where 
\begin{equation}
M(a,b,z) = \frac{\Gamma(b)}{\Gamma(b-a)\Gamma(a)} \int_{0}^{1}{e^{zt}t^{a-1}(1-t)^{b-a-1}dt} 
\label{MHG}
\end{equation}
is Kummer's confluent hypergeometric function \cite{a,Bateman}.

    Let us discuss now how to choose model parameters. One possibility is to make the first level responsible for the long-range BEC correlations (higher order cumulants) by choosing parameters $\epsilon_1 = T\pi \left[\zeta(3/2)/N_{v}\right]^{2/3}$ and $g_1 = 6$ to be equal to their actual values for the first level in the actual box trap, while use the second level with parameters $\epsilon_2$ and $g_2$ to take care of the correct mean value and variance of the total noncondensate occupation:  
$$g_{2}T/\epsilon_2 + g_{1}T/\epsilon_1 = N_c + g_{1}+g_{2}-2 ,$$
\begin{equation}
g_{2}(T/\epsilon_{2})^2 + g_{1}(T/\epsilon_{1})^2 = \sigma^{(\infty)2} .
\label{e1e2g1g2}
\end{equation}
The parameters $g_2$ and $\epsilon_2$ can be easily found from Eq.~(\ref{e1e2g1g2}) by plugging in the quantity $\sqrt{g_2}T/\epsilon_2 = \sqrt{\sigma^{(\infty)2} - g_{1}(T/\epsilon_{1})^2}$ from the second equation into the first equation, that becomes an elementary cubic algebraic equation for $\sqrt{g_2}$. Then, for the chosen $g_2$, the energy of the second level is equal to $\epsilon_2 = T\sqrt{g_2}/\sqrt{\sigma^{(\infty)2} - g_{1}(T/\epsilon_{1})^2}$. Note that the real-valued solution exists for the trap-size parameter $N_v$ larger than about 2000. The result in Eq.~(\ref{rho3infty}) for that choice of parameters is reasonably good in the central part ($-2 < x < 4$) of the critical region. However, there are some deviations because with that choice of parameters $\epsilon_1, g_1 ,\epsilon_2, g_2$ we do match exactly the first two and all high-order cumulants, but match only approximately the third, forth, and other intermediate-order cumulants of the unconstrained probability distribution $\rho_{n}^{(\infty)}$. That mismatch is of the order of $20\%$ for the third and forth cumulants and, what is very important, rapidly decreases for the higher-order cumulants with increasing the cumulant order $m$. 

    Another amazingly accurate and, in fact, sufficient for all practical purposes choice for the four parameters $\epsilon_1, g_1 ,\epsilon_2, g_2$ is to match exactly the first four cumulants to their box trap values at $N = \infty$ (see Sections II and III, in particular, Eq.~(\ref{igcumulants})), 
\begin{equation}
g_{2}T/\epsilon_2 + g_{1}T/\epsilon_1 = N_c + g_{1}+g_{2}-2 ,
\label{4cumulantsmatch}
\end{equation}
$$g_{2}(T/\epsilon_{2})^j + g_{1}(T/\epsilon_{1})^j = \kappa_{j}^{(\infty)}/(j-1)! ,\quad j = 2, 3, 4,$$
and, hence, match only approximately all higher-orders cumulants. Remarkably, in this case the result for the three-level trap model in Eq.~(\ref{rho3infty}) is so perfect that it cannot be discerned from the exact numerical curves for the actual box trap in Figs. 7 and 9, so that we even do not need to plot separate curves for this result. Let us stress that we match the values of the first four cumulants only at one point $N = \infty$ and then the solution in Eq.~(\ref{rho3infty}) with the chosen values of the four parameters $g_1 , \epsilon_1 ,g_2 , \epsilon_2$ yields, with amazing accuracy, the whole functions for these four as well as all other cumulants, moments, and probability distributions for all values of the number of atoms in the central part of the critical region $-3 < x < 6$. We skip here an elementary analysis of the system of algebraic equations (\ref{4cumulantsmatch}), which anyway are very easy to solve numerically. Let us note only two points. First, among the different solutions of Eqs.~(\ref{4cumulantsmatch}), it is necessary to use the one which satisfies the conditions of applicability of the continuous gamma distribution (\ref{rhogammainfty}). Second, the real-valued solutions of Eqs.~(\ref{4cumulantsmatch}) exist only for large enough trap size, namely, for $N_v \geq 9283$. In order to avoid complex-valued solutions for smaller values of the trap-size parameter $N_v < 9283$, one should use the shifted gamma distribution for each excited energy level, i.e. the Pirson distribution of the III type as will be discussed at the end of this Sec. X, or should not insist on the exact matching of the forth cumulant, that is simply omit the last equation in the system of equations (\ref{4cumulantsmatch}). 

	It is straightforward to analyze the asymptotics of the tails and other properties of the exact solution in Eq.~(\ref{rho3infty}) by means of the known asymptotics and properties of the Kummer's confluent hypergeometric function. We skip that straightforward analysis. 
	
	Instead, we focus on the thermodynamic limit of the probability distribution, given by Eqs.~(\ref{rho3infty}) and (\ref{4cumulantsmatch}), in the critical region. The solution to that problem is not trivial and requires a special asymptotics of the Kummer's confluent hypergeometric function $M(g_{1},g_{1}+g_{2},x)$ when $x \sim g_2 \to \infty$ and $x/g_2 \to 1$ that corresponds to the following thermodynamic limit of the three-level trap model parameters 
$$g_1 \to \frac{s_{3}^{4}}{s_{4}^{3}} \approx 14.871; \quad \frac{\epsilon_{1}}{T} \to \frac{s_{3}\pi}{s_{4}}\left[\frac{\zeta(3/2)}{N_v}\right]^{2/3} \approx \frac{7.208}{N_{v}^{2/3}};$$
\begin{equation}
g_2 \to  \frac{\pi^{2}\left[\zeta(3/2)\right]^{4/3}}{s_{2}-s_{3}^{2}/s_{4}} \approx 5.577 N_{v}^{2/3}; \quad \frac{\epsilon_2}{T} \to \frac{g_2}{N_v} \approx \frac{5.577}{N_{v}^{1/3}},
\label{limitparameters}
\end{equation}	
and that, according to \cite{a,Bateman}, was not yet found by standard methods. In the Appendix we find that asymptotics explicitly in terms of a parabolic cylinder function: 
\begin{equation}	
M(g_{1},g_{1}+g_{2},x) \sim \frac{\Gamma(g_{1}+g_{2})e^{\frac{(g_{2}-x)^{2}}{4x}}}{\Gamma(g_{2})x^{g_{1}/2}}D_{-g_1}\left(\frac{g_{2}-x}{\sqrt{x}}\right) 
\label{asymptDHG}
\end{equation}
where $x \sim g_2 \to \infty, \quad x/g_2 \to 1, \quad g_{1}= const \neq \infty$. 

     In Eq.~(\ref{limitparameters}) we use the values $s_m$ given by the generalized Einstein function from Eq.~(\ref{sm}) (see Fig. 10). The thermodynamic-limit value of the energy of the first excited level in the 3-level trap model (\ref{limitparameters}) is only by a numerical factor $s_{3}/s_{4} \approx 1.21$ higher than the energy of the first level in the actual box trap in Eq.~(\ref{e1}), but the degeneracy of the first excited level in the 3-level model $g_1 \approx 15$ is essentially larger than the degeneracy 6 of the first level in the actual box trap. The energy and degeneracy of the second excited level are much larger than their values for the second excited level in the box trap by the factors $\sim N_{v}^{1/3}$ and $\sim N_{v}^{2/3} \to \infty$, respectively, since in that 3-level trap model the second excited level takes the main part of the responsibility for the correct values of the mean occupation and the variance which, contrary to the higher order cumulants, are mostly determined by the contributions from a large number of higher energy levels in the box trap with the energies up to the order of the temperature for the case of the mean occupation \cite{KKS-PRL,KKS-PRA}. Note that in order to keep notations simple we do not introduce new notations for the two excited levels in the 3-level model since they are used only in this Sec. X and there should not be any confusion with the box trap levels. 
    
     Finally, when we use the parameters in Eq.~(\ref{limitparameters}) together with the nontrivial asymptotics of the Kummer's confluent hypergeometric function in Eq.~(\ref{asymptDHG}) (see Appendix) in the exact solution (\ref{rho3infty}), we arrive at the remarkably simple analytical approximation in Eq.~(\ref{rhounivPC}) for the universal unconstrained probability distribution.
     
     We can further improve all these modeling results for the 3-level trap model if introduce an additional parameter $\Delta n$ in the solution (\ref{rho3infty}) that shifts the probability distribution as a whole along the $n$ axis, $n \to n - \Delta n$. 
     
     Again, there are two possibilities to choose the parameters. The first variant is 
$$\frac{\epsilon_1}{T}=\frac{\pi\zeta (\frac{3}{2})^{\frac{2}{3}}}{N_{v}^{2/3}}, g_{1}=6, \frac{g_{1}T}{\epsilon_1}+\frac{g_{2}T}{\epsilon_2}=N_{c}+g_{1}+g_{2}-2-\Delta n,$$
\begin{equation}	
g_{1}\left(\frac{T}{\epsilon_1}\right)^{j}+g_{2}\left(\frac{T}{\epsilon_2}\right)^{j}=\frac{\kappa_{j}^{(\infty)}}{(j-1)!},\quad j = 2, 3,
\label{r3ap}
\end{equation}
that allows us to match exactly the asymmetry cumulant $\kappa_{3}^{(\infty)}$ in addition to the mean value and variance. To solve Eq.~(\ref{r3ap}) for the parameters of the model is simple. In particular, we find 
\begin{equation}	
\frac{\epsilon_2}{T}=2\frac{\sigma^{(\infty)2}-g_{1}(T/\epsilon_{1})^{2}}{\kappa_{3}^{(\infty)}-2g_{1}(T/\epsilon_{1})^{2}}, g_{2}=4\frac{\left[\sigma^{(\infty)2}-g_{1}(T/\epsilon_{1})^{2}\right]^3}{\left[\kappa_{3}^{(\infty)}-2g_{1}(T/\epsilon_{1})^{2}\right]^2}.
\label{r3ap1}
\end{equation}
In the thermodynamic limit it yields an approximation for the universal probability distribution $\rho_{x}^{(univ)}$ as the shifted by $\Delta n$ solution in Eq.~(\ref{rho3infty}) with the parameters 
$$\frac{\epsilon_1}{T}=\pi\left[\frac{\zeta (3/2)}{N_v}\right]^{2/3},\quad g_{1}=6,$$
\begin{equation}	
\frac{\epsilon_2}{T}=\frac{(s_{2}-6)\epsilon_1}{s_{3}T},\quad  g_{2}=\frac{(s_{2}-6)^3}{s_{3}^2}\approx 16.56.
\label{ru3ap}
\end{equation}
Numerically the 3-level trap model with parameters in Eqs.~(\ref{r3ap}) or (\ref{ru3ap}) is certainly more accurate than the two-level trap models. However, due to about 50\% mismatch in the forth cumulant $\kappa_{4}^{(\infty)}$ it is too asymmetric and essentially less accurate than the 3-level trap models (\ref{4cumulantsmatch}) and (\ref{limitparameters}).

     Finally, we obtain the most accurate model with the second choice of the parameters that ensure the exact match of the first five cumulants to their box trap values at $N=\infty$ as follows
$$g_{1}\frac{T}{\epsilon_1}+g_{2}\frac{T}{\epsilon_2}=N_{c}+g_{1}+g_{2}-2-\Delta n,$$ 
\begin{equation}
g_{1}\left(\frac{T}{\epsilon_1}\right)^{j}+g_{2}\left(\frac{T}{\epsilon_2}\right)^{j}=\frac{\kappa_{j}^{(\infty)}}{(j-1)!},\quad j= 2,3,4,5.
\label{r3bp}
\end{equation}
The solution for the probability distribution is the same solution (\ref{rho3infty}) but shifted by the amount $\Delta n$, namely,
\begin{eqnarray}
\rho_{x}^{(3)(\infty)}=\frac{\left(\frac{\epsilon_{1}\sigma^{(\infty)}}{T}\right)^{g_1}\left(\frac{\epsilon_{2}\sigma^{(\infty)}}{T}\right)^{g_2} X^{g_{1}+g_{2}-1}}{\Gamma(g_{1}+g_{2})\exp\left[\epsilon_{2}\sigma^{(\infty)}X/T\right]}\nonumber\\
\times M\left(g_{1},g_{1}+g_{2},\frac{\epsilon_2 - \epsilon_1}{(T/\sigma^{(\infty)})}X\right),
\label{r3b}
\end{eqnarray}
$$X=x+\frac{g_{1}T}{\epsilon_{1}\sigma^{(\infty)}}+\frac{g_{2}T}{\epsilon_{2}\sigma^{(\infty)}}.$$
The solution of the nonlinear system of algebraic equations (\ref{r3bp}) for the parameters of the model can be reduced to the cubic algebraic equation for, e.g., $\epsilon_2$ and done explicitly. We skip these details and present the result only for the universal probability distribution in the thermodynamic limit:
$$\rho_{x}^{(3)(univ)}=\rho_{x}^{(3)(\infty)},\quad \epsilon_{1}=1.0584\epsilon_{1}^{(0)}, \epsilon_{2}=7.2732\epsilon_{1}^{(0)},$$
\begin{equation}
g_{1}=8.5038, g_{2}=473.01, \epsilon_{1}^{(0)}=T\pi \left[\zeta (3/2)/N_{v}\right]^{2/3}.
\label{r3buniv}
\end{equation}
It works perfectly in the whole central part of the critical region and, in fact, cannot be discerned from the exact curve in Fig. 2. Even very sensitive quantities, strongly subject to small discrepancies in approximation of $\rho_{x}^{(univ)}$ and its derivatives, such as specific heat, can be calculated by means of the confluent hypergeometric approximation (\ref{r3buniv}), that is also repeated in Eq.~(\ref{rhounivKummer}), amazingly accurate in the whole central part of the critical region, namely, with accuracy better than a few percents in the interval $-4 < \eta < 10$.     
     
     The analytical solution in terms of the Kummer's confluent hypergeometric function (\ref{rhounivKummer}), or (\ref{r3buniv}), has wider range of validity, $-4 < x < 10$, covers more than 10 orders in $\rho_{x}^{(univ)}$, and is more accurate, but also more complicated than the solution (\ref{rhounivPC}) in terms of the parabolic cylinder function, which is valid in still very wide interval $-3 < x < 6$, that covers more than 6 orders in $\rho_{x}^{(univ)}$ and includes all of the most interesting for critical phenomena central part of the critical region. Both of these solutions well overlap with the asymptotics (\ref{arhox}) and (\ref{rap}) at the wings of the critical region and, thus, both of them yield analytical solution to the problem of universality of critical fluctuations, Gibbs free energy, heat capacity, and other thermodynamic quantities in the BEC phase transition of the ideal gas, as is described in Sections VI, VII, XII-XVI. 

\section{XI. REGULAR SCHEME FOR REFINEMENT OF CONDENSATE STATISTICS APPROXIMATION}

     Exactly solvable two- and three-level trap models of BEC presented in Sections IX and X are very useful since they are physically feasible models and as such automatically include many important physical properties, first of all anomalously large higher order cumulants and associated with them long-range correlations. However, they have only limited number of free parameters to model actual, e.g. box-trap, mesoscopic systems. It would be very useful to have a regular scheme for a further improvement of these or other approximations of the BEC statistics. Here we present such refinement scheme based on the following two ideas.

\subsection{A. Infrared Universality of Higher-Order Cumulants and Method of Superposition}    

     The first idea exploits an observation that in the most interesting mesoscopic systems with the energies of the lowest levels much smaller than temperature, $\epsilon_{\vec{k}} \ll T$, all higher-order cumulants of BEC fluctuations are dominated by a contribution from a few, the most long wavelength modes in the infrared limit of the energy spectrum and almost do not depend on details of all higher parts of the energy spectrum. In other words, all higher-order cumulants are universal for mesoscopic systems with the same infrared limit of the energy spectrum \cite{Koch06}. We used already that infrared universality of the higher-order cumulants in deriving the approximation for the universal characteristic function of noncondensate fluctuations in the thermodynamic limit in Eq.~(\ref{mJTh}). The essence of the infrared universality is in the fast convergence of the sums $s_m$ in Eq.~(\ref{sm}), which determine the generating cumulants $\tilde{\kappa}_{m}^{(\infty)}$ in Eq.~(\ref{igcumulants}) in the thermodynamic limit, to the value $g_1 = 6$, that is equal to the contribution from the lowest 6-fold degenerate energy level $\epsilon_1$, with increasing the order $m$, i.e. $s_m \to 6$, as is shown in Fig. 10. That makes all higher-order generating cumulants $\tilde{\kappa}_{m}^{(\infty)}$ inversely proportional to the m-th power of the energy of the first excited level,  $\tilde{\kappa}_{m}^{(\infty)} \propto (T/\epsilon_{1})^m$, and independent on the energies of all higher levels. 
    
     In a general case of a finite-size, mesoscopic system, i.e. without thermodynamic-limit assumption, we can derive, similar to Eq.~(\ref{mJTh}), an approximation for the characteristic function of the unconstrained probability distribution $\rho_{n}^{(\infty)}$ of the noncondensate occupation
\begin{equation}
\Theta^{(\infty)}(u)\approx\exp \left[\sum_{m=1}^{m^*}{\frac{\tilde{\kappa}_{m}'(e^{iu}-1)^{m}}{m!}}\right]\prod_{j=1}^{J}{\left(\frac{e^{\epsilon_{j}/T}-1}{e^{\epsilon_{j}/T}-e^{iu}}\right)^{g_j}},
\label{mJT}
\end{equation}
where 
\begin{equation}
\tilde{\kappa}_{m}' = \tilde{\kappa}_{m} -(m-1)!\sum_{j=1}^{J}{g_{j}/(e^{\epsilon_{j}/T}-1)^m}
\label{Rc}
\end{equation}
\begin{figure}
\center{\epsfig{file=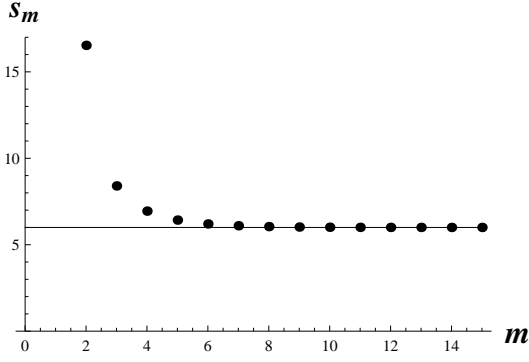,width=7cm}}
\caption{Sum $s_m$, Eq.~(\ref{sm}), as a function of the order $m$ which determines the related generating cumulant $\tilde{\kappa}_{m}^{(\infty)}$, Eq.~(\ref{igcumulants}).} 
\end{figure}
is the residual generating cumulant and now we keep the full function $e^{\epsilon_{\vec{k}}/T}-1$ instead of $\epsilon_{\vec{k}}/T$ when using Eq.~(\ref{igcumulants}) for generating cumulants. The essence of the approximations (\ref{mJT}) and (\ref{mJTh}) is that by exact account for a few ($J$) lowest levels we practically reduce very close to zero the coefficients in front of $(e^{iu}-1)^m$, i.e. the residual generating cumulants, for high orders $m$ so that keeping only finite number of terms $m^*$ in the sum allows us to calculate the unconstrained probability distribution $\rho_{x}^{(\infty)}$ in a finite, wide enough interval $x \in (-X_{1}, X_{2})$ of values of the variable $x=(n-N_{c})/\sigma^{(\infty)}$ with a range of validity $(-X_{1}, X_{2})$ expanding with increasing parameters $m^*$ and $J$. This approach works perfectly for numerical calculations of $\rho_{x}^{(\infty)}$ and all other statistical and thermodynamic quantities. However, analytically that approximation for $\rho_{x}^{(\infty)}$ can be calculated only for two- or three-level trap models ($J = 1$ or $2$) and only for $m^* = 1$ or $2$. Besides, not any, especially so arbitrarily cut function is a characteristic function for some well-defined real-valued positive probability distribution (see Bochner's, Khinchin's, Mathias', and Polya's theorems in probability theory). 

     These considerations lead us to the second idea, namely, to model the residual (after subtraction of fluctuations due to $J$ lowest levels) fluctuations of the noncondensate occupation, i.e. residual generating cumulants $\tilde{\kappa}_{m}'$ in Eq.~(\ref{Rc}), by some well-defined auxiliary stochastic variable that could and usually does have all its generating cumulants not equal to zero, but may be chosen to minimize discrepancy in the lowest order cumulants which become the most important after taking care of the higher-order cumulants via exact account of the fluctuations due to the first $J$ energy levels. The auxiliary stochastic variable should not be necessarily the noncondensate occupation in some physically feasible model of the trap or system, but could be an abstract one just mimicking the residual background fluctuations the detailed physical origin of which is not important due to discussed above infrared universality of long-range BEC correlations. That method of superposition is based on the well-known general property of a superposition of the independent stochastic variables, namely, that any cumulant (or generating cumulant) of order $m$ of the superposition is equal to the sum of the cumulants (or generating cumulants) of the same order $m$ of the superposed independent stochastic variables. 

     We consider below only two simple examples of that regular refinement scheme, namely, for the Poisson and Gaussian background fluctuations. It would be very interesting to find analytical solutions for more complicated examples, especially for the three-level trap model combined with Gaussian background fluctuations. 

\subsection{B. Superposition of Two-Level-Trap and Poisson Fluctuations}

     As a first example of that approach, let us consider a superposition of the two-level trap model (\ref{rho2infty}) and the Poisson distribution
\begin{equation}
\rho_{n}^{(P)(\infty)} = e^{-\kappa}\kappa^{n}/n! , \quad n \in [0, \infty ) ,
\label{Poisson}
\end{equation}
for which all cumulants, including the mean value and variance, are equal to the only parameter of the model $\kappa$, i.e., $\kappa_m = \kappa , \quad m = 1, 2, \ldots$, since its characteristic function is $\Theta^{(P)(\infty)} =\exp \left[\kappa (e^{iu}-1)\right]$. The Poisson distribution is the asymptotics of the probability distribution $\rho_{n}^{(\infty)}$ for the actual box-trap in the small number of atoms region, Eq.~(\ref{Poisson0}), however, here this is not a main point for that general model. The unconstrained probability distribution for that superposition can be calculated explicitly, 
\begin{eqnarray}
\rho_{n}^{(2P)(\infty)} = \sum_{m=0}^{n}{\rho_{n-m}^{(P)(\infty)}\rho_{m}^{(2)(\infty)}}\nonumber\\
= \frac{(1-q)^{g}\kappa^{n+g}e^{-\kappa}}{n!q^g} U(g,n+g+1,\frac{\kappa}{q}),
\label{2Poisson}
\end{eqnarray}
in terms of the confluent hypergeometric function \cite{a,Bateman}
\begin{equation}
U(a,b,z) = \frac{1}{\Gamma(a)}\int_{0}^{\infty}{e^{-zt}t^{a-1}(1+t)^{b-a-1}dt} .
\label{U}
\end{equation}
The cumulative probability distribution for that superposition (\ref{2Poisson}) is
\begin{eqnarray}
P_{N}^{(2P)(\infty)}=\sum_{n=0}^{N}{\rho_{n}^{(2P)(\infty)}} = I_{1-q}(g,N+1)\nonumber\\
-\left(\frac{1-q}{q}\right)^{g}\frac{1}{N!}\int_{0}^{\kappa}{t^{N+g}e^{-t}U(g,N+g+1,\frac{t}{q})dt}.
\label{P2Poisson}
\end{eqnarray}
According to Eq.~(\ref{rhocut}), the actual, cut-off probability distribution for that superposition model is as follows
\begin{equation}
\rho_{n}^{(2P)} = \frac{(1-q)^{g}\kappa^{n+g}e^{-\kappa}}{P_{N}^{(2P)(\infty)}n!q^g}U(g,n+g+1,\frac{\kappa}{q}), n\in [0,N].
\label{rho2Poisson}
\end{equation}
It is straightforward also to generalize that model by an additional shift of the variable $n$ and its mean value by the amount $\Delta n +g-1$ as it is done for the Pirson distribution of the III type in Eq.~(\ref{Ps}). 

     Then, we can choose the four parameters of this model $\Delta n, \kappa , g$, and $q$ to match the cumulants of the model, e.g., the first four cumulants, to their values in the actual box trap. As a result, we get an explicit expression for $\rho_{n}^{(2P)(\infty)}$ in Eq.~(\ref{2Poisson}), that analytically approximates the actual unconstrained probability distribution $\rho_{n}^{(\infty)}$, and, therefore, can immediately calculate all statistical and thermodynamic quantities using the formulas which express the constraint-cut-off mechanism of Sec. II. Numerically this model (\ref{2Poisson}) works very well in the central part of the critical region ($-3 < \eta < 6$), however, it does not give the asymptotics at $|\eta| \to \infty$ correctly. We skip the details of that modeling since it is similar to the analysis of another superposition model that will be discussed in more details in the next subsection. 

\subsection{C. Superposition of Two-Level-Trap and Gaussian Fluctuations}

    Let us consider a superposition of the two-level trap model (\ref{Ps}) with arbitrary shift of the mean value $x_{0}' =(\Delta n - N_{c})/\sigma^{(\infty)}$, i.e. Pirson distribution of the III type 
\begin{equation}
\rho_{x'}^{(Ps)(\infty)} = \frac{(\sigma^{(\infty)}\epsilon/T)^{g}}{\Gamma(g)}(x'-x_{0}')^{g-1} \exp \left[\frac{\sigma_{(\infty)}\epsilon}{T}(x_{0}'-x')\right]
\label{Psx}
\end{equation}
for $x' \in [x_{0}', \infty )$ and $\rho_{x'}^{(Ps)(\infty)} =0$ for $x' < x_{0}'$, and the Gaussian model with arbitrary mean value $x_{0}''$ and variance $\kappa_{2}''/(\sigma^{(\infty)})^2$,
\begin{equation}
\rho_{x''}^{(G)(\infty)} = \frac{\sigma^{(\infty)}}{\sqrt{2\pi \kappa_{2}''}}\exp \left[-\frac{(x''-x_{0}'')^2}{2\kappa_{2}''/\sigma^{(\infty)2}}\right],\quad x''\in (-\infty ,\infty),
\label{Gx}
\end{equation}
which we write, for simplicity's sake, directly in the continuous approximation for the variable (\ref{x}). The cumulants of the noncondensate occupation $n$ for that superposition of two stochastic variables, $\kappa_{m} =\kappa_{m}' +\kappa_{m}'', m = 1, 2,\ldots$, are equal to the sum of the two-level trap cumulants, $\kappa_{1}' =N_c + gT/\epsilon + x_{0}'\sigma^{(\infty)}$ for $m=1$ and $\kappa_{m}' = g(T/\epsilon)^{m}(m-1)!$ for $m \geq 2$, and the Gaussian cumulants, $\kappa_{1}'' = N_c + x_{0}''\sigma^{(\infty)}$ for $m=1$, $\kappa_{2}''$ for $m=2$, and $\kappa_{m}'' =0$ for $m \geq 3$. 

     The probability distribution of the scaled noncondensate occupation $x=(n-N_{c})/\sigma^{(\infty)}$ for this superposition can be calculated exactly. The result is
\begin{eqnarray}
\rho_{x}^{(PsG)(\infty)} = \int_{x_{0}'}^{\infty}{\rho_{x'}^{(Ps)(\infty)}\rho_{x-x'-N_{c}/\sigma^{(\infty)}}^{(G)(\infty)}dx'}\nonumber\\
=\frac{(\epsilon/T)^{g}D_{-g}\left(\frac{\epsilon\sqrt{\kappa_{2}''}}{T}-y\right)}{\sqrt{2\pi \kappa_{2}''}}e^{\frac{\epsilon^{2}\kappa_{2}''}{2T^2}-\frac{1}{4}\left(\frac{\epsilon\sqrt{\kappa_{2}''}}{T}+y\right)^{2}},\quad
\label{PsG}
\end{eqnarray}
where $y=(x-x_{0}'-x_{0}'')/\sqrt{\kappa_{2}''}$ and $D_{-g}$ is the parabolic cylinder function. As expected, the mean value parameters $x_{0}'$ and $x_{0}''$ always come in a sum and provides only one parameter to match the actual mean value of noncondensate occupation in the trap,
\begin{equation}
\kappa_{1}'+\kappa_{1}''=N_{c} \Rightarrow x_{0}'+x_{0}''=-(N_{c}+gT/\epsilon )/\sigma^{(\infty )}.
\label{PsG1}
\end{equation}

     We consider here two possible variants for the choice of the other three parameters of the model, $g, \epsilon/T$, and $\kappa_{2}''$. The first variant is to choose the two-level trap submodel to give precisely fluctuations due to the 6-fold degenerate first energy level (and, therefore, exactly take care of all higher-order cumulants and long-range BEC correlations) and to use the Gaussian variance $\kappa_{2}''$ to match the total second-order cumulant with its actual value  $\kappa_{2}^{(\infty)} \equiv \sigma^{(\infty)2}$ in the box trap:
\begin{equation}
\epsilon =\epsilon_{1},\quad g=g_{1}=6,\quad \kappa_{2}''=\sigma^{(\infty)2}-6(T/\epsilon_{1})^{2}.
\label{PsGa}
\end{equation}
In the thermodynamic limit the result in Eqs.~(\ref{PsG}), (\ref{PsG1}), (\ref{PsGa}) for $\rho_{x}^{(\infty)}$ coincides with the approximation $\rho_{x}^{(univ)}$ in Eq.~(\ref{21rhox}). It works reasonably well in the central part of the critical region, $-1 < x < 5$, although it does not take into account all asymmetry $\kappa_{3}$ and excess $\kappa_{4}$ and, hence, is not as good as the three-level trap model in Eq.~(\ref{rhounivPC}). However, that drawback can be amazingly cured if we implement the second choice for the model parameters. 

     The second variant for the parameters' choice is to match the first four cumulants exactly,
\begin{equation}
\kappa_{2}^{(\infty)}=\kappa_{2}''+g(T/\epsilon)^{2}, \kappa_{3}^{(\infty)}=2g(T/\epsilon)^{3}, \kappa_{4}^{(\infty)}=6g(T/\epsilon)^{4},
\label{PsGb}
\end{equation}
and leave some mismatch for the higher-order cumulants, whatever it is. The formulas for the cumulants $\kappa_{m}^{(\infty)}$ for the box trap are given in Sec. II via the generating cumulants $\tilde{\kappa}_{m}^{(\infty)}$ in Eq.~(\ref{igcumulants}). In the result, we have the following values for the model parameters
\begin{equation}
\frac{\epsilon}{T}=\frac{3\kappa_{3}^{(\infty)}}{\kappa_{4}^{(\infty)}}, g=\frac{27\kappa_{3}^{(\infty)4}}{2\kappa_{4}^{(\infty)3}}, \kappa_{2}''=\sigma^{(\infty)2}-\frac{3\kappa_{3}^{(\infty)}}{2\kappa_{4}^{(\infty)}}.
\label{PsGb2}
\end{equation}
In the thermodynamic limit we find
$$x_{0}'+x_{0}'' = -\frac{s_{3}^3}{\sqrt{s_2}s_{4}^2},\quad \frac{\epsilon}{T} = \frac{s_3}{s_4}\frac{\epsilon_1}{T} ,$$
\begin{equation}
g=\frac{s_{3}^4}{s_{4}^3}\approx 14.8711,\quad \kappa_{2}'' = \sigma^{(\infty)2}\left(1-\frac{s_{3}^2}{s_{2}s_4}\right) , 
\label{aPsGb2}
\end{equation}
where the constants $s_m$ are given in Eq.~(\ref{sm}) and in Fig. 10. Amazingly, the result in Eqs.~(\ref{PsG}) and (\ref{aPsGb2}) yields precisely the same universal unconstrained probability distribution $\rho_{x}^{(univ)}$ as the one found via asymptotics of the Kummer's confluent hypergeometric function in the three-level trap model in Eqs.~(\ref{rho3infty}), (\ref{limitparameters}), and (\ref{asymptDHG}), that is the excellent approximation (\ref{rhounivPC}) in terms of the parabolic cylinder function in the wide interval $-3<x<6$.  

\section{XII. UNIVERSAL STRUCTURE OF THE GIBBS FREE ENERGY IN THE CRITICAL REGION}

     The analytical theory of BEC statistics presented above and, in particular, the knowledge of the constraint-cut-off mechanism of the origin of nonanalyticity in critical fluctuations (see Eq.~(\ref{rhocut})) as well as the explicit analytical formulas for the universal probability distribution $\rho_{x}^{(univ)}$ of the total  noncondensate occupation in Eqs.~(\ref{rhouniv}), (\ref{rhox}), (\ref{rhounivKummer}), (\ref{rhounivPC}), (\ref{arhox}), (\ref{rap}) allow us to find and resolve the universal fine structure of the thermodynamic quantities in the whole critical region around the $\lambda$-point. Let us start with the Gibbs free energy \cite{LLV}
\begin{equation}
F = - T \ln Z ,
\label{Gibbs}
\end{equation}
which is determined by the partition function $Z$, Eq.~(\ref{Z}), of the ideal gas of $N$ atoms in the trap at temperature $T$ in the canonical ensemble. The Gibbs free energy is the basic, generating function for the thermodynamics in the canonical ensemble of particles since its derivatives determine the main thermodynamic quantities. In particular, the average energy $\bar{E}$, the entropy $S$, and the heat capacity $C_V$ are given by its first and second derivatives as follows
\begin{equation}
\bar{E}=-\frac{\partial\ln Z}{\partial (1/T)}=F+TS, S=-\frac{\partial F}{\partial T}, C_{V}=\left(\frac{\partial\bar{E}}{\partial T}\right)_{V}.
\label{EC}
\end{equation}

     The partition function can be represented as a product $Z = Z^{(\infty)} \left\langle \theta(N-\hat{n})\right\rangle^{(\infty)}$ of the unconstrained partition function
\begin{equation}
Z^{(\infty)}=Tr\left\{e^{-\frac{1}{T}\sum_{\vec{k}\neq 0}{\epsilon_{\vec{k}}\hat{n}_{\vec{k}}}}\right\}=\prod_{\vec{k}\neq 0}{\frac{1}{1-e^{-\frac{\epsilon_{\vec{k}}}{T}}}}=\frac{1}{\rho_{n=0}^{(\infty)}}
\label{upf}
\end{equation}
and the average of the cut off constraint $\theta(N-\hat{n})$ over the unconstrained Hilbert space (see Sections II and V) which is precisely the cumulative distribution function $P_{N}^{(\infty)}=\sum_{n=0}^{N}\rho_{n}^{(\infty)}$ of the unconstrained probability distribution $\rho_{n}^{(\infty)}$ of the total noncondensate occupation,
\begin{equation}
P_{N}^{(\infty)}\equiv \left\langle \theta(N-\hat{n})\right\rangle^{(\infty)}= Tr\left\{\theta(N-\hat{n})e^{-\frac{H}{T}}\right\}/Z^{(\infty)}.
\label{step}
\end{equation}

     Thus, the Gibbs free energy is a sum
\begin{equation}
F = F^{(\infty)} - T\ln \left\langle \theta(N-\hat{n})\right\rangle^{(\infty)}
\label{GibbsT}
\end{equation}
of the Gibbs free energy of a system of the unconstrained noncondensate excitations
\begin{equation}
F^{(\infty)} = - T\ln Z^{(\infty)} = T\sum_{\vec{k}\neq 0}{\ln \left(1-e^{-\epsilon_{\vec{k}}/T}\right)}
\label{FI}
\end{equation}
and the contribution from the constraint nonlinearity (see Sec. V). The first term is rather trivial since it describes a deeply condensed regime at very low temperature $T$ and very large number of atoms in the trap $N$, compared to their critical values $T_c$ and $N_c$, when there exists a "reservoir" of the condensate atoms that makes the occupations of all noncondensate (excited) states practically independent stochastic variables. That first term is a constant that does not depend on the number of atoms in the trap and in the thermodynamic limit ($N_v \to \infty$) yields the following expressions for the average energy, $\bar{E}^{(\infty)}$, and the entropy, $S^{(\infty)}$, 
\begin{equation}
F^{(\infty)} = - T\ln Z^{(\infty)} = -\frac{2}{3}\bar{E}^{(\infty)} = - \frac{\zeta (5/2)}{\zeta (3/2)}N_{v}T,
\label{Gibbs1}
\end{equation}
\begin{equation}
S^{(\infty)} = \frac{\bar{E}^{(\infty)}-F^{(\infty)}}{T} = \frac{5}{3}\frac{\bar{E}^{(\infty)}}{T},
\label{S1}
\end{equation}     
which are well-known in statistical physics \cite{LLV}.

     The second, constraint nonlinearity term in Eq.~(\ref{GibbsT}) is responsible for all critical phenomena and peculiarities of the thermodynamic quantities in the phase transition from the condensed phase to the high temperature, or small number of atoms, noncondensed phase. The latter regime is usually described by the grand-canonical-ensemble approximation, Eq.~(\ref{gce-distr}), via the pure exponential distribution of the noncondensate occupation, Eq.~(\ref{gce-ndistr}). 
     
     Thus, in order to resolve the fine universal structure of the Gibbs free energy in the critical region it is necessary to exclude the trivial content (\ref{Gibbs1}), which is by a factor of an order of $N_v \to \infty$ larger than the second term, and introduce the critical function for the Gibbs free energy per unit temperature as a deviation of the Gibbs free energy per unit temperature from its critical-point value:
\begin{equation}
F_F =\frac{F}{T}-\left(\frac{F}{T}\right)_{\eta =0} = \ln P_{N=N_c}^{(\infty)} - \ln P_{N}^{(\infty)}.
\label{GibbsCF}
\end{equation}

     With increasing the trap-size parameter $N_v$, that critical function quickly converges to the following universal function of the Gibbs free energy per unit temperature
\begin{equation}
F_{F}^{(univ)}(\eta) = - \ln P_{\eta}^{(univ)} + \ln P_{\eta =0}^{(univ)} ,
\label{Gibbsuniv}
\end{equation}
which is precisely the absolute value of the logarithm of the universal cumulative distribution function, Eq.~(\ref{Puniv}), for the universal probability distribution $\rho_{x}^{(univ)}$ of the noncondensate occupation (see Eqs.~(\ref{rhox}) and (\ref{rhouniv})) considered relative to its value at the critical point as a function of the universal variable $\eta =(N-N_{c})/\sigma^{(\infty)}$, Eq.~(\ref{eta}). It is depicted in Fig. 11. The important fact of the universality of $F_{F}^{(univ)}(\eta)$ immediately follows from a similar property of the unconstrained probability distribution $\rho_{n}^{(\infty)}$ (see Sec. IV and Figs. 2-4). Analytical formulas for the Gibbs-free-energy universal function obviously follow from the obtained in Sec. IV analytical formulas for $\rho_{x}^{(univ)}$, Eqs.~(\ref{rhouniv}), (\ref{rhox}), (\ref{rhounivKummer}), (\ref{rhounivPC}), (\ref{arhox}), (\ref{rap}), if one uses the Eqs.~(\ref{Gibbsuniv}) and (\ref{Puniv}). 

\begin{figure}
\center{\epsfig{file=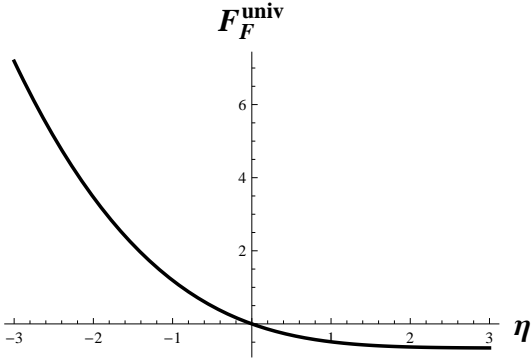,width=7cm}}
\caption{Universal function $F_{F}^{(univ)}$, Eq.~(\ref{Gibbsuniv}), of the Gibbs free energy per unit temperature as the function of $\eta = (N - N_{c})/\sigma^{(\infty)}$ in the critical region.} 
\end{figure}

     In particular, the universal cumulative distribution function in the central part of the critical region $|\eta|<5$ is correctly approximated via the Kummer's confluent hypergeometric function, Eq.~(\ref{rhounivKummer}),
\begin{eqnarray}
P_{\eta}^{(univ)}\approx \frac{e_{1}^{g_1}e_{2}^{g_2}}{\Gamma (g_{1}+g_{2})}\int_{-\infty}^{\eta}{X^{g_{1}+g_{2}-1}e^{-e_{2}X}M(g_{1},g_{1}}\nonumber\\
+g_{2},(e_{2}-e_{1})X)dx,\quad X=x+ g_{1}/e_{1}+ g_{2}/e_{2},\quad
\label{PDHG}
\end{eqnarray}
$$e_{1}\approx 4.303,\quad e_{2}\approx 29.573,\quad g_{1}\approx 8.504,\quad g_{2}\approx 473,$$
or via the parabolic cylinder function, Eq.~(\ref{rhounivPC}),
\begin{equation}
P_{\eta}^{(univ)}\approx \int_{-\infty}^{\eta}{\frac{c^{g}e^{c^{2}/2-Y^2}}{\sqrt{2\pi (1-s_{3}^{2}/(s_{2}s_{4}))}}D_{-g}[2(c-Y)]dx},
\label{PPC}
\end{equation} 
$$Y=\frac{x+s_{3}\sqrt{s_2}/s_4}{2\sqrt{1-s_{3}^{2}/(s_{2}s_{4})}},\quad g=\frac{s_{3}^{4}}{s_{4}^{3}},\quad c=\frac{s_{3}}{s_{4}}\sqrt{s_{2}-\frac{s_{3}^{2}}{s_4}}.$$
The asymptotics at the noncondensed, left wing of the critical region follows from Eq.~(\ref{arhox}),
$$P_{\eta}^{(univ)} \approx \frac{e^{f_0}}{\sqrt{3}}  erfc\left[\frac{s_{2}^{3/4}}{2\sqrt{3}\pi^2}(x_{0}-\eta)^{3/2}\right]$$
$$\approx \frac{2\pi^{3/2}}{s_{2}^{3/4}(x_{0}-\eta)^{3/2}}\exp \left[f_{0}+\frac{s_{2}^{3/2}}{12\pi^4}(\eta -x_{0})^{3}\right]\Big\{1$$
\begin{equation}
+\sum_{m=1}^{\infty}{\frac{(-1)^{m}(2m)!}{2^{m}m!}\left[\frac{6\pi^4}{s_{2}^{3/2}(\eta -x_{0})^{2}}\right]^{m}}\Big\}, -\eta\gg 1.
\label{aP}
\end{equation}
The asymptotics at the condensed, right wing of the critical region, $\eta \gg 1$, follows from the formula $P_{\eta}^{(univ)}=1-\int_{\eta}^{\infty}{\rho_{x}^{(univ)}dx}$ and Eqs.~(\ref{x1})-(\ref{sj1}),(\ref{rap}):
$$P_{\eta}^{(univ)}\approx 1-e^{s_{0}'-s_{0}''}\Big[\frac{\Gamma(6,\eta_{1})}{5!}+\frac{s_{2}'}{12}\Gamma(4,\eta_{1})-\frac{s_{3}'}{6}\Gamma(3,\eta_{1})$$
\begin{equation}
+\left(\frac{s_{2}'^2}{2}+s_{4}'\right)\frac{\Gamma(2,\eta_{1})}{4}-\left(\frac{s_{2}'s_{3}'}{6}+\frac{s_{5}'}{5}\right)e^{-\eta_{1}}\Big],
\label{Pap}
\end{equation}
where $\eta_{1} = \sqrt{s_2}\eta+6-s_{0}'$ and $\Gamma(n,y) =\int_{y}^{\infty}{x^{n-1}e^{-x}dx}$ is the incomplete gamma function. We can use its asymptotics \cite{a}, $\Gamma(n,y) \sim y^{n-1}e^{-y}\left[1+(n-1)/y+(n-1)n/y^{2}+\ldots\right]$, to simplify the asymptotics (\ref{Pap}) as follows
$$P_{\eta}^{(univ)}\approx 1-e^{-\eta_1}\Big[\frac{\eta_{1}^5}{5!}+\frac{\eta_{1}^4}{4!}+\frac{1}{4}\Big(1+\frac{s_{2}'}{3}\Big)\eta_{1}^{3}$$
$$+\frac{1}{2}\Big(\frac{7}{2}+\frac{s_{2}'}{2}-\frac{s_{3}'}{3}\Big)\eta_{1}^{2}+\Big(14+\frac{s_{2}'}{2}-\frac{s_{3}'}{3}+\frac{s_{2}'^2}{8}+\frac{s_{4}'}{4}\Big)\eta_{1}$$
\begin{equation}
+126+\frac{5}{2}s_{2}'-s_{3}'+\frac{s_{2}'^2}{8}+\frac{s_{4}'}{4}-\frac{s_{2}'s_{3}'}{6}-\frac{s_{5}'}{5}\Big].
\label{Pap1}
\end{equation}

\begin{figure}
\center{\epsfig{file=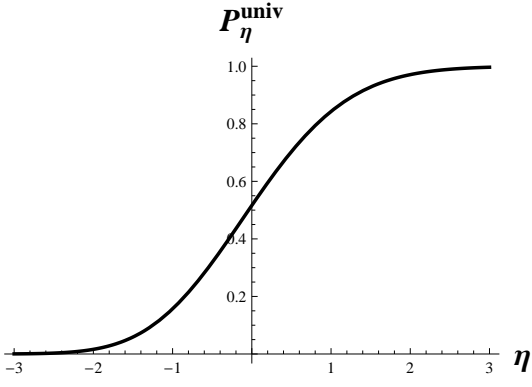,width=7cm}}
\caption{Universal cumulative distribution function $P_{\eta}^{(univ)}$, Eq.~(\ref{Puniv}), as a function of $\eta = (N - N_{c})/\sigma^{(\infty)}$ in the critical region.} 
\end{figure}

     The accuracy of all these approximations (\ref{PDHG}), (\ref{PPC}), (\ref{aP}), (\ref{Pap}), and (\ref{Pap1}) is excellent even at the borders of their validity $|\eta |\sim 3$, so that they have quite good overlapping and, hence, cover the whole critical region, i.e., all infinite interval of the universal variable $\eta \in (-\infty , \infty)$. The exact universal cumulative distribution function $P_{\eta}^{(univ)}$ is plotted in Fig. 12. 

\section{XIII. UNIVERSAL STRUCTURE OF THE AVERAGE ENERGY IN THE CRITICAL REGION}

     The average energy of the ideal Bose gas in the canonical ensemble $\bar{E}= \sum_{\vec{k}\neq 0}{\epsilon_{\vec{k}}\bar{n}_{\vec{k}}}$ can be represented as a sum of the average energy of the system of the unconstrained noncondensate excitations 
\begin{equation}
\bar{E}^{(\infty)}=\sum_{\vec{k}\neq 0}{\epsilon_{\vec{k}}/(e^{\epsilon_{\vec{k}}/T}-1)}=\frac{\epsilon_1}{T}\sum_{\vec{q}\neq 0}{\frac{q^2}{e^{q^{2}\epsilon_{1}/T}-1}}
\label{EI}
\end{equation}
and the contribution from the constrained nonlinearity,
\begin{equation}
\bar{E} = \bar{E}^{(\infty)} + T^2 \frac{\partial}{\partial T} \ln \left\langle \theta (N-\hat{n})\right\rangle^{(\infty)},
\label{ET}
\end{equation}
that follows from Eq.~(\ref{EC}) and the similar representation of the Gibbs free energy in Eq.~(\ref{GibbsT}). Using relation
\begin{equation}
\partial N_v /\partial T = (3/2)N_{v}/T
\label{dNvdT}
\end{equation}
for the trap-size parameter, Eq.~(\ref{Nv}), we can rewrite Eq.~(\ref{ET}) in the equivalent form
\begin{equation}
\frac{\bar{E}}{T} = \frac{\bar{E}^{(\infty)}}{T} + \frac{3}{2}N_{v} \frac{\partial}{\partial N_v} \ln \left\langle \theta (N-\hat{n})\right\rangle^{(\infty)}.
\label{ETT}
\end{equation}

     The second, constraint nonlinearity term is responsible for the fine structure of the average energy in the critical region. However, in the thermodynamic limit ($N_v \to \infty$) it is only of the order of $N_{v}^{1/3}$, that is much less than the trivial, independent on the number $N$ of atoms in the trap first term which has much larger thermodynamic-limit value
\begin{equation}
\frac{\bar{E}^{(\infty)}}{T} = \frac{3\zeta (5/2)}{2 \zeta (3/2)} N_v , \quad N_v \to \infty .
\label{E1}
\end{equation}

     Therefore, for the purpose of the analysis of the universal structure of the average energy in the critical region we should choose the infinitely large constant $\bar{E}^{(\infty)}/T$ to be a reference level and introduce a critical function by scaling the average energy per unit temperature to the zeroth order in $N_v$ function as follows
\begin{equation}
F_{E}=\frac{1}{\sqrt{\sigma^{(\infty)}}}\left(\frac{\bar{E}}{T}-\frac{\bar{E}^{(\infty)}}{T}\right)=\frac{3N_v}{2\sqrt{\sigma^{(\infty)}}}\frac{\partial\ln P_{N}^{(\infty)}}{\partial N_v}.
\label{ECF}
\end{equation}
In the thermodynamic limit ($N_v \to \infty$) we have $\partial /\partial N_{v} \approx -(1/\sigma^{(\infty)})\partial /\partial \eta$ and, using an obvious relation $\partial (\ln P_{\eta}^{(\infty)})/\partial \eta = \rho_{\eta}^{(\infty)}/P_{\eta}^{(\infty)}$, we find that the critical function $F_E$ in Eq.~(\ref{ECF}) converges, with increasing trap-size parameter $N_v$, to the following universal function of the scaled average energy per unit temperature
\begin{equation}
F_{E}^{(univ)}(\eta)= -\frac{3\pi^{3/2}\zeta (3/2)\rho_{\eta}^{(univ)}}{2s_{2}^{3/4}P_{\eta}^{(univ)}} ,
\label{Euniv}
\end{equation}
\begin{figure}
\center{\epsfig{file=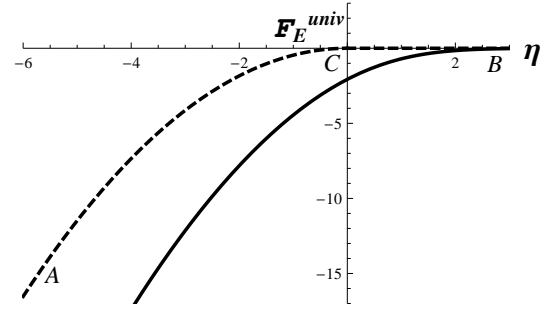,width=7cm}}
\caption{Universal function $F_{E}^{(univ)}(\eta)$, Eq.~(\ref{Euniv}), of the scaled average energy per unit temperature as a function of $\eta = (N - N_{c})/\sigma^{(\infty)}$ in the critical region. The inverse parabola AC continued by the horizontal line CB represents the prediction of the standard Landau mean-field theory.} 
\end{figure}
which is again explicitly given by the analytical formulas for the universal probability distribution in Eqs.~(\ref{rhouniv}), (\ref{rhox}), (\ref{rhounivKummer}), (\ref{rhounivPC}), (\ref{arhox}), (\ref{rap}), and Eqs.~(\ref{Puniv}), (\ref{PDHG}), (\ref{PPC}), (\ref{aP}), (\ref{Pap}). It is plotted in Fig. 13. In the deeply condensed region $\eta \to \infty$, i.e. $N-N_{c} \gg \sigma^{(\infty)}$, it tends to zero that corresponds to the constant energy per unit temperature given in Eq.~(\ref{E1}). That saturation of the average energy at the level (\ref{E1}) is well-known in the BEC thermodynamics \cite{LLV,Pathria,Ziff} and originates from the fact that all extra atoms loaded into the trap in excess of the critical number of atoms $N_{c}\approx N_v$ go into the condensate, i.e., into the ground level with zero energy, and, therefore, do not contribute to the average energy of the system.

     The inverse-parabola asymptotics of the universal function 
\begin{equation}
F_{E}^{(univ)}(\eta) \approx -\frac{3\zeta (3/2)s_{2}^{3/4}}{8\pi^{5/2}}\eta^{2},\quad \eta \to -\infty ,
\label{aEuniv}
\end{equation}
in the opposite limit of very high temperature and very small number of atoms in the trap, that follows from the asymptotics of $\rho_{\eta}^{(univ)}$ given in Eq.~(\ref{arhox}), also exactly matches the known from the grand-canonical-ensemble approximation asymptotics of the average energy \cite{Wang2004,LLV,Pathria,Ziff}
\begin{equation}
\bar{E} \approx \frac{3\zeta (\frac{5}{2})}{2\zeta (\frac{3}{2})}TN_{c}-T\sqrt{\sigma^{(\infty)}}\frac{3\zeta (\frac{3}{2})s_{2}^{3/4}}{8\pi^{5/2}}\eta^{2},\quad \eta \to -\infty .
\label{aE}
\end{equation}
Thus, similar to the BEC order parameter in Eq.~(\ref{n0eta}), the average energy also has the universal fine structure in the critical region. This smooth structure was not resolved by the standard mean-field theory that predicted only asymptotics AC and BC in Fig. 13 in the grand-canonical-ensemble approximation. 

     Finally, we stress again that the obtained universal functions of the Gibbs free energy (\ref{Gibbsuniv}) and the average energy (\ref{Euniv}) contain no any physical parameters and, hence, are pure mathematical, truly universal functions.

\section{XIV. HEAT CAPACITY OF A MESOSCOPIC IDEAL GAS IN THE CANONICAL ENSEMBLE}

     The specific heat, that is the heat capacity per particle, is the mostly often addressed thermodynamic quantity in the theoretical studies of BEC phase transition for it is directly measured in experiments and has a subtle structure near the critical point resembling the Greek letter $\lambda$ (hence the name $\lambda$-point for the critical point). That structure originates due to contribution of the second derivative of the Gibbs free energy to the specific heat which is defined via a derivative of the average energy as per Eq.~(\ref{EC}). The $\lambda$-point structure of the specific heat of the ideal gas can be described by a universal function that surprisingly was not yet analytically found despite many studies since the original works by Bose and Einstein in 1924. Here we present an explicit analytical solution to that long-standing problem. 

\subsection{A. Grand-Canonical-Ensemble Approximation in the Thermodynamic Limit}

  First of all, we briefly summarize what are the related results of the standard in the statistical physics mean-field theory in the grand-canonical-ensemble approximation following textbooks \cite{LLV,Pathria} and a very useful paper \cite{Wang2004}. Similar to the classical gas, for very high temperatures, $T\gg T_c$, or small numbers of atoms in the trap, $N\ll N_c$, the specific heat is a constant equal to 3/2, i.e., $C_{V}=(3/2)N$. In the condensed region, when $T<T_c$ or $N>N_c$, the heat capacity is an independent on the number of atoms in the trap constant 
\begin{equation}
C_{Vc}=\frac{\partial}{\partial T} \sum_{\vec{k}\neq 0}\frac{\epsilon_{\vec{k}}}{e^{\epsilon_{\vec{k}}/T}-1}\equiv \left(\frac{\epsilon_1}{T}\right)^{2}\sum_{\vec{q}\neq 0}{\frac{q^{4}e^{q^{2}\epsilon_{1}/T}}{(e^{q^{2}\epsilon_{1}/T}-1)^2}},
\label{CVc}
\end{equation}
that tends in the thermodynamic limit to a value 
\begin{equation}
C_{Vc} \to \frac{15\zeta (5/2)}{4\zeta (3/2)}N_{c},\quad N>N_{c}, N_{v} \to \infty ,
\label{CVcT}
\end{equation}
which is determined by the critical number of atoms $N_c$. The latter tends to the trap-size parameter $N_v$ in accord with Eqs.~(\ref{Nc}) and (\ref{Nv}). The thermodynamic-limit value of the specific heat at the critical point
\begin{equation}
\frac{C_{Vc}}{N_c} = \frac{15\zeta (5/2)}{4\zeta (3/2)} \approx 1.92567 > 3/2
\label{Ccr}
\end{equation}
is larger than the classical gas specific heat 3/2 and, hence, with decrease of the number of atoms in the trap the specific heat should decrease from 1.92567 to 1.5. With increase of the number of atoms above critical value at $N>N_c$, the specific heat also decreases, namely, inversely proportional to N as per Eq.~(\ref{CVc}). Thus, the standard grand-canonical-ensemble approximation predicts the $\lambda$-point structure of the specific heat as follows \cite{Wang2004}
\begin{equation}
C_{V}/N \approx \left[(15/4)\zeta (5/2)/\zeta (3/2)\right]N_{c}/N ,\quad N > N_{c},
\label{Cgce}
\end{equation}
$$\frac{C_{V}}{N}\approx 2c_{1}-c_{2}+(c_{1}-c_{2})\frac{N}{N_c}+\left(2c_{2}-\frac{4c_1}{3}\right)\frac{N^2}{N_{c}^2}, N<N_{c},$$
$$c_{1}=(9/4)\zeta(5/2)/\zeta(3/2),\quad   c_{2}=(3/8)(\zeta(3/2))^{2}/\pi .$$ 
However, that result does not resolve the fine universal structure of the $\lambda$-point reducing it, instead, only to a discontinuity with a jump $(9/(8\pi))(\zeta(3/2))^{2}\sigma^{(\infty)}/N_{v}$ in its first derivative with respect to $\eta=(N-N_{c})/\sigma^{(\infty)}$ at the critical point $\eta =0$. In the critical region the prediction of the grand-canonical-ensemble approximation (\ref{Cgce}) in the thermodynamic limit is reduced to 
\begin{equation}
\frac{C_{V}}{N}\approx \frac{15\zeta (5/2)}{4\zeta (3/2)}\left(1-\frac{\eta\sigma^{(\infty)}}{N_{v}}\right),\quad \eta >0,
\label{aCgce}
\end{equation}
$$\frac{C_{V}}{N}\approx\frac{15\zeta (5/2)}{4\zeta (3/2)} +\left[\frac{9(\zeta(3/2))^{2}}{8\pi}-\frac{15\zeta (5/2)}{4\zeta(3/2)}\right]\frac{\eta\sigma^{(\infty)}}{N_{v}}, \eta <0,$$
but, in fact, gives correctly only the slopes of the asymptotics and the maximum value in the main order of magnitude $\sim N_{v}^{0}$. It does not predict correctly the shifts of the asymptotics along the axis $\eta$ and the shift of the maximum value as well as the whole fine structure in the vicinity of the $\lambda$-point which all are quantities of the next order of magnitude $\sim N_{v}^{-1/3}$. 

\subsection{B. Universal $\lambda$-Point Structure}

     In accord with Eqs.~(\ref{EC}), (\ref{dNvdT}), and (\ref{ETT}), the heat capacity is equal to
\begin{equation}
C_{V}=C_{Vc}+\frac{15N_v}{4}\frac{\partial\ln P_{N}^{(\infty)}}{\partial N_v}+\frac{9N_{v}^2}{4}\frac{\partial^{2}\ln P_{N}^{(\infty)}}{\partial N_{v}^2},
\label{CNv}
\end{equation}
where the standard heat capacity in the condensed regime $C_{Vc}$ is a constant given in Eq.~(\ref{CVc}). The result in Eq.~(\ref{CNv}) means that the subtle $\lambda$-point structure of the heat capacity is all due to the constraint nonlinearity contribution coming via the first and second derivatives of the logarithm of the cumulative distribution function $P_{N}^{(\infty)}$, Eq.~(\ref{step}), with respect to the trap-size parameter $N_v$. 

     Let us now calculate the heat capacity in the thermodynamic limit in the critical region where the function $P_{N}^{(\infty)}$ and the unconstrained probability distribution $\rho_{N}^{(\infty)}$ are the universal functions of the only one universal variable $\eta =(N-N_{c})/\sigma^{(\infty)}$ as it was discussed in Sections IV-VII, XII, and XIII. Hence, we can evaluate the derivatives in Eq.~(\ref{CNv}) via the derivatives with respect to the universal variable $\eta$ and the relations
$$\frac{\partial\eta}{\partial N_v}=-\frac{1}{\sigma^{(\infty)}}-\frac{2(\eta-\eta_{c})}{3N_v},$$
\begin{equation}
\frac{\partial^{2}\eta}{\partial N_{v}^2}=\frac{4}{3N_v}\left[\frac{1}{\sigma^{(\infty)}}+\frac{5(\eta-\eta_{c})}{6N_v}\right],
\label{detadNv}
\end{equation}
where $\eta_{c}=(N_{v}-N_{c})/\sigma^{(\infty)}\ll N_{v}^{1/3}$ is very slowly (compared to $N_{v}^{1/3}$) growing function of $N_v$ that describes how close is the exact critical number of atoms $N_c$ in Eq.~(\ref{Nc}) to its continuous-limit approximation $N_v$ in Eq.~(\ref{Nv}). The result is
\begin{equation}
C_{V}=C_{Vc}-\frac{3N_{v}\partial\ln P_{\eta}^{(\infty)}}{4\sigma^{(\infty)}\partial \eta}+\left[\frac{3N_{v}}{2\sigma^{(\infty)}}+\eta -\eta_{c}\right]^{2}\frac{\partial^{2}\ln P_{\eta}^{(\infty)}}{\partial \eta^2},
\label{CVeta}
\end{equation}
where obviously 
\begin{equation}
\frac{\partial\ln P_{\eta}^{(\infty)}}{\partial \eta}=\frac{\rho_{\eta}^{(\infty)}}{P_{\eta}^{(\infty)}},    \frac{\partial^{2}\ln P_{\eta}^{(\infty)}}{\partial \eta^2}=\frac{\partial\rho_{\eta}^{(\infty)}/\partial\eta}{P_{\eta}^{(\infty)}}-\left(\frac{\rho_{\eta}^{(\infty)}}{P_{\eta}^{(\infty)}}\right)^{2}.
\label{dln}
\end{equation}
In Eq.~(\ref{CVeta}) the first term $C_{Vc}$ has the highest order of magnitude $\sim N_v$. The next to it is the term $(9/4)(N_{v}/\sigma^{(\infty)})^2$ which is of order of $N_{v}^{2/3}$. All other terms are infinitesimally small and do not contribute to the universal function of the specific heat. Thus, the final formula for the heat capacity in the critical region is
\begin{equation}
C_{V}=C_{Vc}+\left(\frac{3N_{v}}{2\sigma^{(\infty)}}\right)^{2}\frac{\partial^{2}\ln P_{\eta}^{(\infty)}}{\partial \eta^2} ,
\label{CVeta1}
\end{equation}
where $C_{Vc}$ is given by the discrete sum in Eq.~(\ref{CVc}).

     In order to resolve the fine universal structure of the $\lambda$-point we have to scale ("magnify") properly a deviation of the specific heat from its value at the critical point. This can be achieved by means of the following critical function
\begin{equation}
F_{C} = \sqrt{\sigma^{(\infty)}}\left(\frac{C_{V}}{N}-\frac{C_{V}(N=N_{c})}{N_c}\right) ,
\label{FCV}
\end{equation}
where $C_V$ is given by Eq.~(\ref{CNv}). With increasing trap-size parameter $N_v$, that critical function quickly converges to the following universal function of the scaled specific heat
\begin{equation}
F_{C}^{(univ)}(\eta)=\frac{9\zeta(3/2)}{4s_{2}^{3/4}\pi^{-\frac{3}{2}}}\frac{\partial^{2}\ln P_{\eta}^{(univ)}}{\partial \eta^2}-c_{0}-\frac{15\zeta(\frac{5}{2})s_{2}^{3/4}\eta}{4\pi^{\frac{3}{2}}(\zeta(\frac{3}{2}))^2},
\label{FCVuniv}
\end{equation}
where the value of the constant 
\begin{equation}
c_{0}=\frac{9\pi^{3/2}\zeta(3/2)}{4s_{2}^{3/4}}\left[\frac{\partial^{2}\ln P_{\eta}^{(univ)}}{\partial \eta^2}\right]_{\eta =0}\approx -2.79
\label{C0}
\end{equation}
\begin{figure}
\center{\epsfig{file=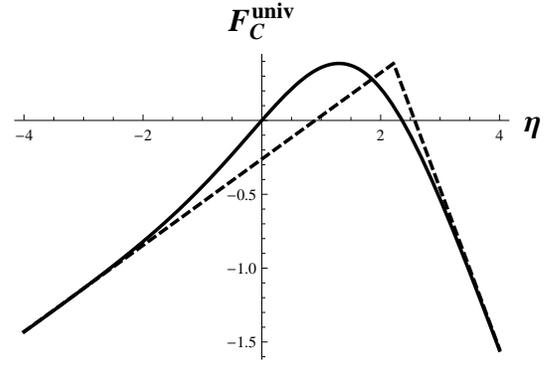,width=7cm}}
\caption{Universal function $F_{C}^{(univ)}(\eta)$, Eq.~(\ref{FCVuniv}), of the scaled specific heat as a function of $\eta = (N - N_{c})/\sigma^{(\infty)}$ in the critical region. The slopes of its linear asymptotics on both wings of the critical region match the ones predicted by the standard grand-canonical-ensemble approximation.} 
\end{figure}
is determined by the exact function $\rho_{\eta}^{(univ)}$, Eq.~(\ref{rhouniv}). The universal function $F_{C}^{(univ)}(\eta)$ is again explicitly given by the analytical formulas for the universal probability distribution in Eqs.~(\ref{rhouniv}), (\ref{rhox}), (\ref{rhounivKummer}), (\ref{rhounivPC}), (\ref{arhox}), (\ref{rap}), and Eqs.~(\ref{Puniv}), (\ref{PDHG}), (\ref{PPC}), (\ref{aP}), (\ref{Pap}) via Eq.~(\ref{dln}). It is plotted in  Fig. 14 and, as it should be, has a familiar $\lambda$-shape. The value of the specific heat at the critical point $\eta =0$ differs from the standard value in Eq.~(\ref{Ccr}) by corrections of order $1/\sqrt{\sigma^{(\infty)}}\sim N_{v}^{-1/3}\ll 1$,
\begin{equation}
\frac{C_{V}(N=N_{c})}{N_c} = \frac{C_{Vc}}{N_{c}} + \frac{c_0}{\sqrt{\sigma^{(\infty)}}}.
\label{CVcr}
\end{equation}
     
     The most remarkable consequence of the existence of the universal functions for all statistical and thermodynamic quantities in the critical region, including the one for the specific heat in Eq.~(\ref{FCVuniv}) and all others discussed in the previous sections, is their dependence on only one universal variable $\eta =(N-N_{c})/\sigma^{(\infty)}$ that means definite self-similarity of all curves for a given statistical or thermodynamic quantity as a function of any given physical variable at fixed values of other parameters of the system. For example, we can immediately find and plot the dependence of the specific heat on temperature $t=T/T_c$ in the critical region at any particular values of the number of atoms in the trap $N$ and volume $V$ as the function $F_{C}^{(univ)}(\eta(t))/\sqrt{\sigma^{(\infty)}(t)}$ where the self-similar variable $\eta (t)=(N-N_{c}(t))/\sigma^{(\infty)}(t)$ depends on temperature via the dependences of the critical number $N_c(t)$, Eq.~(\ref{Nc}), and the dispersion $\sigma^{(\infty)}(t)$, Eq.~(\ref{sigma}), on the reduced temperature $t=T/T_c$. It is especially simple if one does not need to take into account relatively small finite-size effects of deviation of the values of the exact discrete sums for the critical number $N_c$ in Eq.~(\ref{Nc}) and for the dispersion $\sigma^{(\infty)}$ in Eq.~(\ref{sigma}) from their continuous approximations in Eqs.~(\ref{Nv}) and (\ref{sigmainfinity}), respectively. Then we have a simple self-similarity 
$$N_{v}(t) = Nt^{3/2},\quad \sigma^{(\infty)}(t) = [s_{2}^{1/2}N^{2/3}/(\pi (\zeta(3/2))^{2/3})]t ,$$
\begin{equation}
\eta(t) = \pi [\zeta(3/2)]^{2/3}s_{2}^{-1/2}N^{1/3}(1-t^{3/2})/t ,
\label{etat}
\end{equation}
where we assume that $N_v \gg 1$ and $|t-1| \ll 1$. It fully specifies the temperature dependence of the specific heat near its maximum (i.e., near the $\lambda$-point) as the function $F_{C}^{(univ)}(\eta(t))/\sqrt{\sigma^{(\infty)}(t)}$ for any particular values of the number of atoms in the trap $N$ as is shown in Fig. 15 for $N = 10^{4}$. That shape of the specific heat near the $\lambda$-point is the same as the one numerically calculated in \cite{Kleinert2007} from the exact recursion relation. It is worth to note that for such, still not very large number of atoms there is noticeable shift of the critical point of order $-\eta_{c}=(N_{v}-N_{c})/\sigma^{(\infty)}$. That finite-size effect is well-known from numerical calculations (see nice graphs in \cite{Kleinert2007}) and can be easily taken into account via accurate value of $N_c$ in the self-similar variable $\eta (t)$. For example, in the case of $N=N_v = 10^4$ we have $N_c \approx 8663$ and $\sigma^{(\infty)} \approx 300$, so that $-\eta_c \sim 4$ and the critical temperature shift is $\Delta t_c \sim 0.1$. That same self-similarity can be used to find and plot temperature dependences of all other statistical and thermodynamic quantities in the critical region at particular values of the number of atoms $N$ and volume $V$ from their universal functions, e.g., temperature dependence of the BEC order parameter on the basis of its universal function in Eq.~(\ref{n0eta}). The self-similarity, even in the continuous approximation (\ref{etat}), was not discovered in the large amount of papers devoted to the numerical studies of various dependences of the statistical and thermodynamic quantities at numerous possible combinations of the parameters in the finite-size systems (see, e.g., \cite{Koch06,Ziff,Kleinert2007,KKS-PRA,CNBII,Ketterle1996,BagnatoPRA1997}).

\begin{figure}
\center{\epsfig{file=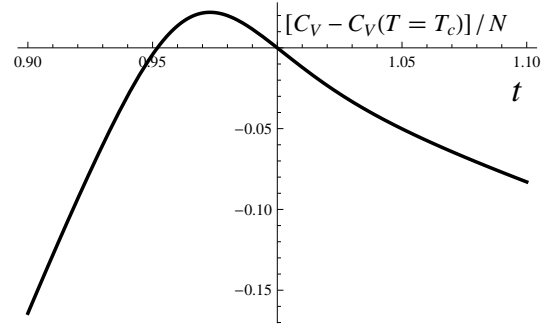,width=7cm}}
\caption{Deviation of the specific heat from its critical value, $[C_{V} - C_{V}(T=T_{c})]/N$, as a function of reduced temperature $t=T/T_c$. The graph is plotted by means of the scaled universal function $F_{C}^{(univ)}(\eta (t))/\sqrt{\sigma^{(\infty)}(t)}$ in Eq.~(\ref{FCVuniv}) and self-similar substitution (\ref{etat}) for the mesoscopic system with $N=10^4$ atoms in the trap.} 
\end{figure}
     
\subsection{C. Asymptotics at the Wings of the Critical Region}

    The only nontrivial term in the universal function of the specific heat in Eq.~(\ref{FCVuniv}) at the wings of the critical region is the second derivative $\partial^{2}\ln P_{\eta}^{(univ)}/\partial \eta^2$ that can be immediately found in accord with Eq.~(\ref{dln}) from the asymptotics of $\rho_{\eta}^{(univ)}$ in Eqs.~(\ref{arhox}) and (\ref{rap}) and of $P_{\eta}^{(univ)}$ in Eqs.~(\ref{aP}) and (\ref{Pap}). 
    
    In the noncondensed, high temperature or small number of atoms regime, i.e., at the left wing of the critical region, $-\eta\gg 1$, we find
$$\frac{\partial^{2}\ln P_{\eta}^{(\infty)}}{\partial\eta^2}\approx \frac{\sqrt{3}s_{2}^{9/4}}{8\pi^{13/2}}(\eta -x_{0})^{5/2}\left[1-\frac{2\pi^4}{s_{2}^{3/2}(\eta -x_{0})^3}\right]B$$
\begin{equation}
-\frac{3s_{2}^{3/2}}{4\pi^5}(x_{0}-\eta)B^{2},\quad -\eta\gg 1,
\label{aD2lnP}
\end{equation}   
$$B=\exp \left[\frac{s_{2}^{3/2}}{12\pi^4}(\eta -x_{0})^{3}\right]/ erfc\left[\frac{s_{2}^{3/4}}{2\sqrt{3}\pi^2}(x_{0}-\eta)^{3/2}\right].$$   
The leading term of that asymptotics yields the asymptotics of the universal function of the specific heat at $-\eta\gg 1$ as a linear function of $\eta =(N-N_{c})/\sigma^{(\infty)}$,
\begin{equation}
F_{C}^{(univ)}(\eta)\approx \frac{s_{2}^{3/4}}{\pi^{3/2}\zeta(\frac{3}{2})}\left[\frac{9(\zeta(\frac{3}{2}))^2}{8\pi}-\frac{15\zeta(\frac{5}{2})}{4\zeta(\frac{3}{2})}\right]\eta-c_{0}.
\label{aFCVuniv}
\end{equation}  

     In the condensed regime, i.e., at the right wing of the critical region, $\eta \gg 1$, the second derivative 
\begin{equation}
\frac{\partial^{2}\ln P_{\eta}^{(\infty)}}{\partial\eta^2}\approx \left[\frac{\partial\ln \alpha_{5}(\sqrt{s_2}\eta)}{\partial\eta}-\sqrt{s_2}\right]\rho_{\eta}^{(univ)} ,
\label{D2lnPap}
\end{equation}     
where functions $\rho_{\eta}^{(univ)}$ and $\alpha_{5}(s_{2}^{1/2}\eta)$ are given in Eqs.~(\ref{rap}) and (\ref{a5s}), respectively, is exponentially small. Hence, the asymptotics is determined by the last two terms in Eq.~(\ref{FCVuniv}) and again is a linear function but with a different slope,
\begin{equation}
F_{C}^{(univ)}(\eta)\approx -\frac{s_{2}^{3/4}15\zeta(5/2)}{4\pi^{3/2}(\zeta(3/2))^2}\eta - c_{0},\quad \eta \gg 1 .
\label{FCVunivap}
\end{equation}  

     For both asymptotics the slopes are shown in Fig. 14 and are exactly the same as the ones given by the standard grand-canonical-ensemble approximation in Eq.~(\ref{aCgce}). The results (\ref{aFCVuniv}) and (\ref{FCVunivap}) predict also the exact positions of these linear asymptotic lines relative to the critical point $\eta =0$ in the $\eta$-scale. These positions were not given right by the standard grand-canonical-ensemble approximation \cite{LLV,Pathria}. 

\subsection{D. Strong Finite-Size and Discreteness Effects \\
in the Asymptotics of the Specific Heat \\
in the Small Number of Atoms Regime}

     Outside the critical region, in the small number of atoms regime ($N_{c}-N \sim N_{c}$), there is no universality anymore, as is discussed in the last subsection of Sec. IV, and we must return to the general formula for the heat capacity in Eq.~(\ref{CNv}). The most interesting is the region adjacent to the end point $n=0$ (i.e., the end point $\eta_{0}$ in Eq.~(\ref{endpoint}), see Fig. 7) where the finite-size and discreteness effects are especially pronounced. We can use in that region the asymptotics (\ref{rhoBa}) of the unconstrained probability distribution $\rho_{n}^{(\infty)}$, valid for $n \ll 3\sqrt{N_v}$, to calculate explicitly the corresponding asymptotics of the heat capacity. 
     
     To simplify the analysis let us use only its leading term, that is the Poisson distribution (\ref{Poisson0}). Straightforward calculation of Eq.(\ref{CNv}) yields 
\begin{equation}
C_{V} \approx \frac{15}{4}B_{1} - \frac{(15B_{1}+9N)B_{1}^{N}e^{-B_{1}}}{4\Gamma(N+1, B_{1})} ,
\label{aCV}
\end{equation} 
that gives asymptotics of the specific heat in the thermodynamic limit, $N_{v} \gg 1$, as follows 
\begin{equation}
\frac{C_{V}}{N}\approx \frac{3}{2}+\frac{9\zeta(3/2)(N-2/3)}{4N_{v}}\approx \frac{3}{2}+\frac{9\zeta(3/2)}{4}\left(\frac{T_c}{T}\right)^{3/2},
\label{aCVN}
\end{equation} 
when $1 \ll N \ll 3\sqrt{N_v}$. This is the ultimate law with which the specific heat tends to its classical value 3/2 in the dilute high temperature ideal gas with decreasing the number of atoms in the trap ($N\to 1$) or with increasing the temperature ($T/T_{c}\to \infty$). Like the universal asymptotics (\ref{aFCVuniv}) at the left wing of the critical region, the asymptotics (\ref{aCVN}) is also a linear function of $N/N_v$, but its slope $9\zeta(3/2)/4 =5.878$ differs from the slope $9(\zeta(3/2))^{2}/(8\pi)-15\zeta(5/2)/(4\zeta(3/2)) =0.518$ given by Eq.~(\ref{aFCVuniv}) at the wing of the critical region. 

     Specific heat for the trap with only a few atoms $N = 1, 2, 3, 4, 5$ demonstrates a strong discreteness effect and can be easily found from explicit formulas (\ref{rho0}), (\ref{rho1}), (\ref{r2345}) and Eq.~(\ref{CNv}). Further details will be presented elsewhere. 

\section{XV. CRITICAL EXPONENTS AND FINITE-SIZE SCALING FUNCTIONS NEAR CRITICAL POINT: COMPARISON WITH THE RENORMALIZATION-GROUP THEORY}

     The modern theory of the second order phase transitions is based on the phenomenological renormalization-group approach and is focused on the calculation of the universal features of phase transitions, such as the critical exponents, which are the same for all phase transitions within a given universality class (see reviews \cite{Fisher1974,Fisher1986,PatPokr,LLV,Kleinert1989,Gasparini} and references therein). The analytical microscopic theory of the second order phase transitions in the mesoscopic systems yields the full quantum-statistical description of the critical fluctuations phenomena and allows us to find both universal quantities (critical exponents) and nonuniversal quantities (in particular, scaling functions and metric amplitudes) which were introduced in the renormalization-group theory for a close vicinity of the critical point. In this section, we compare the universal scaling given by the analytical microscopic theory against the finite-size scaling given by the phenomenological renormalization-group theory for the mesoscopic ideal gas in the canonical ensemble and present the explicit results for the scaling functions and critical exponents. 
     
     Let us consider the BEC phase transition in the ideal gas trapped in the three-dimensional box. It is known \cite{Fisher1972,Fisher1973,Fisher1974,Fisher1986,Pathria} that this transition belongs to the universality class of the Gaussian complex-field model (spherical model) and, hence, the correlation-length exponent $\nu$ is equal to the condensate-fraction exponent $\textit{v}$, $\nu = \textit{v} = 1$, and the specific-heat exponent is $\alpha = -1$. The BEC phase transition in the weakly interacting Bose gas belongs to the three-dimensional XY, or O(2), universality class \cite{Fisher1986,Kleinert1989,Pollock1992,Schultka1995,Wang2009}, which has different critical exponents ($\nu \approx 0.6717$ and $\alpha \approx -0.015$, see \cite{Svistunov2006,Campostrini2006,Gasparini} for the most recent numerical and experimental data) and will be discussed elsewhere. The specific-heat and correlation-length exponents are related via the hyperscaling relation $\alpha = 2 - d \nu$ and the condensate-fraction (superfluid-stiffness) and correlation-length exponents are related via the Josephson scaling relation $\textit{v} = (d-2)\nu$, so that the condensate-fraction and specific-heat exponents are also directly related, $\textit{v} = (d-2)(2-\alpha)/d$. Here $d$ is the dimensionality that is equal to 3 in the case of the box. 
     
     The renormalization-group theory analyzes the finite-size scaling in the close vicinity of the critical point $T = T_c$ on the basis of a power-law ansatz for a critical ("singular") part of a physical quantity as a function of the reduced temperature $\Delta t = (T-T_{c})/T_c$ and the size of the system $L = V^{1/3}$. In the present case of the BEC phase transition, for any physical quantity it is convenient to introduce its properly normalized value, say $y(\Delta t,L)$, which is finite in the thermodynamic (bulk) limit, $y(0,\infty) = y^{(c)}$ at $L \to \infty$, at the critical point $T = T_c$ , i.e. at $N = N_{c}$. In this case, the renormalization-group ansatz can be cast into the following form \cite{Schultka1995,Privman1984,Pollock1992,Gasparini}: 
\begin{equation}
y(\Delta t, L, N) = y^{(c)} + \left|\Delta t\right|^{\zeta_y} g_{y}\left[\frac{L}{\xi (\Delta t)}\right], \xi (\Delta t) = \xi_0 |\Delta t|^{-\nu}, 
\label{RGansatz}
\end{equation}
where $\zeta_y$ and $g_y$ are the critical exponent of the physical quantity $y$ and its universal scaling function, respectively, and $\xi(\Delta t) = \xi(\Delta t, L=\infty)$ is the correlation length $\xi(\Delta t, L) = |\Delta t|^{-\nu} f_{\xi}(L/\xi(\Delta t))$ for the infinite size system $L = \infty$.    
     
     As the examples of the physical quantity $y$, let us consider the condensate fraction ${\bar n}_{0}/N$ and specific heat $c_{V} = C_{V}/N$. It was found in the previous sections that such quantities are actually described by the universal functions of the self-similar variable $\eta = (N-N_{c})/\sigma^{(\infty)}$ in Eq.~(\ref{eta}) rather than the renormalization-group scaling variable $L/\xi(\Delta t)$. However, it is easy to see that these two variables are in fact proportional to each other in the close vicinity of the critical point for large enough systems (to the first order), 
\begin{equation}
\eta \approx -\frac{3N_v}{2\sigma^{(\infty)}}\Delta t = -\frac{3\pi [\zeta(3/2)]^{2/3}}{2\sqrt{s_2}}N_{v}^{1/3}\Delta t \propto \Delta tL , 
\label{eta-RG}
\end{equation}
since $\Delta t \equiv T/T_c -1 = (N_{v}/N)^{2/3}-1$, $N_c \approx N_v \propto L^3$ in accord with Eq.~(\ref{Nv}), and  $\sigma^{(\infty)} \propto L^{\Delta_{\sigma}}$ in accord with Eq.~(\ref{sigma}) in the large-size limit $L \to \infty$, where the scaling dimension $\Delta_{\sigma}$ of the unconstrained dispersion $\sigma^{(\infty)}$ depends on the trap and is equal to 2 for the box and to 3/2 for the harmonic trap \cite{KKS-PRL,KKS-PRA,Koch06}. Hence, the exact universal structure of any physical quantity $y$ in the critical region, which is described by the appropriate (found in the previous sections) universal function of the true universal, self-similar variable $\eta$ and can be written in the form 
\begin{equation}
y(T, L, N) = y^{(c)} + \left(\sigma^{(\infty)} \right)^{-\zeta_y /(\nu \Delta_{\sigma})} f_{y}(\eta) , 
\label{eta-ansatz}
\end{equation}
is reduced to the renormalization-group ansatz in Eq.~(\ref{RGansatz}) in the close vicinity of the critical point for large enough systems with the correlation-length critical exponent $\nu = 1$ if one identifies the product of the function $g_{y}(L/\xi(\Delta t))$ and the pre-factor $\propto (\Delta t L^{1/\nu})^{\zeta_y}$ in Eq.~(\ref{RGansatz}) with the universal function $f_{y}(\eta)$ in Eq.~(\ref{eta-ansatz}). Thus, just the knowledge of the true universal, self-similar variable in Eq.~(\ref{eta}), without any specific information on the universal functions, is enough to calculate the correlation-length exponent. 

     For the condensate fraction $y = {\bar n_0}/N$, the thermodynamic-limit value at the critical temperature is zero, $y^{(c)} \equiv \bar{n}_{0}(T=T_{c})/N \to 0$, and the analytical microscopic theory in Eq.~(\ref{n0eta}) yields the universal scaling in the form of Eq.~(\ref{eta-ansatz}),  
\begin{equation}
{\bar n}_{0}(T, L, N)/N = \left(\sigma^{(\infty)} \right)^{-\textit{v}/(\nu \Delta_{\sigma})} f_{n_{0}}(\eta) , 
\label{n0-eta-RG}
\end{equation}     
with the explicit formula for the condensate universal function as a regular function of $\eta$ and the critical exponent $\textit{v} = 1$ derived from Eq.~(\ref{sigma}). It is exactly the same value of the critical exponent that can be directly obtained via the Josephson scaling relation $\textit{v} = (d-2)\nu$ from the correlation-length critical exponent $\nu = 1$ derived from Eq.~(\ref{eta-RG}). The phenomenological renormalization-group theory uses the scaling ansatz in Eq.~(\ref{RGansatz}), i.e.
\begin{equation}
{\bar n}_{0}(\Delta t, L, N)/N =  \left|\Delta t\right|^{\textit{v}} g_{n_{0}}(L/\xi(\Delta t)) , 
\label{n0-RG}
\end{equation}
which is different from that in Eq.~(\ref{n0-eta-RG}). 
     
     For the specific heat $y = C_{V}/N \equiv c_V$, the thermodynamic-limit value at the critical temperature, $y^{(c)} = C_{Vc}/N_c$ in Eq.~(\ref{Ccr}), is not zero and the analytical microscopic theory in Eqs.~(\ref{FCV}) and (\ref{FCVuniv}) yields the universal scaling 
 \begin{equation}
c_{V}(T, L, N) =  C_{Vc}/N_c + \left(\sigma^{(\infty)} \right)^{\alpha /(\nu \Delta_{\sigma})} f_{CV}(\eta) , 
\label{cv-eta-RG}
\end{equation}     
with the explicit formula for the specific-heat universal function in Eq.~(\ref{FCVuniv}) and the critical exponent $\alpha = -1$. As it should be, the critical exponent value is exactly the same as the value derived from the hyperscaling relation $\alpha = 2 - d \nu$. The renormalization-group theory uses a different scaling ansatz in Eq.~(\ref{RGansatz}), namely, 
\begin{equation}
c_{V}(t, L, N) = C_{Vc}/N_c + \left|\Delta t\right|^{-\alpha} g_{CV}(L/\xi(t)) . 
\label{cv-RG}
\end{equation}

		In a similar way, it is straightforward to derive the universal functions and critical exponents for the Gibbs free energy, average energy as well as higher moments and cumulants of BEC fluctuations from the analytical universal functions obtained in the present paper. Note also that the critical exponents for the Gibbs free energy per particle and average energy per particle are obviously equal to $2-\alpha$ and $1-\alpha$, respectively, since the average energy and heat capacity are determined by the first and second derivatives of the Gibbs free energy with respect to temperature, respectively. 

	   For the discussed above and other physical quantities, the phenomenological renormalization-group theory does not give any explicit formulas for the universal functions, which, instead, are usually discussed on the basis of numerical, first of all Monte Carlo, simulations and some numerical fits for the few first terms in their Taylor series including corrections from some irrelevant scaling field (see, e.g., \cite{Fisher1986,Pollock1992,Schultka1995,Ceperley1997,Holzmann1999,Wang2009} and references therein). However, as we found in the previous sections, the universal functions are highly nontrivial functions even in the case of the ideal gas and the true universal, self-similar variable $\eta$ given in Eq.~(\ref{eta}) is different from the usually assumed one, $\Delta t L^{1/\nu}$. Besides, such direct simulations of the finite-size versions of the universal functions for relatively small systems are greatly subject to the finite-size effects which are difficult to separate from the universal part of the functions without knowing the universal constrain-cut-off mechanism that basically controls the critical phenomena in the second order phase transitions as is described in Sec. V. All these reasons prevented the renormalization-group approach from finding the full fine structure of the universal functions in the critical region. The analytical solution to this nontrivial problem, obtained in the present paper, becomes possible only due to first calculating analytically the universal unconstrained probability distribution of the noncondensate occupation in Eq.~(\ref{rhouniv}) and then using it for the analytical calculation of the universal functions of the particular physical quantities via the exact formulas which express the constraint-cut-off mechanism. 
     
     Thus, the analytical microscopic theory of the second order phase transitions in the mesoscopic systems yields the explicit formulas for the true universal, self-similar variable in Eq.~(\ref{eta}) and the universal functions such as the ones given in Eqs.~(\ref{n0eta}), (\ref{Meta}), (\ref{Ceta}), (\ref{Gibbsuniv}), (\ref{Euniv}), and (\ref{FCVuniv}). They describe a nontrivial structure of the whole critical region in all details, not just the linear terms in the very vicinity to the critical point which were usually discussed and numerically fitted in the renormalization-group analysis. The analytical microscopic and renormalization-group theories coincide only in the first order near the critical point where the true universal variable $\eta$ is reduced to the finite-size scaling variable $L/\xi(\Delta t) \propto \Delta tL$ via Eq.~(\ref{eta-RG}). It is worth to note that a deviation of any physical quantity, scaled with a positive critical exponent, $\zeta_y > 0$, from its bulk critical value $y^{(c)}$ in general is not zero as one could naively expect from the renormalization-group ansatz $|\Delta t|^{\zeta_y} g_{y}(L/\xi(\Delta t))$ at $\Delta t = 0$. This is the case, for instance, for the condensate fraction (critical exponent $\textit{v} = 1$) and for the specific heat capacity (critical exponent $\alpha = -1$). Indeed, according to Eq.~(\ref{eta-ansatz}), the critical exponent determines actually not the power of the reduced temperature in front of the universal function, but rather the power of scaling of the deviation of the physical quantity from its bulk critical value in terms of the dispersion $\sigma^{(\infty)}$ of BEC fluctuations which can scale with anomalous dimension in the large-size limit, $\sigma^{(\infty)} \propto L^{\Delta_{\sigma}}$. The dispersion $\sigma^{(\infty)}$ depends on the temperature and size of the trap, as well as on its shape and boundary conditions, and in the present case of the ideal gas in the box scales as $\sigma^{(\infty)} \propto N_{v}^{2/3} \propto L^2$.

\section{XVI. CONCLUSIONS AND DISCUSSIONS}

	We conclude that the probability distribution of the condensate occupation, the order parameter, and all moments and cumulants of the BEC statistics as well as the thermodynamic quantities in the canonical ideal gas for any finite-size mesoscopic system of atoms in the trap can be scaled to the appropriate regular nontrivial critical functions which resolve the structure of the BEC phase transition in the critical region and converge fast to the corresponding universal functions in the thermodynamic limit with increasing the size of the mesoscopic system (see also \cite{KKD-RQE}). We find exact analytical formulas for these universal functions in the whole critical region and their amazingly simple explicit approximations via the confluent hypergeometric and parabolic cylinder functions in the central part of the critical region. The universal scaling and structure of the BEC statistics originate from the constraint cut off of the noncondensate occupation probability distribution and are controlled by the anomalously large dispersion of the condensate fluctuations. That dispersion determines the main scale of the evolution of the noncondensate probability distribution when the number of atoms in the trap is changing around the critical value. 
	
	    We identify the constraint-cut-off mechanism as the universal reason of the strongly non-Gaussian properties of the BEC fluctuations in the critical region as well as of the nonanaliticity and all other unusual critical properties of the BEC phase transition in the ideal gas. In terms of the Landau function \cite{Goldenfeld,Sinner}, i.e., the logarithm of the probability distribution of the order parameter, all these critical features originate from the constraint nonlinearity $\theta (N-\hat{n})$, i.e. due to many-body Fock space cut off in the canonical ensemble (Sec. II), that is responsible for the infinite potential wall in the effective fluctuation Hamiltonian and makes the Hamiltonian strongly asymmetric and nonanalytical even in the ideal gas, i.e., without any interparticle interaction. 
	
	  The obtained explicit formulas for the solution to the problem of critical behavior of the statistical and thermodynamic quantities in the BEC phase transition for the mesoscopic ideal gas in the trap allows one to study analytically all dependences and asymptotics without being content with ad hoc numerical simulations and being forced to keep and present a vast number of particular graphs for numerous particular combinations of different parameters of the finite-size systems. All this huge graphical database accumulated in the numerous previous works on the mesoscopic ideal gas systems can be completely understood and classified in terms of the self-similar, universal functions, given by the derived above explicit formulas, and finite-size corrections to them. At the  same time, the obtained analytical formulas allow one to plot any statistical or thermodynamic quantity for the mesoscopic ideal gas in the trap as a function of any parameter, e.g. temperature, number of atoms or volume, in a few minutes using PC and standard code packages, like Mathematica. The finite-size corrections are also described by the formulas for the constraint-cut-off mechanism and can be taken care of via an accurate, beyond the continuous approximation, calculation of related discrete sums, such as the sums for the critical number of atoms $N_c$ in Eq.~(\ref{Nc}), dispersion $\sigma^{(\infty)}$ in Eq.~(\ref{sigma}), Gibbs free energy $F^{(\infty)}$ in Eq.~(\ref{FI}), average energy $\bar{E}^{(\infty)}$ in Eq.~(\ref{EI}), and heat capacity $C_{Vc}$ in Eq.~(\ref{CVc}).
	
	The main idea that makes it possible for us to solve analytically the problem of critical fluctuations in the ideal gas is to calculate first the universal probability distribution of the noncondensate occupation and then to use it for the analytical calculation of the universal functions for the particular physical quantities via the exact formulas which express the constraint-cut-off mechanism. The point is that this constraint is directly related, through the Noether's theorem, to the symmetry to be broken in the second order phase transition and, hence, is the main reason for the phase transition and critical phenomena themselves. The nonanalyticity imposed by the constraint cut off is so important, cannot be taken into account perturbatively, and makes the universal functions of physical quantities in the critical region so nontrivial that it is very difficult to find them directly, without explicit knowledge of the constraint-cut-off mechanism. It is worth to stress that the constraint-cut-off solution for the probability distribution in Eq.~(\ref{rhocut}) satisfies exactly to the well-known recursion relation, as is proven in Sec. V, and is the exact, rigorous solution for the BEC statistics in the ideal gas in the canonical ensemble. It is not an approximation or a model. We prove the exposed remarkable universality of the mesoscopic BEC statistics and thermodynamics also by the exact numerical simulations for a wide range of the numbers of atoms in the trap, $N < 10^5$. 

     A similar universality is valid for the ideal gas in any trap and for other than periodic boundary conditions. The particular shapes of the universal functions depend on a trap energy spectrum. The standard mean-field theory does not resolve the structure of the BEC phase transition at all. The described universality provides amazingly complete and clear picture of the BEC statistics and thermodynamics in the mesoscopic systems that makes basically unneeded numerous numerical graphs for the particular values of the parameters to which the analysis of the finite-size, mesoscopic BEC statistics and thermodynamics was reduced in the most previous papers. All previous attempts did not result in the full analytical solution to the problem in the whole critical region. Only some fragments were discussed in the literature. In particular, the coefficient $s_{2}^{3/2}/(12\pi^4)$ in front of the cubic term in the exponent of the asymptotics (\ref{arhox}) was correctly obtained in \cite{Sinner}. The finite-size corrections to the universal functions are relatively pronounced only for very small traps and numbers of atoms ($N, N_v < 10^2$) and quickly disappear with increase of the system size as is discussed in Sections IV, VI-X, and XIV. We find analytical formulas for the universal functions of the Gibbs free energy, average energy, and specific heat in the whole critical region as well as their asymptotics which match the known asymptotics far from the critical point. Hence, we solve the problem of resolving the universal structure of specific heat of the ideal gas near $\lambda$-point. 
     
     Thus, we conclude that the long-standing problem of finding the universal structure of the critical region for the ideal Bose gas has a full analytical solution. 
     
   We find the generic two-level and three-level trap models of BEC and their exact analytical solutions which allow us to analyze all details of the mesoscopic BEC statistics. As an example, we presented the detailed analysis of the BEC statistics in the box trap. The two-level and three-level trap models provide the basic blocks which, together with the property of the infrared universality of the higher-order cumulants, allows us to formulate a regular scheme for the refinement of the condensate statistics approximations in any actual traps, including correct description of all higher-order moments. Note that higher-order moments are not given accurately by the quasithermal ansatz \cite{CNBII} since it is reduced to the cut-off Gaussian statistics in the bulk limit. We consider two examples of that regular refinement scheme, namely, Poisson and Gaussian background fluctuations. It would be very interesting to find analytical solutions for more complicated examples, especially for the 3-level trap model combined with Gaussian background fluctuations. 
   
    Among the most important results of the paper are remarkably accurate analytical solutions that we find using a three-level trap model with matching the first five or four cumulants in Eqs.~(\ref{r3b}), (\ref{r3buniv}) and in Eqs.~(\ref{rho3infty}), (\ref{4cumulantsmatch}), respectively. Their thermodynamic limit yields very important analytical approximations (\ref{rhounivKummer}) and (\ref{rhounivPC}) for the universal unconstrained probability distribution $\rho_{x}^{(univ)}$ of the total noncondensate occupation for the whole central part of the critical region in terms of the confluent hypergeometric function and the parabolic cylinder function, respectively. It is amazing that precisely the same analytical formula (\ref{rhounivPC}) results also from completely different model of superposition of two-level trap and Gaussian background fluctuations solved in Sec. XI. Another, very important remark is that the universal constraint-cut-off mechanism so much dominates the origin of the strong non-Gaussian effects in the BEC fluctuations that it correctly predicts the qualitative behavior of all, including higher-order, moments in the critical region even on the basis of the simplest two-level trap model (quasithermal ansatz) or cut-off Gaussian model.  
   
   An interesting conclusion from the analysis of the higher, non-Gaussian cumulants $\kappa_m$ (first of all, asymmetry, $m=3$, and excess, $m=4$) summarized in Sec. VII is that they have much stronger dependence on the proximity to the critical point than the mean value (order parameter) and dispersion of its fluctuations. Hence, measurements of non-Gaussian features of the order parameter statistics and related to them quantities for the mesoscopic systems near the critical point is important source of information on the interactions and many-body processes. A possible experiment on such measurements of the non-Gaussian statistics near QCD critical point was discussed recently for the phase transition in the quark-gluon plasma produced in the process of collisions of the relativistic heavy ions in the accelerators \cite{StephanovPRL2009}. 
  
     It is worth to emphasize the important fact that the described above universal functions of the order parameter, all moments and cumulants as well as of all thermodynamic quantities remain regular and nontrivial in the bulk limit. Thus, a simulation of these functions for the relatively small mesoscopic systems with a finite number of atoms constitutes a very effective tool for the studies of BEC phase transition in the macroscopic systems in the thermodynamic limit. In fact, the information on the universal functions is extracted usually by fitting the simulations (such as Monte Carlo simulations) or experimental data on the finite-size systems (for the related to BEC examples, see \cite{Fisher1986,Pollock1992,Schultka1995,Ceperley1997,Holzmann1999,Svistunov2006,Campostrini2006,Wang2009,Gasparini}) to a renormalization-group, finite-size scaling ansatz that typically includes only two or a few first terms of the universal function's Taylor expansion in the close vicinity of the critical point. That procedure has certain problems in providing the form of the universal functions relatively far from the critical point, i.e. in the whole critical region, and with high enough accuracy. In this respect, the analytical formulas for the true scaling, self-similar variable in Eq.~(\ref{eta}), or Eq.~(\ref{x}), and universal functions are crucially important. As is discussed in Sec. XV, the analytical microscopic theory of the BEC and other second order phase transitions in the mesoscopic systems goes beyond the known renormalization-group theory and yields the results both for the universal (critical indices) and nonuniversal (universal functions and metric amplitudes) quantities. 
    
     The constraint-cut-off mechanism discussed in the present paper is also generic for other second order phase transitions and, in particular, works in any interacting gas. The exactly solvable Gaussian model of BEC in a degenerate interacting gas, Eq.~(\ref{gauss}), is a particular example, that also elucidates on how the unconstrained Gaussian statistical distribution $\rho_n \sim \exp \left[-H(n)/T\right]$ with vanishing higher-order cumulants ($\kappa^{(\infty)}_m = 0$ for all $m \neq 2$) results, via the constraint-cut-off mechanism, in the strongly non-Gaussian BEC fluctuations. In a general case of the interacting gas, the shapes of the universal functions for the moments and cumulants of the BEC fluctuations and the thermodynamic quantities, in addition, depend on a deformation of the statistical distribution due to a feedback of the order parameter on the quasiparticle energy spectrum and correlations. These effects can be taken into account in a quite general, nonperturbative-in-fluctuations way using a theorem on the nonpolynomial averages in statistical physics and appropriate diagram technique \cite{KochLasPhys2007,KochJMO2007}. Thus, the formulated above analytical theory of the critical phenomena and universal functions in the mesoscopic BEC statistics and thermodynamics can be directly generalized to the case of the weakly interacting gas. An appropriate analysis is outside the scope of this paper and will be published in a separate paper. Presented in this paper full analytical theory of critical fluctuations in statistics and thermodynamics of mesoscopic BEC phase transition in the ideal Bose gas sets a reference level of completeness and detailness which a theory of critical fluctuations in the interacting systems may try to make a reach for. 
     
\section{ACKNOWLEDGMENTS} 

\noindent
The support from RFBI (grant 09-02-00909-a) and Council on grants of the President of the Russian Federation for support of the leading scientific schools of the Russian Federation (grant HIII-4485.2008.2) is acknowledged.

\section{APPENDIX. \quad DERIVATION OF THE ASYMPTOTICS OF THE KUMMER'S CONFLUENT HYPERGEOMETRIC FUNCTION}

     To derive the asymptotics (\ref{asymptDHG}) we use a binomial theorem $(1-t)^{g_{1}-1}=\sum_{j=0}^{g_{1}-1}{C_{g_{1}-1}^{j}(-t)^j}$ in the integral representation (\ref{MHG}) to represent Kummer's confluent hypergeometric function as a sum of incomplete gamma functions:
\begin{eqnarray}
M(g_{1},g_{1}+g_{2},x)=\frac{e^{x}\Gamma(g_{1}+g_{2})}{\Gamma(g_{1})\Gamma(g_{2})}\sum_{j=0}^{g_{1}-1}{(-1)^{j}C_{g_{1}-1}^{j}}\nonumber\\
\times G(g_{2}+j,x),\quad G(g_{2}+j,x)=\frac{\gamma(g_{2}+j,x)}{x^{g_{2}+j}}.
\label{Mgamma}
\end{eqnarray}
Now let us use a Taylor series 
\begin{equation}
G(g_{2}+j,x) = \sum_{k=0}^{\infty}{\frac{j^k}{k!}\frac{\partial^{k}G(g_{2},x)}{\partial g_{2}^k}}
\label{Gj}
\end{equation}
and take into account that for each term of Eq.~(\ref{Gj}) with $k = 0, 1,\ldots , g_{1}-2$ the sum over $j$ in Eq.~(\ref{Mgamma}) is equal to zero. Then we find that the first nonzero term in Eq.~(\ref{Gj}), $k= g_{1}-1$, makes a leading contribution to $M(g_{1},g_{1}+g_{2},x)$ in Eq.~(\ref{Mgamma}) and neglect all higher order terms, $k>g_{1}-1$, in Eq.~(\ref{Gj}). In this way we find from Eq.~(\ref{Mgamma}) the following result
\begin{equation}
M(g_{1},g_{1}+g_{2},x)\approx \frac{e^{x}\Gamma(g_{1}+g_{2})}{\Gamma(g_{1})\Gamma(g_{2})}(-1)^{g_{1}-1}\frac{\partial^{g_{1}-1}G(g_{2},x)}{\partial g_{2}^{g_{1}-1}}.
\label{MG}
\end{equation}

     The next step is to use, first, the asymptotics of the incomplete gamma function in terms of a complementary error function \cite{a},
\begin{equation}
\gamma(g_{2},x)\approx \frac{1}{2}\Gamma(g_{2}) erfc\left(-\frac{y}{\sqrt{2}}\right),\quad y=\frac{x-g_{2}}{\sqrt{x}},
\label{agamma}
\end{equation}
and, second, an obvious asymptotics
\begin{equation}
\Gamma(g_{2})x^{-g_{2}}\approx \sqrt{2\pi /g_{2}} \exp (-x+y^{2}/2) , 
\label{aexp}
\end{equation}
both of which are valid at $x \sim g_{2} \to \infty$, to find the following asymptotics
\begin{equation}
\frac{\gamma(g_{2},x)}{x^{g_{2}}} \approx \sqrt{\frac{\pi}{2}}\frac{e^{-x}}{\sqrt{x}}\exp \left(\frac{y^{2}}{2}\right) erfc\left(-\frac{y}{\sqrt{2}}\right) . 
\label{agamma1}
\end{equation}
Then we can calculate the derivative in Eq.~(\ref{MG}), $\partial^{g_{1}-1}\left[\gamma(g_{2},x)/x^{g_{2}}\right]/\partial g_{2}^{g_{1}-1}$, by means of a well-known formula \cite{a}
\begin{equation}
\frac{d^{k}[e^{z^2}erfc(z)]}{dz^k} = (-1)^{k}2^{(k+1)/2} k! e^{z^{2}/2} D_{-k-1}(z\sqrt{2})
\label{Derfc}
\end{equation}
that completes the derivation of the asymptotics of the Kummer's confluent hypergeometric function (\ref{asymptDHG}) via the parabolic cylinder function $D_{-g_{1}}\left[(g_{2}-x)/\sqrt{x}\right]$.

\end{document}